\documentclass[letterpape,11pt]{article}
\usepackage{jheppub}

\usepackage{amsmath}
  \usepackage{bm}
\usepackage{slashed}
\usepackage{amssymb}
\usepackage{mathtools}
\usepackage{graphicx}
\usepackage{subfigure}
\usepackage{setspace}
\usepackage{sidecap}
\usepackage{color}
\usepackage{hyperref}
\usepackage{tabu} % Custom tables
\usepackage{hyperref}
\usepackage[export]{adjustbox}
\usepackage{tensor}
\usepackage{dcolumn}% Align table columns on decimal point

     \usepackage{pifont}
      \usepackage{multirow,boldline}

\newcommand{\al}{\alpha}

\newcommand{\ga}{\gamma}
\newcommand{\Ga}{\Gamma}

\newcommand{\la}{\lambda}

\newcommand{\nn}{\nonumber}

\newcommand{\ot}{\leftarrow}

\renewcommand{\(}{\left(}
\renewcommand{\)}{\right)}
\renewcommand{\[}{\left[}
\renewcommand{\]}{\right]}

\newcommand{\eq}[1]{eq.~\eqref{eq:#1}}

\newcommand{\vecb}[1]{\mbox{\boldmath $#1$}}

\newcommand{\specialcellleft}[2][l]{\begin{tabular}[#1]{@{}l@{}}#2\end{tabular}}

\title{W-boson production in TMD factorization}
\date{\today}
\author[a]{Daniel Gutierrez-Reyes}
\author[b]{Sergio Leal-Gomez}
\author[a]{and Ignazio Scimemi}
\affiliation[a]{ Universidad Complutense de Madrid (UCM), Departamento de F\'{i}sica Te\'{o}rica and IPARCOS, Plaza Ciencias~1, E-28040 Madrid, Spain}
\affiliation[b]{Wien Universit\"at, Faculty of Physics, Boltzmanngasse 5, A-1090 Vienna, Austria}
\emailAdd{dangut01@ucm.es}
\emailAdd{sergiol95@univie.ac.at}
\emailAdd{ignazios@ucm.es}
\preprint{UWThPh-2020-26}
\abstract{
  At hadron colliders, the differential cross section for $W$ production can be factorized and it is sensitive transverse momentum dependent distributions (TMD) for low boson transverse momentum. While, often, the corresponding non-perturbative QCD contributions  are extrapolated from $Z$ boson production, here we use an existing extraction (based on the code Artemide) of TMD which includes data coming from Drell-Yan and semi-inclusive deep inelastic scattering, to provide checks and predictions for the $W$ case. Including fiducial cuts with different configurations and kinematical power corrections, we consider transverse momentum dependent cross sections within several intervals of the vector boson transverse mass. We perform the same study for the $p_T^{W^-}/p_T^{W^+}$ and $p_T^Z/p_T^W$ distributions. We compare our predictions with recent extractions of these quantities at ATLAS and CMS and results from TeVatron. The results encourage a broader experimental and phenomenological work, and a deeper study of TMD for the $W$ case.
}

\begin{document}

\maketitle

%***************************************************************************************************%
\section{Introduction}
Vector boson production is very relevant at hadron colliders and its measurement has achieved an increasing precision  in the latest years
\cite{Ito:1980ev,McGaughey:1994dx,Aidala:2018ajl,Antreasyan:1981uv,Albajar:1988ka,Alitti:1990cn,Abbott:1998jy,Abbott:1999wk,Abbott:1999yd,Abazov:2007ac,Abazov:2010kn,Abazov:2020moo,Abe:1991rk,Abe:1995nm,Affolder:1999jh,Aaltonen:2012fi,Aad:2014xaa,Aad:2015auj,Chatrchyan:2011wt,Khachatryan:2015oaa,Khachatryan:2016nbe,Aaij:2015gna,Aaij:2015zlq,Aaij:2016mgv,Aaboud:2017svj}.
The experimental results  have allowed  the extraction of important Standard Model quantities like collinear parton densities and the mass of the $W$-boson~\cite{Buckley:2014ana,Harland-Lang:2014zoa,Ball:2017nwa,Boughezal:2017nla}. %~\cite{Aaboud:2017svj}
 The precise measurements of $W$ cross section and similar observables need a good determination of QCD non-perturbative inputs which can be partially extracted from neutral boson mediated processes.
Because we are interested in transverse momentum dependent observables it is mandatory to consider the
extraction of transverse momentum dependent parton distribution functions (TMD) that have used the data of the neutral bosons at low transverse momentum~\cite{DAlesio:2014mrz,Bacchetta:2017gcc,Scimemi:2017etj,Bertone:2019nxa,Bacchetta:2019sam,Scimemi:2019cmh}. 

In~\cite{Hautmann:2020cyp} it is  pointed out that the low-transverse  momentum spectrum for $Z$-boson production  at LHC cannot be fully understood without the introduction of TMD, and further  study is considered as necessary (see also \cite{Ridder:2016nkl}). The $W$-production is a natural test for the TMD factorization, and an explicit evaluation of the cross section within this approach is yet missing in the literature to our knowledge. 
On the other side there exist codes  and predictions
\cite{Catani:2009sm,Karlberg:2014qua,Gavin:2012sy,Li:2012wna,CarloniCalame:2005vc,CarloniCalame:2006zq,CarloniCalame:2007cd,Golonka:2005pn,Alioli:2008gx,Barze:2012tt,Barze:2013fru,Bernaciak:2012hj,Dittmaier:2001ay,Arbuzov:2005dd,Arbuzov:2007db,Hoeche:2014aia,Placzek:2003zg,Placzek:2013moa,Baur:1998kt,Baur:2001ze,Baur:2004ig,Alioli:2016fum,Bizon:2019zgf,Ebert:2020dfc,Becher:2020ugp}
for the $W$ spectrum that include non-perturbative TMD effects only in a limited way.

The $W$-spectrum is also interesting in order to establish some properties of the TMD like their flavor dependence.
 In  fact, out of all the extractions cited above, only \cite{Scimemi:2019cmh} can be sensitive  to some non-trivial flavor structure of TMD because it includes also data from semi-inclusive deep inelastic scattering (SIDIS)  from Hermes~\cite{Airapetian:2012ki} and Compass experiments~\cite{Aghasyan:2017ctw}, despite the fact that it analyzes only processes mediated by neutral vector bosons. 
Some preliminary work in this sense has been performed in \cite{Signori:2013mda} and later in \cite{Bacchetta:2018lna} where the authors  conclude that LHC data on $W$-boson production must be sensitive to the non-trivial flavor structure of TMD.
This statement can partially be tested in the present work. 
We also consider of major importance to well establish the flavor dependence of the TMD before the starting of the Electron Ion Collider (EIC).

The results that we present want to provide a
 set of predictions for $W$-production which include the latest information from TMD studies. 
 We revise the kinematics of $W$-boson transverse momentum distribution in order to include kinematical power suppressed terms as in \cite{Scimemi:2019cmh}  and to implement the fiducial cuts typical of these processes.
 We pay particular attention to errors in our predictions as coming from scale variations, PDF, TMD parameterization whose present knowledge is here described in detail.
 %This study  results necessary both for a precise measurement of the $W$-mass and also for the flavor-dependence problem of the TMD.
 We recall that
the TMD factorization applies to Drell-Yan (DY) type processes for values of the transverse momentum of the  vector boson ($q_T$) much lower then its virtual mass ($Q$).
The factorization theorem  concludes that  the non-perturbative parts of the $q_T$-differential cross-sections of  boson production  when $q_T\ll Q$
are included in 
 the transverse momentum dependent (TMD) distributions and, separately, 
their evolution kernel \cite{Collins:1989gx,Bacchetta:2006tn,Bacchetta:2008xw,Becher:2010tm,Collins:2011zzd,GarciaEchevarria:2011rb,Echevarria:2012js,Echevarria:2014rua,Chiu:2012ir,Vladimirov:2017ksc,Scimemi:2018xaf}.
Phenomenologically it has been found that the TMD factorization at leading order works for $q_T/ Q\lesssim 0.2$ \cite{Scimemi:2017etj,Bacchetta:2019sam,Scimemi:2019cmh} and we consider this range of validity  also here.
A theoretical estimate of this range has been provided in \cite{Balitsky:2017gis}.
% The extraction of these non-perturbative (NP) elements from data is then a major challenge for modern phenomenology~\cite{Angeles-Martinez:2015sea}. 

The theoretical  perturbative calculations for TMD distributions at small-$b$ performed in recent  years is highly significative and recently it has reached the N$^3$LO 
precision\footnote{As usual NLO stands for next-to-leading order, NNLO for next-to-next-to-leading order, and so on.}
 \cite{Gehrmann:2014yya,Echevarria:2015byo,Echevarria:2015usa,Echevarria:2016scs,Li:2016ctv,Vladimirov:2016dll,Luo:2019hmp,Luo:2019szz,Ebert:2020yqt}. Results of the same N$^3$LO already exist for the
 universal QCD anomalous dimensions \cite{Gehrmann:2010ue,Baikov:2016tgj,Moch:2017uml,Moch:2018wjh,Lee:2019zop}, so that finally one has an extremely accurate perturbative input. 
 At present the non-perturbative parts of the TMD are extracted at NNLO (which means N$^3$LL in the evolution kernel and NNLO in the matching of the TMD to PDF) so that for consistency we will use the perturbative results up to this order which are included in the code Artemide \cite{web} and the non-perturbative parameters as extracted in \cite{Scimemi:2019cmh}. 
The theoretical settings of this work are very similar to the ones of \cite{Scimemi:2019cmh} 
and we make explicit use  of the $\zeta$-prescription \cite{Scimemi:2018xaf,Vladimirov:2019bfa}. 
The main difference with respect to the neutral boson case is represented by the  particular kinematics of the $ W$-production, which we study in detail in the next sections.
Thanks to our explicit study of the leptonic tensor and  implementation of fiducial cuts we can compare our predictions with existing results from LHC (ATLAS, CMS) and TeVatron (CDF, D$\slashed 0$).
We explore  explicitly new regions of the measured phase space where TMD effects can be significative and experimentally testable, especially for di-lepton masses below the $W$-mass peak.

In order to establish some notation we start writing  the $W$ mediated reaction
%\begin{align}
%p+p\rightarrow  \ell^{\pm}+\nu+X
%\end{align}
\begin{eqnarray}
\label{eq:genreact}
h_1 h_2 \to W^+(W^-) \to l^+(l^-)+\nu_l (\bar \nu_l),
\end{eqnarray}
where $h_{1,2}$ are hadrons (typically protons and antiprotons) and $l^\pm=e^\pm, \mu^\pm$ and $\nu_l(\bar \nu_l)$ are their corresponding (anti)neutrinos.
We consider a cross section differential in the vector boson transverse mass ($m_T$) and  transverse momentum ($q_T$). The typical TMD condition $q_T\ll Q$ becomes in this case $q_T\ll m_T$,
while the relation  among $m_T$ and lepton momenta is highly non-trivial as outlined in the next sections.  For our predictions, we consider the case of LHC experiments at  $\sqrt{s}=13$ TeV, assuming that lepton cuts and fiducial cross sections are similar to the case of 
$\sqrt{s}=7$ TeV \cite{Aaboud:2017svj}.
In principle this discussion can be extended to other center of mass energies.  We consider several intervals of $m_T$ and $q_T$ which are relevant both for the TMD flavor determination and the mass of the $W$-boson.  In this sense  we discover that several intervals of these variables can be interesting for QCD studies and we investigate them in detail.  We consider also the $p_T^Z/p_T^W$ and $p_T^{W^-}/p_T^{W^+}$ observables, 
 in different ranges of the transverse invariant mass of the $W$. %We estimate the errors due to scale uncertainties and PDF. 

The paper is prepared as in the following. In sec.~\ref{sec:xsecW} we establish the notation and write the cross section of the $W$ distributions with TMD factorization  and we explain how fiducial cuts are implemented. 
The kinematical relations are here described in detail (and we are not aware of a similar detailed description in the literature).
In sec.~\ref{sec:err} we list the source of errors for the various observables examined in this paper. 
In sec.~\ref{sec:Wspeck} we provide our prediction for  $W$, $p_T^{Z}/p_T^W$, $p_T^{W^-}/p_T^{W^+}$ transverse momentum differential distributions.
In sec.~\ref{sec:comp} we compare the result of our code with existing theoretical and experimental results.
 We conclude  in sec.~\ref{sec:sum} and we provide some details of our calculations in the appendix.

%The consistent composition of all elements is made employing the $\zeta$-prescription \cite{Scimemi:2018xaf,Vladimirov:2019bfa}. The $\zeta$-prescription is essential for current and future TMD phenomenology 
%because it grants a unified approach to observables irrespectively of the order of perturbative matching. So, the collinear matching procedure that is fundamental for resummation approaches (such as in refs.~\cite{Landry:2002ix,Qiu:2000hf,Bozzi:2008bb,Becher:2010tm,Catani:2012qa,Bizon:2018foh,Bizon:2019zgf}) or
% $b^*$-like prescriptions (such as in refs.\cite{Collins:1981va,Collins:2011zzd,Aybat:2011zv,Bacchetta:2017gcc}), is considered just as  part of the model for a TMD distribution in the $\zeta$-prescription. 

%%%%%%%%%%%%%%%%%%%%%%%%%%%%%%%%%%%%%%%%%%%%%%%%%%%%%%%%%%%%%%%%%%%%%%%%%%%%%%%%
\section{$W$-boson cross section in TMD factorization}
\label{sec:xsecW}
%%%%%%%%%%%%%%%%%%%%%%%%%%%%%%%%%%%%%%%%%%%%%%%%%%%%%%%%%%%%%%%%%%%%%%%%%%%%%%%%
In this section we present the cross section for charged Drell-Yan (DY) process in TMD factorization. Because the derivation of the factorized cross section is analogous to the neutral Drell-Yan case
\cite{Collins:1989gx,Bacchetta:2006tn,Bacchetta:2008xw,Becher:2010tm,Collins:2011zzd,GarciaEchevarria:2011rb,Echevarria:2012js,Echevarria:2014rua,Chiu:2012ir,Vladimirov:2017ksc,Scimemi:2018xaf,Scimemi:2019cmh},
we omit a detailed discussion of the factorization, concentrating on the inputs relevant to us.
%We apply this framework to describe $W^\pm$ boson production through the processes
%\begin{eqnarray}
%\label{eq:genreact}
%h_1 h_2 \to W^+(W^-) \to l^+(l^-)+\nu_l (\bar \nu_l),
%\end{eqnarray}
%where $h_{1,2}$ are hadrons (typically protons and antiprotons) and $l^\pm=e^\pm, \mu^\pm$ and $\nu_l(\bar \nu_l)$ are their corresponding (anti)neutrinos.

%%%%%%%%%%%%%%%%%%%%%%%%%%%%%%%%%%%%%%%%%%%%%%%%%%%%%%%%%%%%%%%%%%%%%%%%%%%%%%%%
\subsection{Charged DY cross section with transverse variables}
\label{sec:kinvar}
%%%%%%%%%%%%%%%%%%%%%%%%%%%%%%%%%%%%%%%%%%%%%%%%%%%%%%%%%%%%%%%%%%%%%%%%%%%%%%%%
Let us consider a charged DY process as in \eq{genreact} in which two hadrons with momenta $P_1$ and $P_2$ lead to a lepton with momentum $l$ and an (anti)neutrino with momentum $l'$. We approximate the interaction among protons and the lepton-neutrino pair through the production of a $W^\pm$ boson with momentum $q=l+l'$. In this paper we do not consider masses of the initial hadrons nor masses of final-state particles, so
%\begin{eqnarray}
$
P_1^2=M_1^2 \approx 0, \;P_2^2=M_2^2 \approx 0, \; l^2 =m_l^2 \approx 0, \; l'^2=m_{\nu_l}^2 \approx 0. 
$
%\end{eqnarray}

The combination of these momenta leads to the definition of the relevant kinematical variables of the DY process
\begin{eqnarray}
s=(P_1+P_2)^2, \qquad q^2=Q^2, \qquad y=\frac{1}{2}\ln \(\frac{q^0+q^z}{q^0-q^z}\),
\end{eqnarray}
where $s$ represents the square of the center of mass energy, $Q^2$ is the invariant mass of the lepton-neutrino pair and $y$ is the rapidity of the produced $W$ boson. As the neutrino is a non-detected particle the quantity $Q^2$ is not measurable anymore and one introduces the transverse mass \cite{Bizon:2019zgf,Aad:2011fp,Aad:2014nim,Aaboud:2016btc,Aaboud:2017svj,Khachatryan:2016nbe} as a measurable quantity in processes with invisible particles in final state as
 \begin{eqnarray}
\label{mt}
m_T^2=2|l_T||l_T'| (1-\cos \phi_{l\nu}),
\end{eqnarray}
where $l_T$ and $l_T'$ are the transverse parts of the lepton and neutrino momenta respectively and $\phi_{l\nu}$ is the relative angle between both particles. 

The cross section of a general DY process mediated by a gauge-boson $G$ can be written as in \cite{Scimemi:2019cmh}
\begin{eqnarray}
\label{eq:xsec1}
d\sigma=\frac{\alpha_{\rm{em}}^2}{2s}d^4 q \sum_{GG'} L_{GG'}^{\mu \nu} W_{\mu\nu}^{GG'} \Delta_G(q) \Delta^*_{G'}(q),
\end{eqnarray}
where $\alpha_{\rm{em}}=e^2/4\pi$, being $e$ the electron charge. This cross section is written in terms of a product of a lepton tensor $L_{GG'}^{\mu \nu}$ and a hadron tensor $ W_{\mu\nu}^{GG'}$ defined as
\begin{eqnarray}
L_{\mu\nu}^{GG'}&=&e^{-2} \langle 0 | J_\mu^G (0) |l,l' \rangle \langle l,l' | J_\nu^{G'\dagger}(0) |0\rangle, \\
W_{\mu\nu}^{GG'}&=&e^{-2} \int \frac{d^4 x}{(2\pi)^4} e^{-i(x \cdot q)}\sum_X \langle P_1, P_2 | J_\mu^{G\dagger} (x) |X \rangle \langle X| J_\nu^{G'}(0) |P_1,P_2 \rangle,
\end{eqnarray}
where $J_\mu^G$ is the current for the production of a gauge boson G.

On the other hand, $\Delta_G$ is the (Feynman) propagator of the gauge boson, in this case $G=W$ and
\begin{eqnarray}
\label{prop}
\Delta_G(q)=\frac{1}{Q^2-M_W^2+i \Ga_W M_W} \delta_{GW},
\end{eqnarray}
where $M_W$ and $\Gamma_W$ are the mass and width of the $W$ boson given in \cite{Tanabashi:2018oca}, respectively.  We do not include electroweak (EW) corrections, which are however calculated in \cite{Kuhn:2007qc}.

%The principal difference between neutral and charged boson cross section is that in the second case we need it as differential in  the boson transverse mass. 
The $W$-production is usually expressed in terms of the transverse mass.
The relation between the invariant and  transverse mass in a general case is non-trivial, so we describe it here in detail.
We have
\begin{eqnarray}
Q^2=m_T^2+f(l,l'),
\end{eqnarray}
and
 %of the cross section of a DY process with invisible particles in the final state respect to the neutral case in which a pair of detectable leptons is produced is that in the first case the cross section should be rewritten in terms of the transverse mass. 
 %So, we find a relation between the invariant mass of the lepton-neutrino pair invariant mass and the transverse mass in terms of a function of lepton and neutrino momentum i.e. $Q^2=m_T^2+f(l,l')$. This function can be written as
\begin{eqnarray}
f(l,l')=2 \[(l_T^2+l_z^2)^{1/2}(l_T'^2+l_z'^2)^{1/2}-l_zl_z'\]-2|l_T| |l_T'|\ .
\end{eqnarray}
In order to  include this function in the cross section we introduce the identity
\begin{eqnarray}
\label{eq:1mt}
1=\int d m_T^2 \, \delta (Q^2-m_T^2-f(l,l')),
\end{eqnarray}
and the cross section in \eq{xsec1} can be rewritten as
\begin{eqnarray}
\label{xsec1W}
d\sigma=\frac{\alpha_{\rm{em}}^2}{2s}d^4q \, dm_T^2 \sum_{GG'} L_{GG'}^{\mu \nu} W_{\mu\nu}^{GG'} \Delta_G(q) \Delta^*_{G'}(q).
\end{eqnarray}
Note that the Dirac delta in \eq{1mt} is relocated inside the definition of the lepton tensor which simplifies the practical evaluation of the integral. 

The next two sections are devoted to the factorization of the hadronic tensor and the final expression for the lepton tensor affected by transverse mass and fiducial cuts. This information will lead us to the final form of the cross section of charged DY within TMD factorization.
%%%%%%%%%%%%%%%%%%%%%%%%%%%%%%%%%%%%%%%%%%%%%%%%%%%%%%%%%%%%%%%%%%%%%%%%%%%%%%%%
\subsection{Factorization of the hadronic tensor}
\label{sec:kinvar}
%%%%%%%%%%%%%%%%%%%%%%%%%%%%%%%%%%%%%%%%%%%%%%%%%%%%%%%%%%%%%%%%%%%%%%%%%%%%%%%%
In this paper we use the unpolarized part of the hadron tensor from \cite{Scimemi:2019cmh} neglecting the masses of both hadrons. Thus, for a generic gauge boson $G$ we omit power suppressed higher-twist TMD contributions and we write
\begin{eqnarray}
\label{eq:Wten}
W_{\mu\nu}^{GG'}&&=-\frac{g_T^{\mu\nu}}{\pi N_c} |C_V(Q^2,\mu)|^2 \sum_{ff'} z_{ff'}^{GG'} \int \frac{d^2 \vecb b}{4\pi} e^{i (\boldsymbol{q} \cdot \boldsymbol {b})} f_{1,f\ot q}(x_1,\vecb b,\mu,\zeta) f_{1,f'\ot q}(x_2,\vecb b,\mu,\zeta),
\nn\\
&&
\end{eqnarray}
where $C_V$ is the matching coefficient for vector current to collinear/anti-collinear vector and $z_{ff'}^{GG'}$ are the EW factors that will be defined later. On the other hand $f_1$ is the unpolarized TMDPDF defined as
\begin{align}
f_{1,f\ot h}&(x,\vecb b,\mu,\zeta)=\\
&\int \frac{d\lambda}{2\pi} e^{-i x\lambda p^+} \sum_X \langle h(p)|\bar q (n\la + \vecb b ) W_n^\dagger (n\la + \vecb b) \frac{\ga^+}{2} |X \rangle \langle X | W_n(0) q(0) |h(p) \rangle \nn,
\end{align}
where $p$ is the momentum of the hadron, $W_n(x)$ is a Wilson line rooted at $x$ and pointing along vector $n$ to infinity. The momentum fraction $x$ is defined in DY kinematics as
\begin{align}
x_{1,2}=\frac{\sqrt{Q^2+q_T^2}}{\sqrt{s}}e^{\pm y}.
\end{align}

 In the small-$b$ limit, the TMDPDF can be re-factorized in terms of matching coefficients (calculated up to NNLO in \cite{Echevarria:2016scs}) and integrated PDFs. 
 In order to write a complete TMD, we consider also
  a function modeling non-perturbative effects that are not included into collinear PDFs 
\begin{align}\label{eq:tmd1}
f_{1,f\ot h}&(x,\vecb b,\mu,\zeta)= C(x,\vecb b,\mu,\zeta) \otimes f_1(x,\mu) f_{NP}(x,\vecb b),
\end{align}
where the symbol $\otimes$ represents the Mellin convolution in the $x$ variable. We use the ansatz for $f_{NP}$ suggested in \cite{Scimemi:2019cmh}
\begin{align}
\label{eq:fnp}
f_{NP}(x,\vecb b)=\exp \(-\frac{\la_1(1-x)+\la_2 x +x(1-x) \la_5}{\sqrt{1+\la_3 x^{\la_4}\vecb b^2}}\),
\end{align}
where the parameters $\la_1, \text{...}, \la_5$ are extracted from a combined DY+SIDIS fit to data in \cite{Scimemi:2019cmh}. 
In the TMD parameterization of eq.~\ref{eq:tmd1} the flavor dependence is in principle contained in the parameters $\lambda_i$ and the PDF.
The extraction of  \cite{Scimemi:2019cmh} has found that, given actual SIDIS data it is sufficient to count flavor dependence  in PDF only. While we make predictions using the values
of $\lambda_i$ as in \cite{Scimemi:2019cmh}, it is possible that the $W$-boson measurements that we propose will be able to achieve a better control of these parameters, including a flavor dependence.

%One of the goals of this paper is to investigate if it is necessary to modify the non-perturbative ansatz given in \eq{fnp} in order to reproduce the experimental data given for $W$ boson transverse momentum spectrum \cite{Aaboud:2017svj,Khachatryan:2016nbe}.

The angular part of the Fourier integral in \eq{Wten} can be solved analytically and we can rewrite the hadron tensor
\begin{eqnarray}
\label{eq:Wten1}
W_{\mu\nu}^{GG'}&&=-\frac{g_T^{\mu\nu}}{\pi N_c} |C_V(Q^2,\mu)|^2 \sum_{ff'} z_{ff'}^{GG'} W^{ff'}_{f_1 f_1}(Q, q_T,x_1,x_2, \mu, \zeta),
\end{eqnarray}
where
\begin{align}
W^{ff'}_{f_1 f_1}(Q, q_T,x_1,x_2, \mu, \zeta)=\int \frac{|\vecb b|d |\vecb b|}{2} J_0(|\vecb b| |\vecb q|) f_{1,f\ot q}(x_1,\vecb b,\mu,\zeta) f_{1,f'\ot q}(x_2,\vecb b,\mu,\zeta).
\end{align}
%%%%%%%%%%%%%%%%%%%%%%%%%%%%%%%%%%%%%%%%%%%%%%%%%%%%%%%%%%%%%%%%%%%%%%%%%%%%%%%%
\subsection{Lepton tensor and fiducial cuts}
\label{sec:lt}
%%%%%%%%%%%%%%%%%%%%%%%%%%%%%%%%%%%%%%%%%%%%%%%%%%%%%%%%%%%%%%%%%%%%%%%%%%%%%%%%
The lepton tensor that enters in the cross section is integrated over the lepton and neutrino momenta, thus contracting it with the Lorentz structure that comes from the hadron tensor in \eq{Wten} we obtain,
\begin{align}
\label{eq:LcutsW}
-g_T^{\mu\nu}L_{\mu\nu}^{GG'({\rm cuts})}&=32 z_{ll'}^{GG'}\int \frac{d^3l}{2E}\frac{d^3l'}{2E'} [ll'-(ll')_T]\theta({\rm cuts}) \delta^{(4)}(l+l'-q) \delta(Q^2-m_T^2-f(l,l')), \nn\\
&
\end{align}
where the extra label \textit{cuts} represents the fiducial cuts introduced in different experiments. These cuts are implemented over the momenta of lepton and neutrino and the rapidity of the lepton
\begin{align}
\eta_{\rm{min}}< \eta <\eta_{\rm{max}}, \qquad l_T^2 > p_1^2, \qquad l_T'^2 >p_2^2.
\end{align}
On the other hand note that the difference of this lepton tensor with the one for neutral DY production \cite{Scimemi:2017etj,Scimemi:2019cmh} is the extra dependence on the transverse mass through the delta function introduced in \eq{1mt}. The integral in \eq{LcutsW} 
\begin{align}
\label{IW}
I_W(Q^2,m_T^2,q_T)&=\int \frac{d^3l}{2E}\frac{d^3l'}{2E'} [ll'-(ll')_T]\theta({\rm cuts}) \delta^{(4)}(l+l'-q) \delta(Q^2-m_T^2-f(l,l')), \nn \\
&
\end{align}
cannot be solved analytically and only numerical results can be obtained. A detailed discussion about this integral can be found in appendix \ref{app:iw}. 
%%%%%%%%%%%%%%%%%%%%%%%%%%%%%%%%%%%%%%%%%%%%%%%%%%%%%%%%%%%%%%%%%%%%%%%%%%%%%%%%
\subsection{Final expression of the cross section}
\label{sec:kinvar}
%%%%%%%%%%%%%%%%%%%%%%%%%%%%%%%%%%%%%%%%%%%%%%%%%%%%%%%%
Once we have precise definitions for hadron and lepton tensor and taking into account that
\begin{align}
d^4 q= \frac{\pi}{2} dQ^2 dy d q_T^2,
\end{align}
we get the desired differential cross section as
\begin{align}
\label{eq:xsec3W}
\frac{d\sigma}{dm_T^2 dy d q_T^2}&=\int_0^\infty dQ^2 \frac{8}{N_c} \frac{\al^2_{\rm{em}}}{s}  I_W(Q^2,q_T, m_T^2)\nn \\
&\times\sum_{ff'}\sum_{GG'}z_{ll'}^{GG'} z_{ff'}^{GG'} \Delta_G (q) \Delta_{G'}^* (q) W_{f_1f_1}^{ff'} (Q^2, q_T,x_1,x_2).
\end{align}
where, in the case of $W$ boson production, the product of EW factors for hadronic and leptonic part and propagators can be written as
\begin{eqnarray}
\label{eq:DDWW}
z_{ll'}^{GG'} z_{ff'}^{GG'}\Delta_G(q)\Delta_{G'}^*(q)=\delta_{GW}\delta_{G'W} |V_{ff'}|^2 \frac{e_f e_{f'}}{Q^4} \frac{Q^4}{(Q^2-M_W^2)^2+\Ga_W^2 M_W^2} z_l^{WW} z_q^{WW},
\end{eqnarray}
where $V_{ff'}$ are the elements of CKM matrix that mixes flavors, $e_f$ are the quark charges in terms of the electron charge and
\begin{eqnarray}
\label{zzW}
z_l^{WW} z_q^{WW}=\(\frac{1}{4s_W^2}\)^2.
\end{eqnarray}

Thus, the final expression for the desired cross section is
\begin{align}
\label{eq:xsec5W}
\frac{d\sigma}{dm_T^2 dy d q_T^2}&= \int_0^\infty \frac{dQ^2}{Q^4}\frac{8}{N_c} \frac{\al^2_{\rm{em}}}{s} I_W(Q^2,q_T,m_T^2) \frac{1}{(4s_W^2)^2} \frac{Q^4}{(Q^2-M_W^2)^2+\Ga_W^2 M_W^2}\nn\\
&\times\sum_{ff'} |V_{ff'}|^2 e_f e_{f'} W_{f_1f_1}^{ff'} (Q^2, q_T,x_1,x_2).
\end{align}
The plots of next sections are done using the PDF set of 
NNPDF31\_nnlo\_as\_0118~\cite{Ball:2017nwa}.
The numerical inputs that we have used are
\begin{align}\nn
M_Z&=91.1876\text{ GeV},\quad M_W=80.379\text{ GeV},\quad \Gamma_Z=2.4952\text{ GeV},\quad \Gamma_W=2.085\text{ GeV},\quad 
\end{align}
and $G_F=1.1663787\times 10^{-5}$ GeV.
%%%%%%%%%%%%%%%%%%%%%%%
\section{Sources of errors}
\label{sec:err}
%%%%%%%%%%%%%%%%%%%
 In the course of the paper  we consider several sources of errors that  come from the way we write and parameterize the cross section. We list them  here as a reference for the following sections.

  \begin{description}
  \item{\bf Scale variations.}
  In the present work we use the $\zeta$-prescription defined in \cite{Scimemi:2018xaf} and we include the non-perturbative fixing of the $\zeta$-scale as defined in \cite{Vladimirov:2019bfa,Scimemi:2019cmh}. As a result the scale variation is done changing the parameters $c_{2,4}\in [0.5,2]$. These  parameters define the uncertainty in the hard matching scale and in the matching of the TMD on the collinear PDF respectively.

\item{\bf Error from a reference PDF set using replicas.} 
The central value of  our predictions is deduced using the PDF set NNPDF31\_nnlo\_as\_0118~\cite{Ball:2017nwa}, which includes also LHC data.  The error on each bin is evaluated taking  the variance over  1000 replicas.  
%This  estimation is particularly useful for the precise measurements of the $p_T^Z/p_T^W$ spectrum see fig.~\ref{fig:ratioZWrep}.

\item{\bf Predictions from different PDF sets.}
In \cite{Scimemi:2019cmh}  the authors have performed fits of DY data at low and high energy with different sets of PDF and we report their results in appendix~\ref{sec:inputs}. We consider here the same sets of PDF and we show  how the predictions change with these different sets.

\item{\bf Error from TMD parametrization.}
 In \cite{Scimemi:2019cmh} the authors provide a set of replicas of their TMD parameterization and we  provide the main estimation of this error using them. Alternatively, one  could use a more standard error propagation coming from the estimate of the non-perturbative parameters. We have used this second method as a check, and we provide more details in appendix~\ref{app:errors}.

%\item{\bf Error from TMD parameters.}\textcolor{blue}{
%It is difficult to establish an error for the TMD non-perturbative parameters.  In \cite{Scimemi:2019cmh} the authors have an error coming from data which is probably underestimated. In order to provide a more reasonable error we do the following. We consider the average value of the constants as coming from different PDF sets and its relative variance. In this way we obtain that these parameters belong to the intervals reported in table~\ref{tab:lerr}.
%Then, using the central  PDF of NNPDF31\_nnlo\_as\_0118~\cite{Ball:2017nwa} and  using the the correlation matrix provided in  \cite{Scimemi:2019cmh}
%we obtain how the estimate of the non perturbative parameters affects our predictions.
%}

\end{description}
 
 %%%%%%%%%%%%%%%%%%%%%%
 \section{Observables  in $W$ production with TMD }
\label{sec:Wspeck}
%%%%%%%%%%%%%%%%%%%%%
In this section we concentrate on some observables that  can be, in principle,   sensitive to TMD effects and/or are relevant to establish some important properties of the $W$ as its mass. 
We consider the $W$-boson differential cross section, the ratios for $p_T^Z/p_T^W$, and $p_T^{W^-}/p_T^{W^+}$ all of them as a function of the boson transverse momentum distribution and different intervals of the transverse mass.
The fiducial cuts have been set as the ATLAS experiment  \cite{Aaboud:2017svj}  but for $\sqrt{s}=13$ TeV.
 We have identified three interesting  intervals of $m_T$, namely [50,66] GeV, [66,99] GeV, [99,120] GeV corresponding to $m_T$ values below, around, above the $W$-mass, with typical cuts on lepton momenta.
The $m_T$-interval which is mostly studied in the  literature  is the one with $m_T\in [66,99] $ GeV, however this is not the only interesting one from the perspective of studying the TMD properties.
This is because non-perturbative TMD effect are expected  to be more relevant for low values of  $m_T$, similarly to  Drell-Yan processes.
The $q_T$ interval that we have considered, that is $q_T/m_T\lesssim 0.2$ is consistent with the application for TMD factorization.
In order to have a reference setup for PDF we have chosen NNPDF31\_nnlo\_as\_0118~\cite{Ball:2017nwa} as in \cite{Scimemi:2019cmh} as it is one of the sets that include LHC data. Nevertheless we have also examined the results taking into account different sets. 
We have considered a $q_T$ binning of 1 GeV, which much probably exceeds the possibilities of current experiments. We motivate this choice because it has been used also by other authors, see f.i. \cite{Bizon:2019zgf}, and also because it is allows to better evidence the TMD effects.  

\subsection{Spectra of the $W^\pm$}

\begin{figure}
\begin{center}
\includegraphics[width=0.3\textwidth]{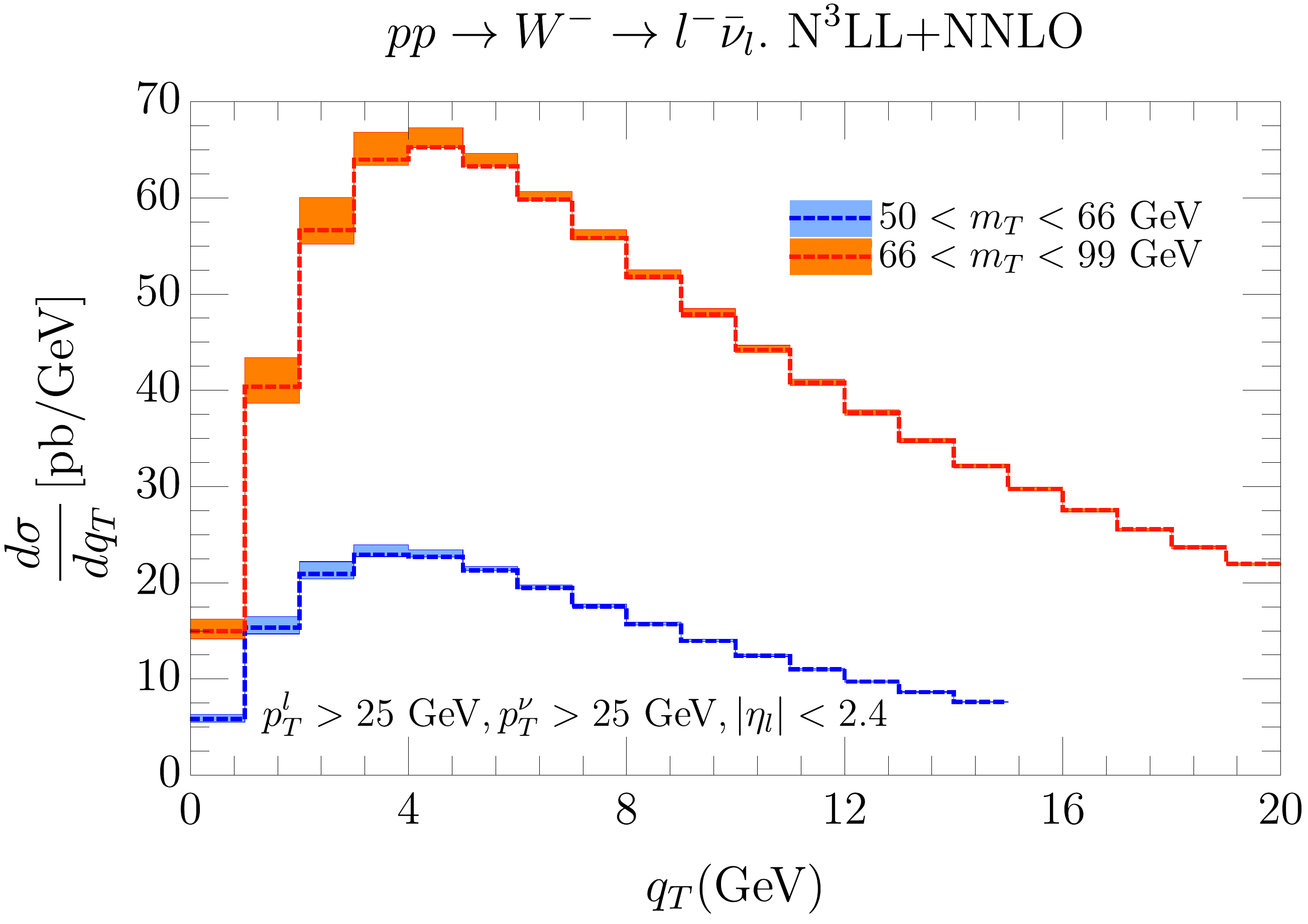}
\includegraphics[width=0.3\textwidth]{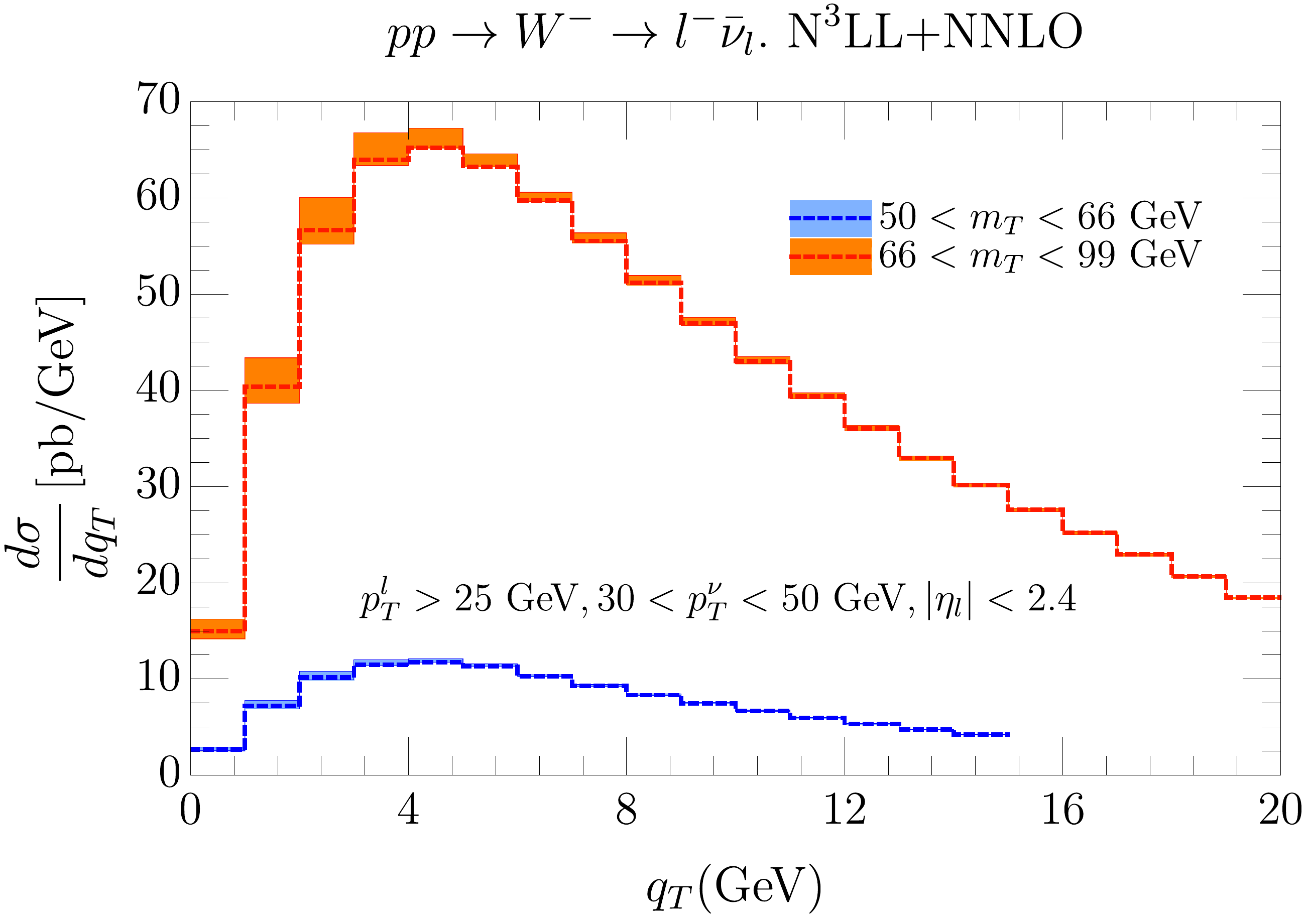}
\includegraphics[width=0.3\textwidth]{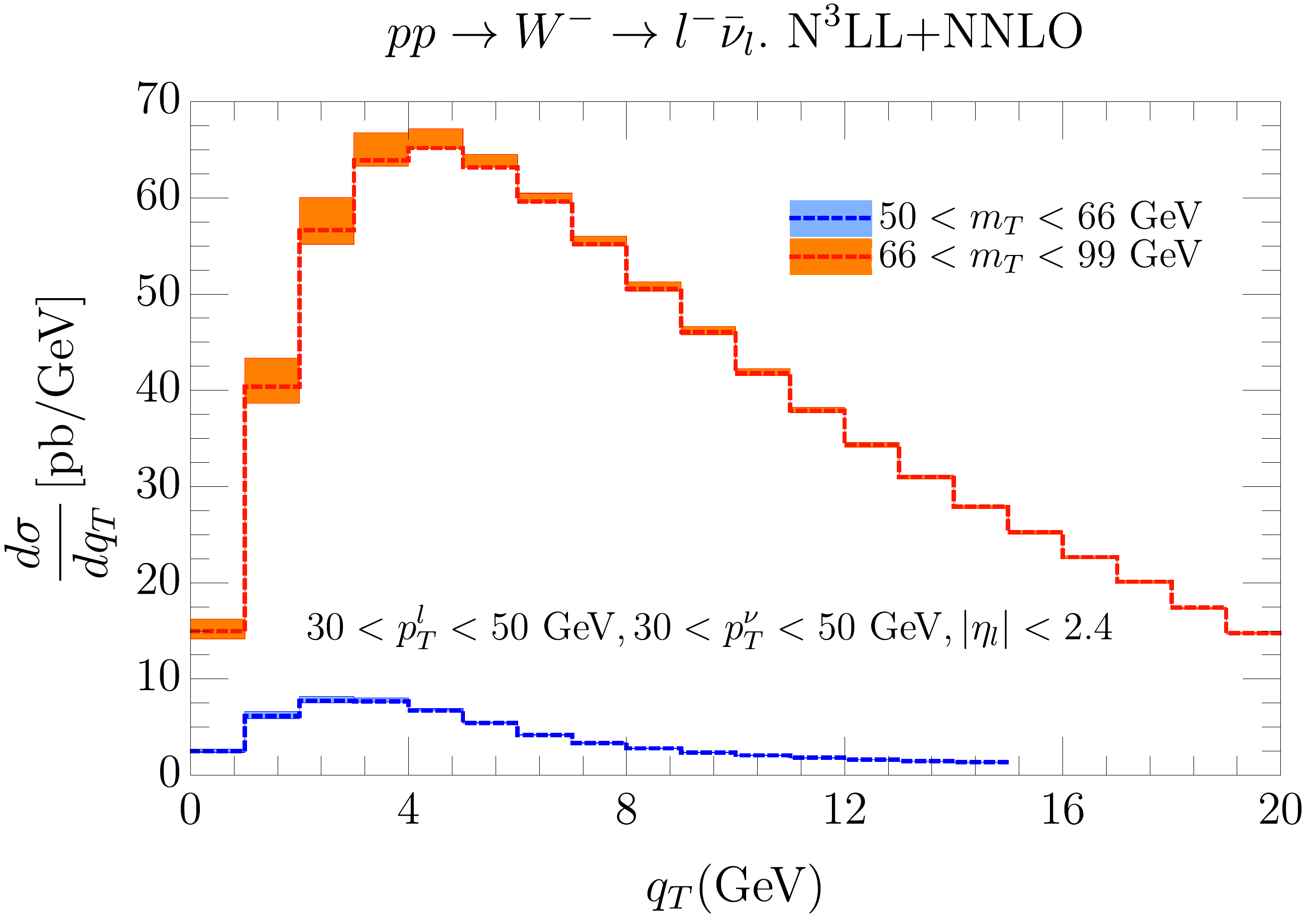}
\\
\includegraphics[width=0.3\textwidth]{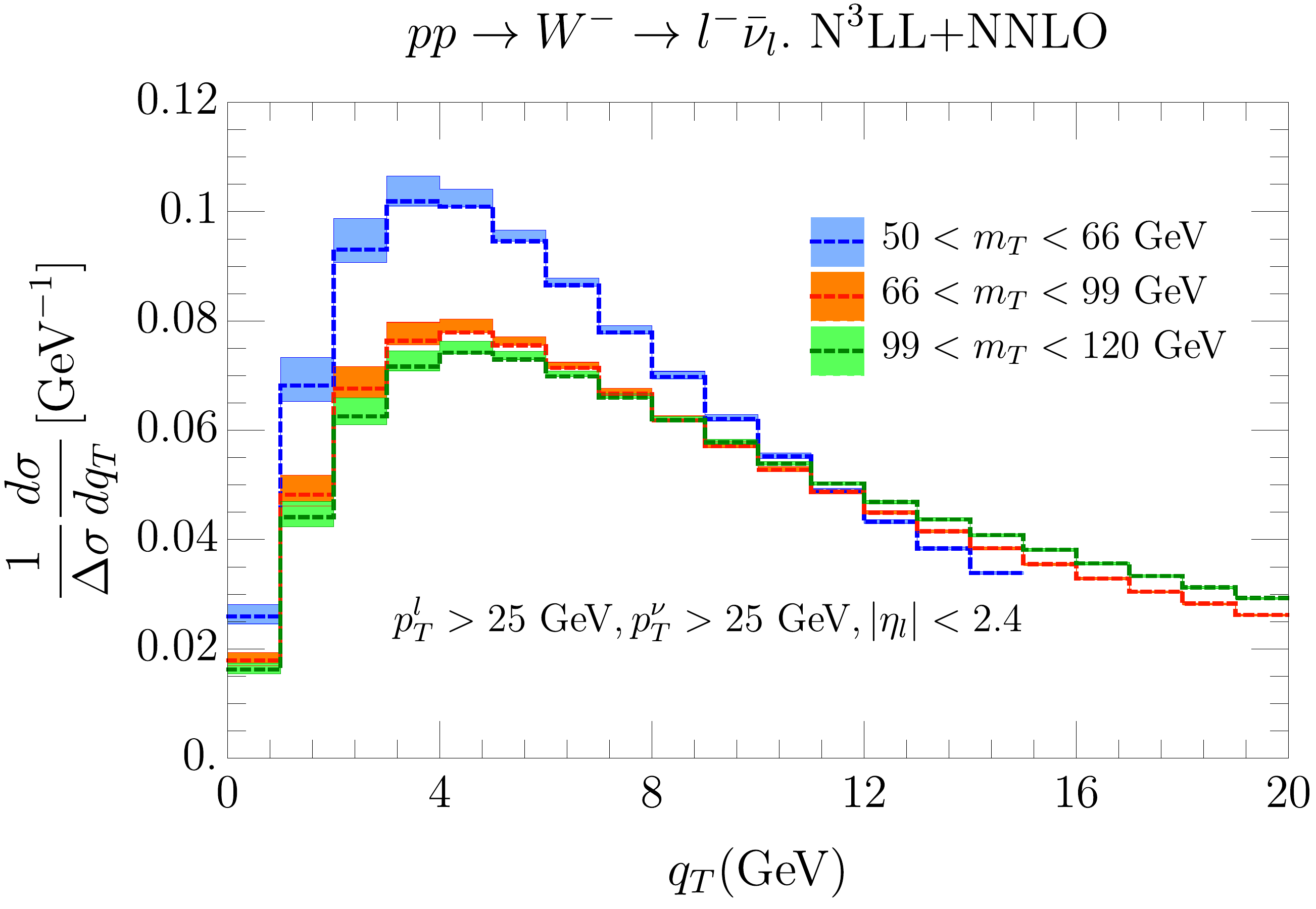}
\includegraphics[width=0.3\textwidth]{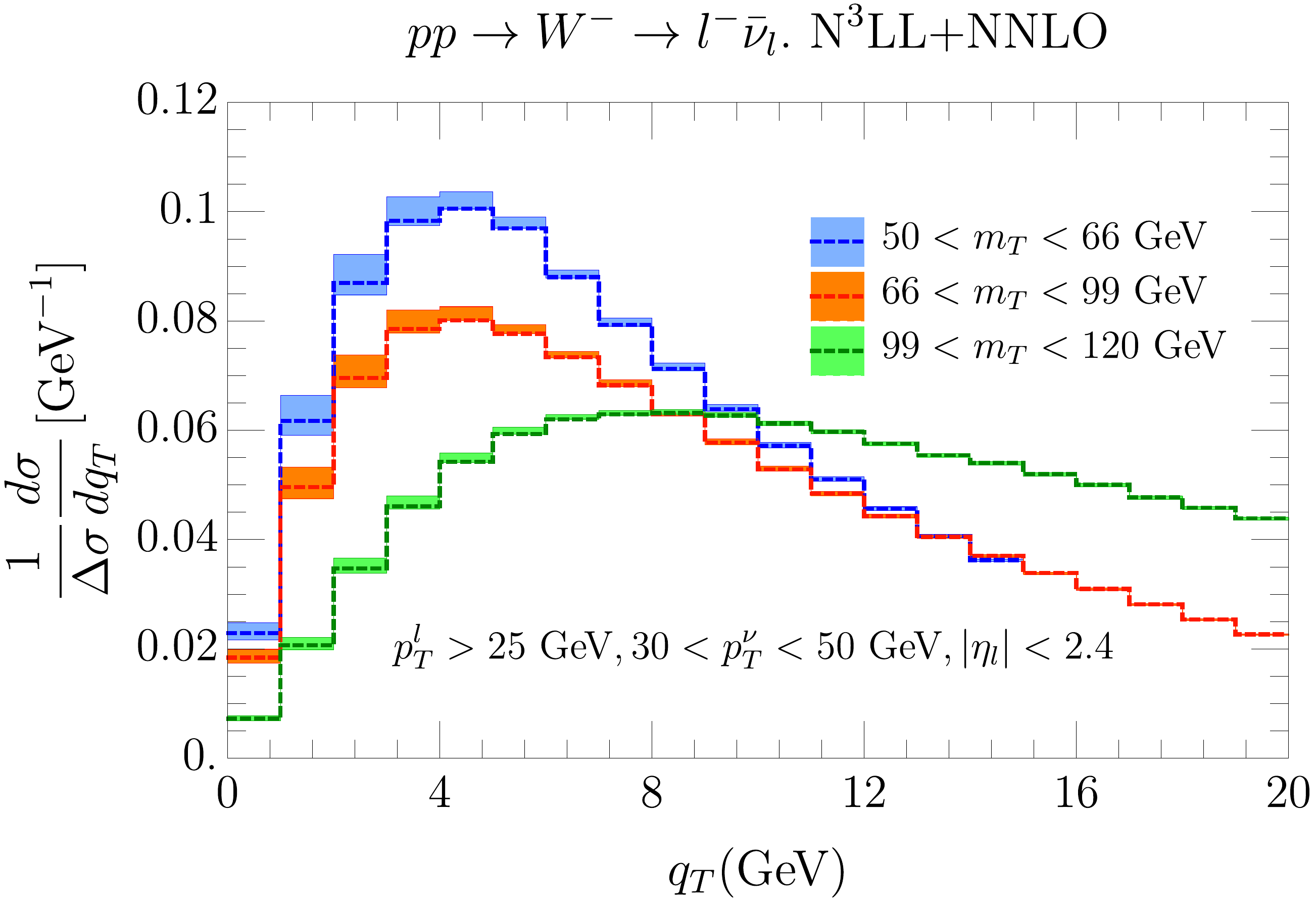}
\includegraphics[width=0.3\textwidth]{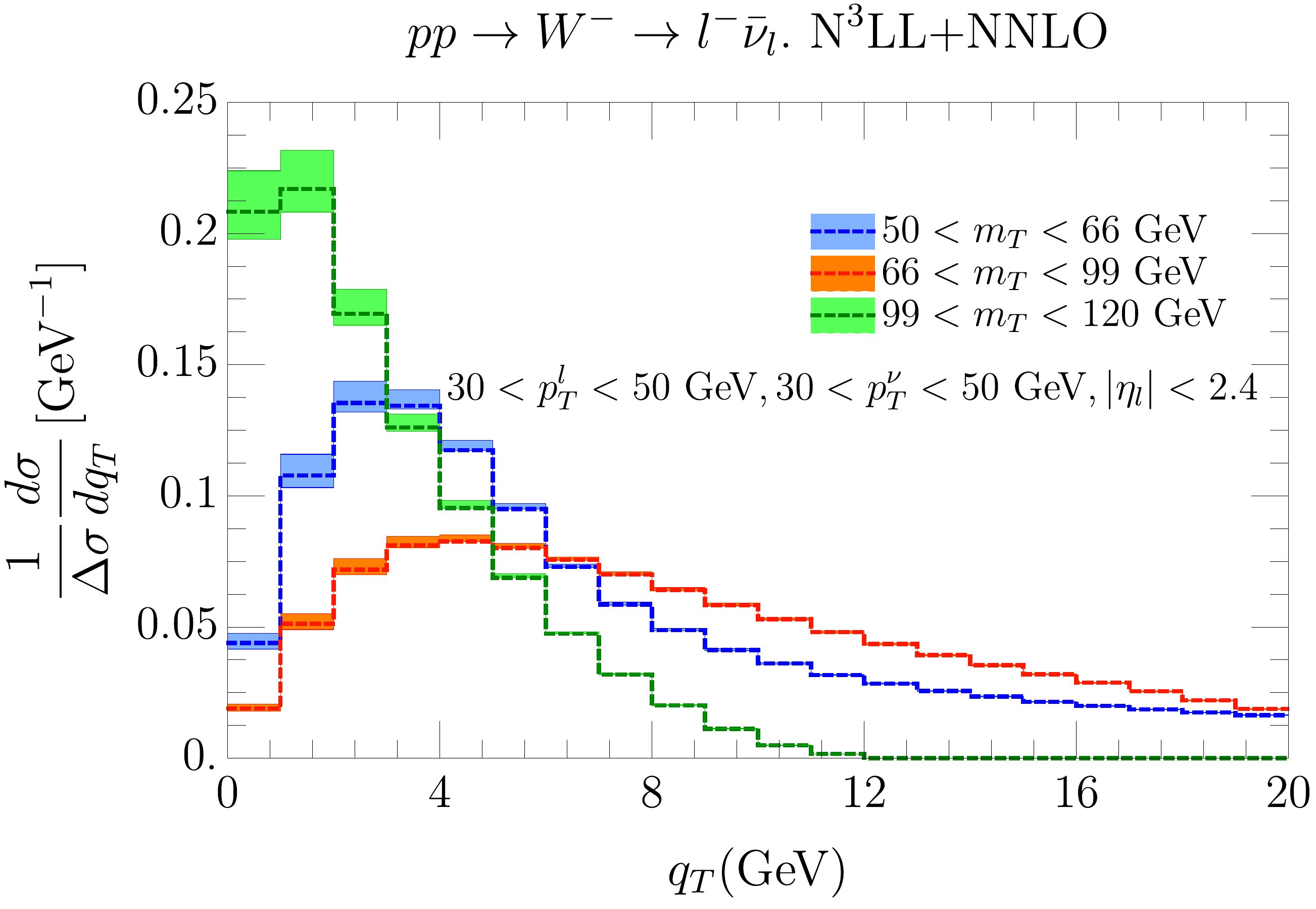}
%\vspace{-3cm}
\caption{\label{fig:Wspectra} (top) Unnormalized cross sections for $W^-$ production with different intervals of lepton cuts. (bottom)
Normalized cross sections for $W^-$ production with different intervals of lepton cuts.
}
\end{center}
\end{figure}
We have considered the spectra of the $W^\pm$ and in 
 fig.~\ref{fig:Wspectra} we show the case for the $W^-$ (similar plots  are obtained for $W^+$). 
In the upper panels of  fig.~\ref{fig:Wspectra} we have considered three possible cuts for lepton momenta as reported in the same figures. In the upper panels of  fig.~\ref{fig:Wspectra} the cross section for $m_T\in [99,120]$ GeV is not reported because it is much smaller than the others. In the same figure we have included error bands corresponding to scale variations. The figures in these panels show that  the cuts have marginal impact for $m_T\in [66,99]$ GeV while they can suppress the cross section in the  lower interval $m_T\in [50,66]$ GeV. 
The difference in the values of the cross sections however is about a factor 4. %, which suggests that the data for the lower interval  $m_T\in [50,66]$ GeV can be significative.
 The interval $m_T\in [99,120]$ GeV instead results to be even more suppressed an it is not shown in the figure. Given this suppression and the fact at high values of $m_T$ all TMD effects are washed out  we neglect it in the rest of the figures.

In the lower panels of fig.~\ref{fig:Wspectra}  we normalize the cross section to  its value integrated in $q_T$   in the interval  shown in the figure. The shape of the curves now changes and in the left and central panels the more peaked cross sections is the one obtained  for $m_T\in [50,66]$ GeV, which is partly due to the fact that we always select $q_T/m_T\leq 0.2$.

In order to drive  a conclusion from these plots we have to recall the previous experience of the fit of \cite{Scimemi:2019cmh,Bacchetta:2019sam} and also the results of \cite{Hautmann:2020cyp}.  In these works it shown that in TMD analyses increasing the value of $Q$ the non-perturbative  QCD effects are washed out. As a a result having data below the $W,\, Z$ boson peaks is extremely useful for this kind of research and it can provide valuable information.
 The plots shown in fig.~\ref{fig:Wspectra} actually show that within the current facility it is possible to achieve this goal. The bands shown in fig.~\ref{fig:Wspectra}  come from scale variations.
 
\begin{figure}
\begin{center}
\includegraphics[width=0.3\textwidth]{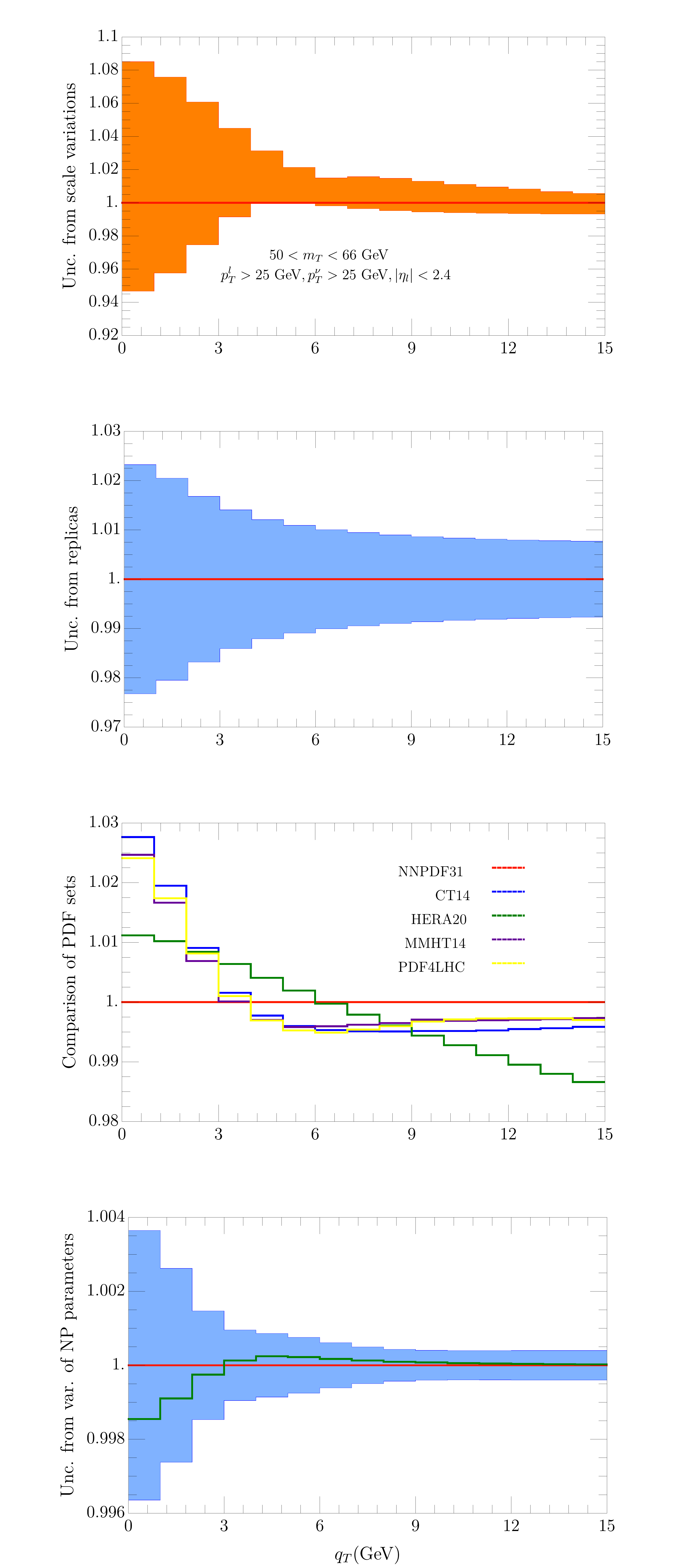}
\includegraphics[width=0.3\textwidth]{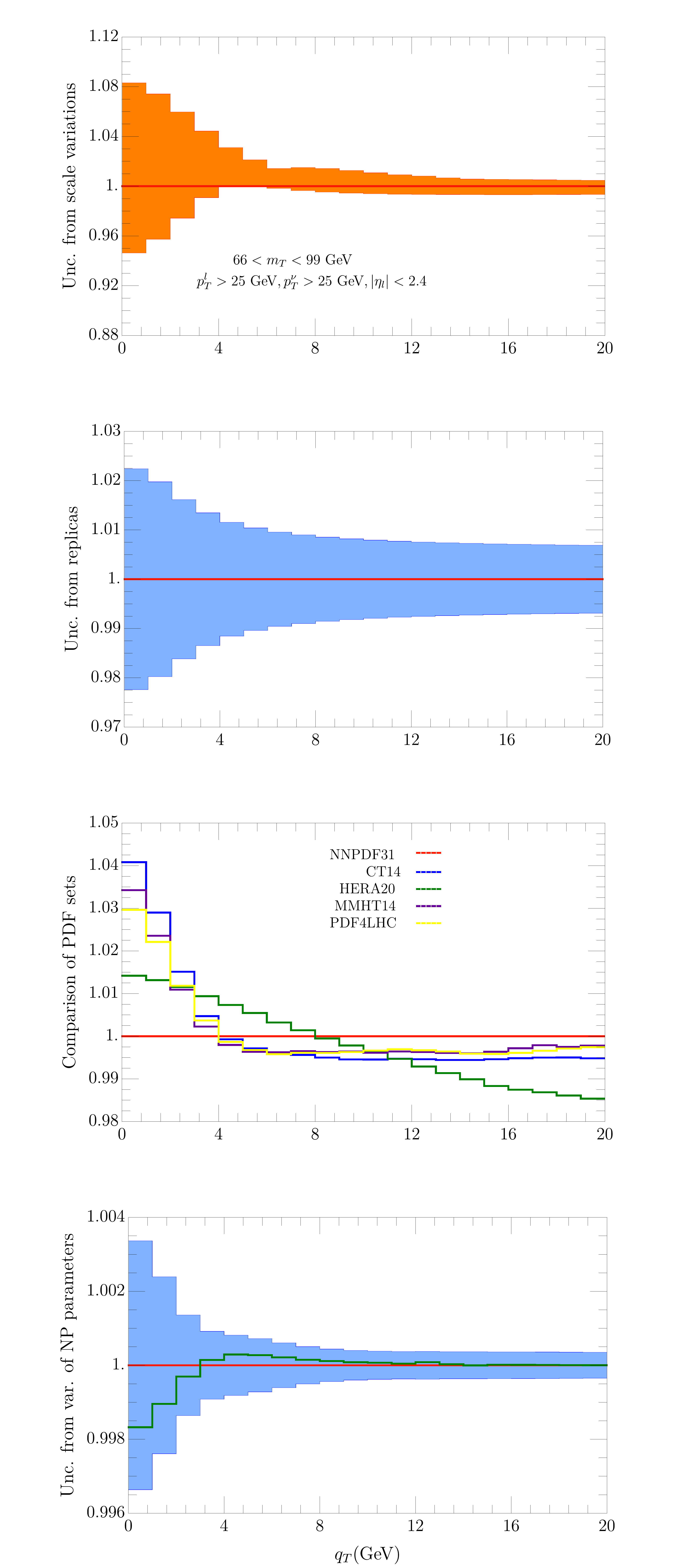}
\includegraphics[width=0.3\textwidth]{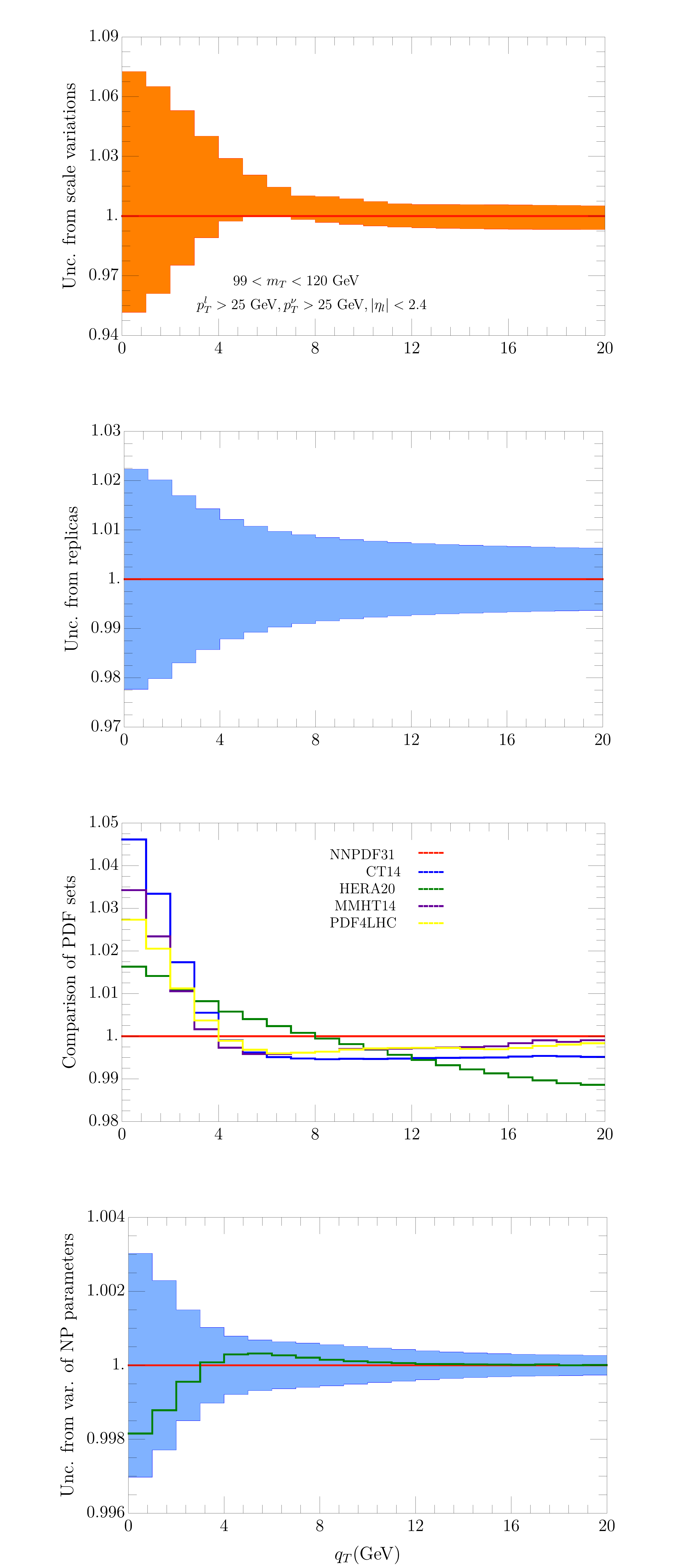}
%\includegraphics[width=0.3\textwidth]{Figures/Replica_plots/ReplicaNYM.pdf}
%\includegraphics[width=0.3\textwidth]{Figures/Replica_plots/ReplicaYYM.pdf}\\
%\includegraphics[width=0.3\textwidth]{Figures/Replica_plots/ReplicaNNM.pdf}
%\includegraphics[width=0.3\textwidth]{Figures/Replica_plots/ReplicaNYL.pdf}
%\includegraphics[width=0.3\textwidth]{Figures/Replica_plots/ReplicaYYL.pdf}\\
%\includegraphics[width=0.3\textwidth]{Figures/Replica_plots/ReplicaNNH.pdf}
%\includegraphics[width=0.3\textwidth]{Figures/Replica_plots/ReplicaNYH.pdf}
%\includegraphics[width=0.3\textwidth]{Figures/Replica_plots/ReplicaYYH.pdf}
%\vspace{-3cm}
\caption{\label{fig:Wspeck} Error for  $W^-$ cross section for $p_T^{l,\nu}>25$ GeV, $|\eta_l|<2.4$ and the  $m_T$ intervals  $[50,66]$ GeV (left column), $[66,99]$ GeV (middle column), $[99,120]$ GeV (right column).
In the first line  we report the theoretical error from scale uncertainties as explained in the text. In the second line we have
the error  calculated as a variance in each bin of 100 replicas of the set NNPDF31\_nnlo\_as\_0118~\cite{Ball:2017nwa}.
The uncertainty is referred to the average value of each bin (red line). The value of the cross section given by the central replica is represented by the green line.
On the third line we represent the value of each been with different sets of PDF.
On the fourth row we have the uncertainty due to non-perturbative parameters. The central value is given by the the central replica of NNPDF31\_nnlo\_as\_0118~\cite{Ball:2017nwa}.
% The cross section for the central replica is shown in fig.~\ref{fig:Gehrmann}.
}
\end{center}
\end{figure}
\begin{figure}
\begin{center}
\includegraphics[width=0.3\textwidth]{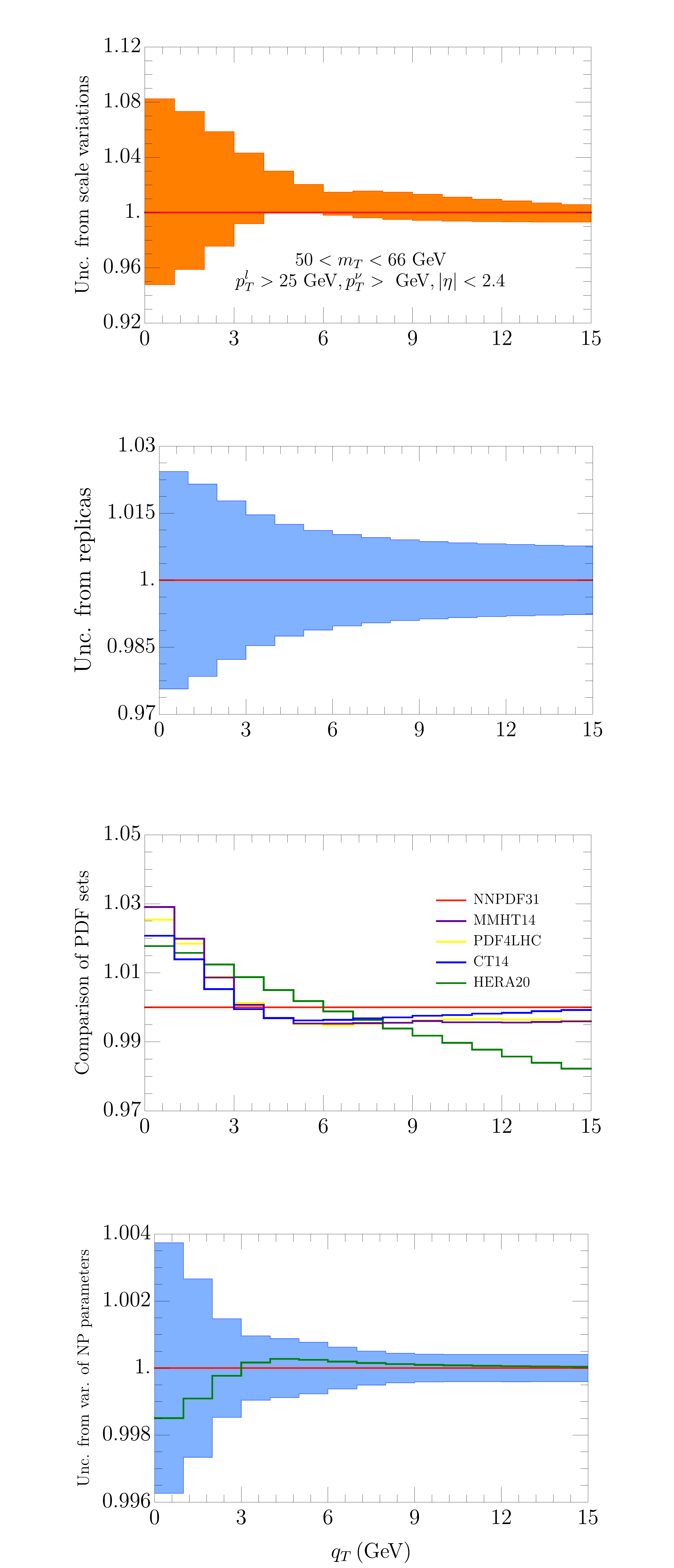}
\includegraphics[width=0.3\textwidth]{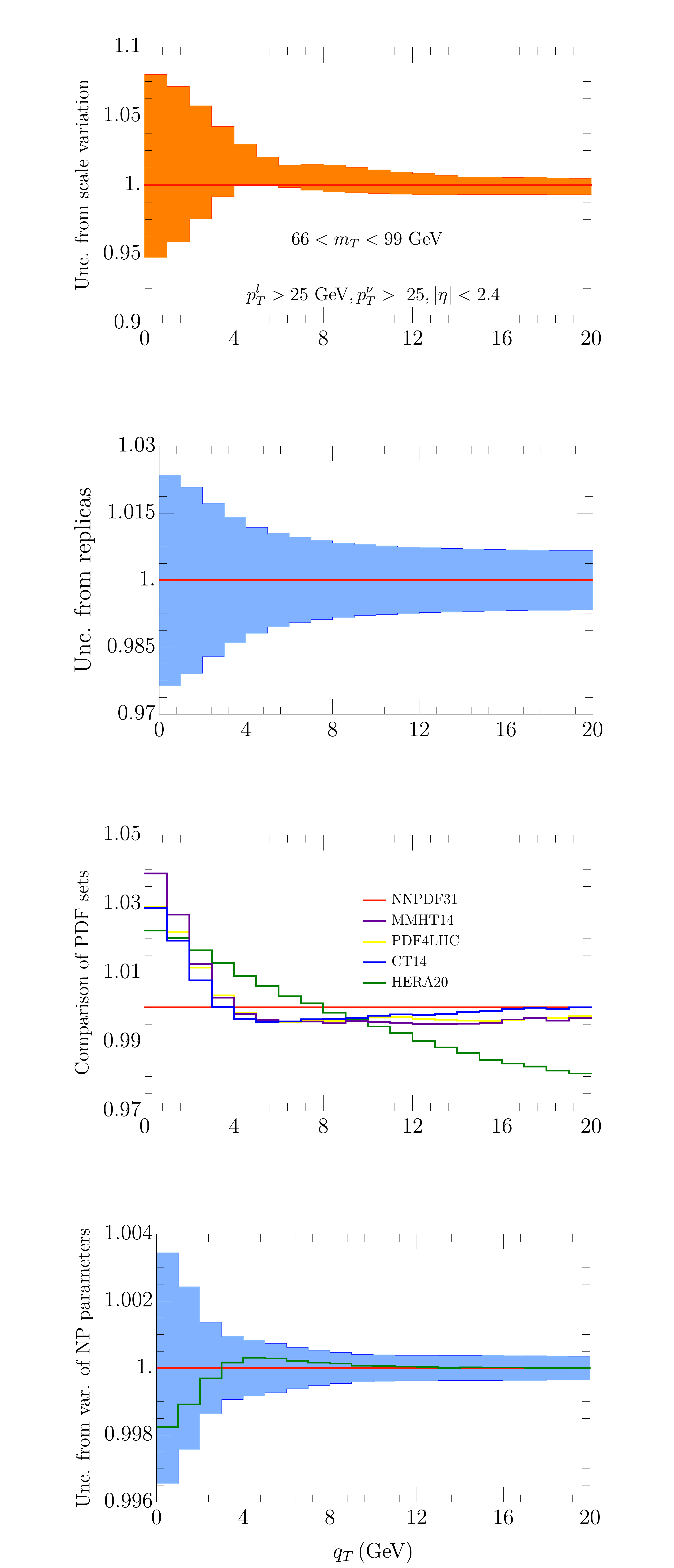}
\includegraphics[width=0.3\textwidth]{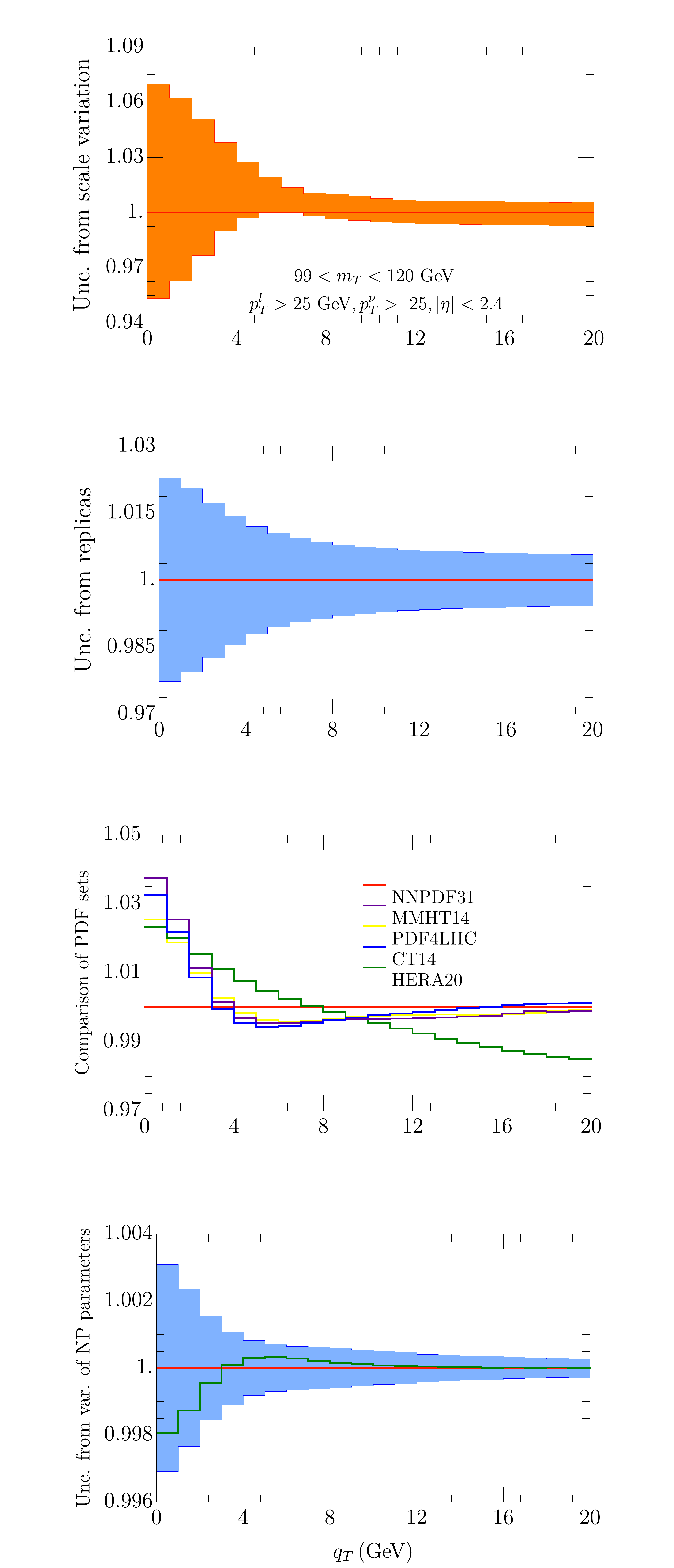}
%\includegraphics[width=0.3\textwidth]{Figures/Replica_plots/ReplicaNYM.pdf}
%\includegraphics[width=0.3\textwidth]{Figures/Replica_plots/ReplicaYYM.pdf}\\
%\includegraphics[width=0.3\textwidth]{Figures/Replica_plots/ReplicaNNM.pdf}
%\includegraphics[width=0.3\textwidth]{Figures/Replica_plots/ReplicaNYL.pdf}
%\includegraphics[width=0.3\textwidth]{Figures/Replica_plots/ReplicaYYL.pdf}\\
%\includegraphics[width=0.3\textwidth]{Figures/Replica_plots/ReplicaNNH.pdf}
%\includegraphics[width=0.3\textwidth]{Figures/Replica_plots/ReplicaNYH.pdf}
%\includegraphics[width=0.3\textwidth]{Figures/Replica_plots/ReplicaYYH.pdf}
%\vspace{-3cm}
\caption{\label{fig:Wspeckplus} Error for  $W^+$ cross section for $p_T^{l,\nu}>25$ GeV, $|\eta_l|<2.4$ and the  $m_T$ intervals  $[50,66]$ GeV (left column), $[66,99]$ GeV (middle column), $[99,120]$ GeV (right column).
In the first line  we report the theoretical error from scale uncertainties as explained in the text. In the second line we have
the error  calculated as a variance in each bin of 100 replicas of the set NNPDF31\_nnlo\_as\_0118~\cite{Ball:2017nwa}.
The uncertainty is referred to the average value of each bin (red line). The value of the cross section given by the central replica is represented by the green line.
On the third line we represent the value of each been with different sets of PDF.
On the fourth row we have the uncertainty due to non-perturbative parameters. The central value is given by the the central replica of NNPDF31\_nnlo\_as\_0118~\cite{Ball:2017nwa}.
 The cross section for the central replica is shown in fig.~\ref{fig:Gehrmann}.
}
\end{center}
\end{figure}
 In fig.~\ref{fig:Wspeck}-\ref{fig:Wspeckplus} we  have studied the errors as listed in sec.~\ref{sec:err} for one particular set of fiducial cuts for $W^-$ and $W^+$ respectively. Similar errors are obtained also in the other cases.
 Going from the top row to the bottom one in   fig.~\ref{fig:Wspeck}-\ref{fig:Wspeckplus}  one finds:  1) the theoretical error from scale uncertainties; 2)
the error  calculated as a variance in each bin of 1000 replicas of the set NNPDF31\_nnlo\_as\_0118~\cite{Ball:2017nwa}.  Here
the bands are referred to the average value of each bin (red line). The value of the cross section given by the central replica is represented by the green line;
3) the uncertainty as coming from different sets of PDF; 4) the uncertainty due to non-perturbative parameters: in this case the central line is given by the the central replica of NNPDF31\_nnlo\_as\_0118~\cite{Ball:2017nwa}.

There is a series of observations that one can  make on these plots.  As a first the smallest variations are provided by replicas and non-perturbative parameters and both of them are below $1\%$. The scale variation is certainly the biggest source of error: it is  in range 2-9$\%$ for  $q_T\leq$ 4 GeV, and about 1-2$\%$ for $q_T\geq$ 4 GeV. We recall that the present analysis uses N$^3$LL for TMD evolution and NNLO for all the rest. 
In the introduction it was pointed out that recently higher order perturbative calculations have been performed. It will be interesting to observe how the scale variation changes when including this higher order term. On the other side  before doing this one should have a TMD extraction of the same order.

In exploring different PDF extractions we have considered the ones for which a TMD extraction has been provided in  \cite{Scimemi:2019cmh} . We recall that the HERA PDF
\cite{Abramowicz:2015mha} actually provided the best fit in that case. Observing  fig.~\ref{fig:Wspeck}-\ref{fig:Wspeckplus} one sees that actually this set behaves differently with respect to the others, however the difference with NNPDF31 is below 1-2 $\%$ an all over the range of $qT$ that we have considered.  For all the other sets, there is a major difference with NNPDF31  for $q_T\leq$4 GeV while they agree within 1$\%$ for  greater values of $q_T$.

The uncertainties due to non-perturbative parameters result to be almost negligeable. The non-perturbative parameters have uncertainties coming  from  their extraction from data, that is from a $\chi^2$ analysis of fits. Despite the fact that some of these uncertainties may be significative on each single parameter the impact of each single variation is just a fraction of the impact of the whole non-perturbative TMD structure because their value is significantly different from zero. It is also possible that the errors on each parameter is under-estimated as suggested in \cite{Scimemi:2019cmh}. 
%%%%%%%%%%%%%%%%%%%%%%%%%%%%%%% 
\subsection{Ratio $p_T^Z/p_T^W$}

%%%%%%%%%%%%%%%%%%%%%%%%%%%%%%%%%

\begin{figure}[h]
\begin{center}
\includegraphics[width=0.35\textwidth]{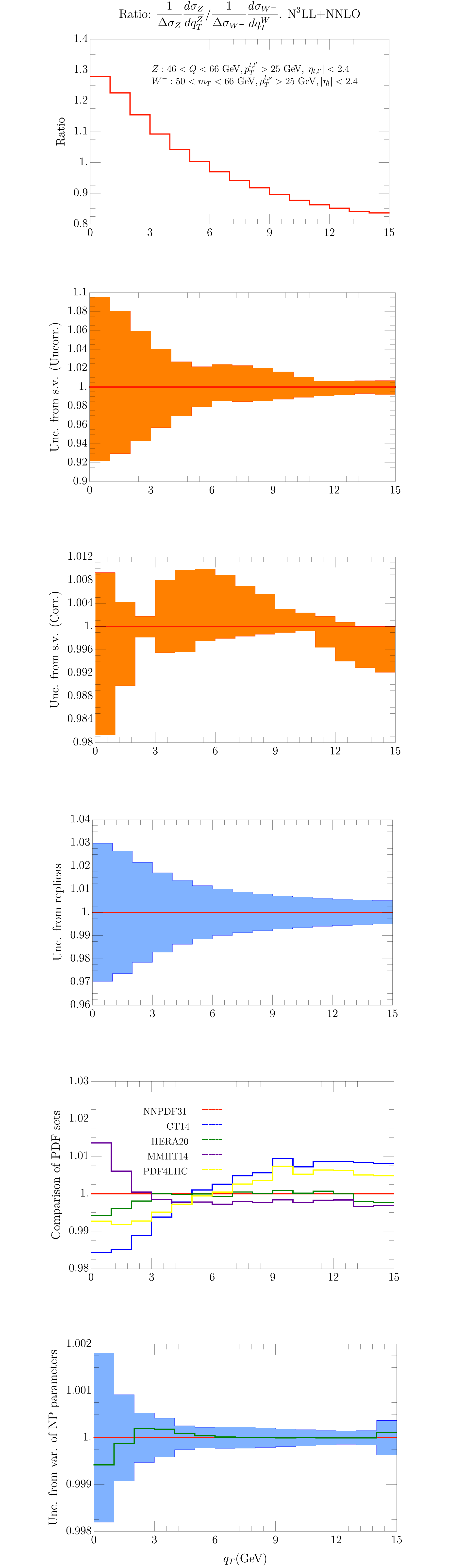}
\includegraphics[width=0.35\textwidth]{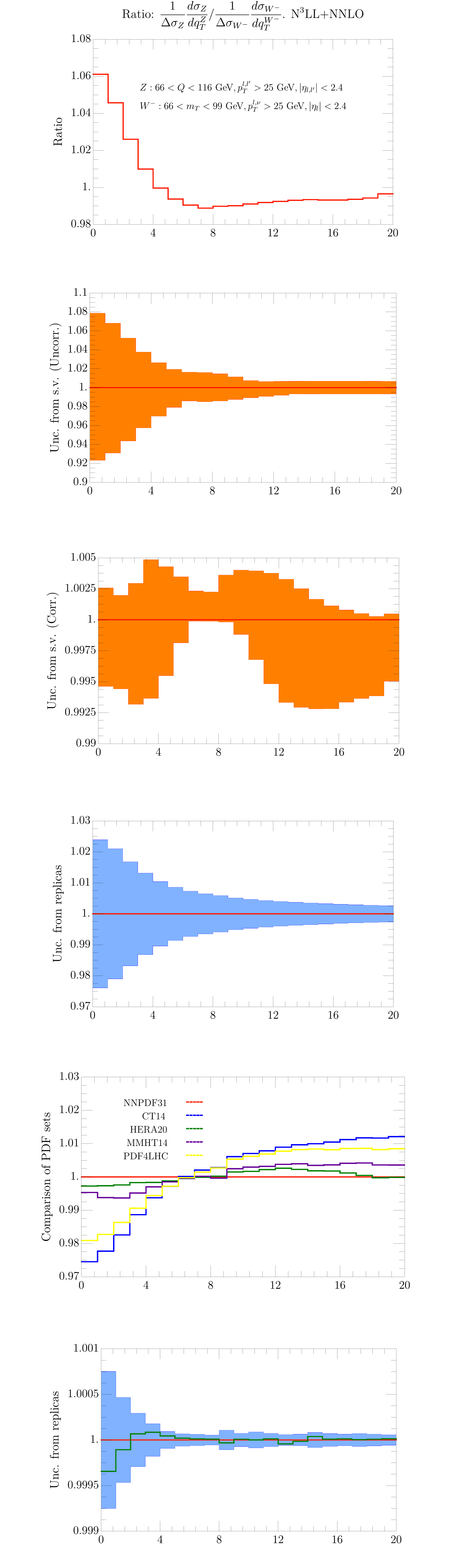}
%\vspace{-3cm}
\caption{\label{fig:ratioZW} Ratio of $Z/W^-$ spectrum for $m_T\in [50,66]$ GeV (left column) , and $m_T\in [66,99]$ GeV (right column).
On the first row we have the  $Z/W^-$ spectrum using the central replica of NNPDF31\_nnlo\_as\_0118~\cite{Ball:2017nwa}.
Uncorrelated and correlated  theoretical uncertainties are given in second and third row respectively. 
In the third line we have
the error  calculated as a variance in each bin of 100 replicas of the set NNPDF31\_nnlo\_as\_0118~\cite{Ball:2017nwa}.
The uncertainty is referred to the average value of each bin (red line). The value of the observable given by the central replica is represented by the green line.
On the fourth line we represent the value of each been with different sets of PDF.
On the fifth row we have the uncertainty due to non-perturbative parameters. %The central value is given by the the central replica of NNPDF31\_nnlo\_as\_0118~\cite{Ball:2017nwa}.
}
\end{center}
\end{figure}

\begin{figure}[h]
\begin{center}
\includegraphics[width=0.35\textwidth]{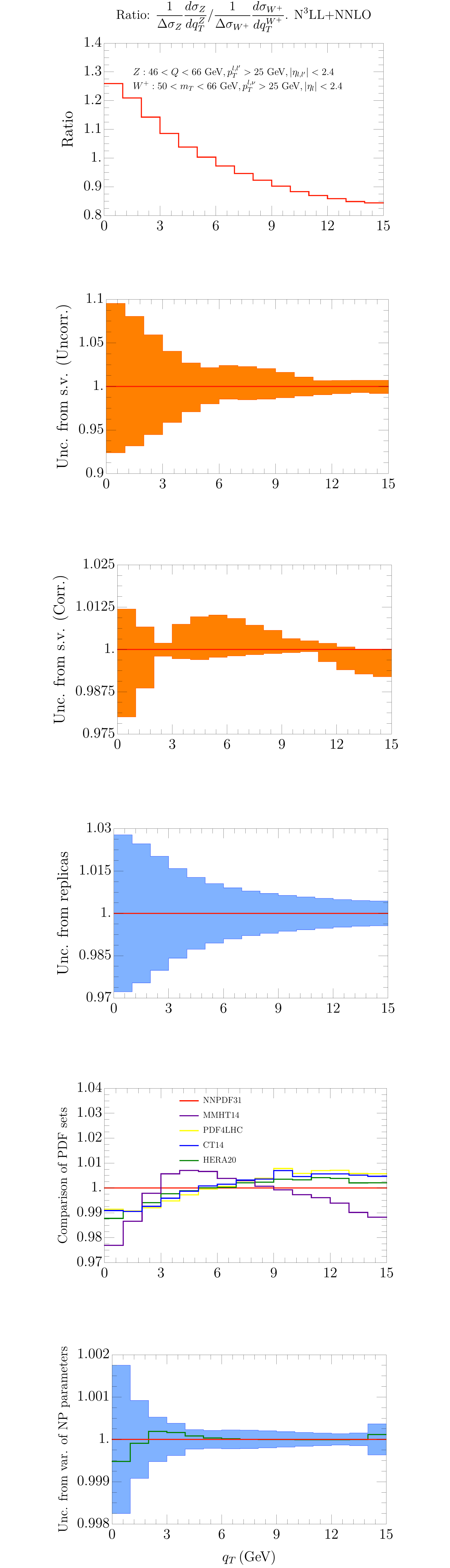}
\includegraphics[width=0.35\textwidth]{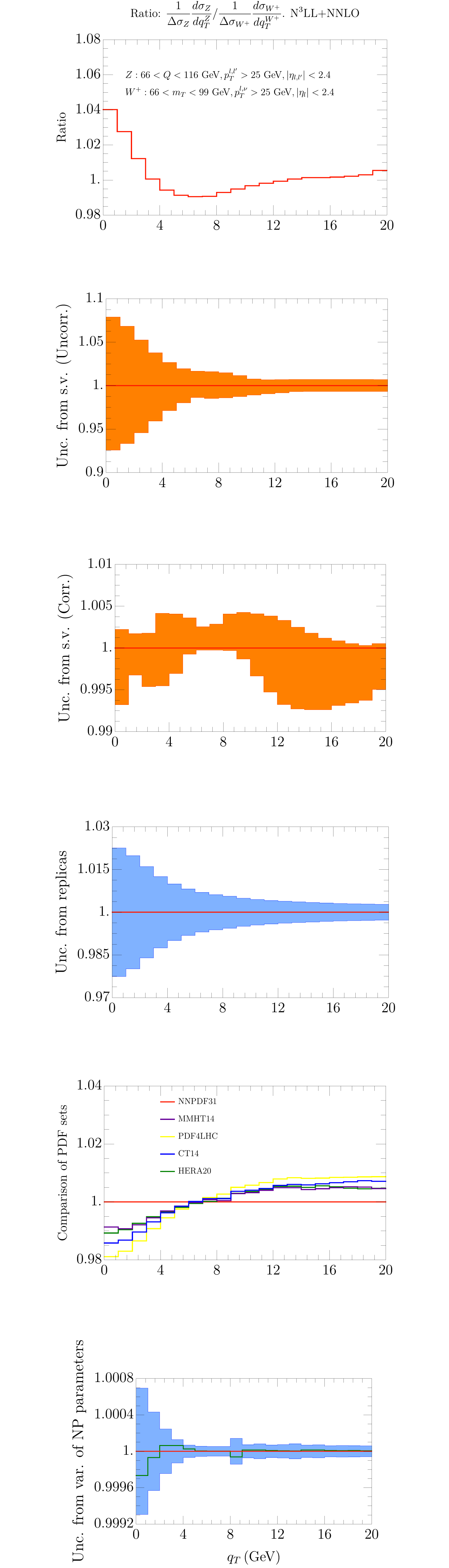}
%\vspace{-3cm}
\caption{\label{fig:ratioZWp} Ratio of $Z/W^+$ spectrum for $m_T\in [50,66]$ GeV (left column) , and $m_T\in [66,99]$ GeV (right column).
On the first row we have the  $Z/W^-$ spectrum using the central replica of NNPDF31\_nnlo\_as\_0118~\cite{Ball:2017nwa}.
Uncorrelated and correlated  theoretical uncertainties are given in second and third row respectively. 
In the third line we have
the error  calculated as a variance in each bin of 100 replicas of the set NNPDF31\_nnlo\_as\_0118~\cite{Ball:2017nwa}.
The uncertainty is referred to the average value of each bin (red line). The value of the observable given by the central replica is represented by the green line.
On the fourth line we represent the value of each been with different sets of PDF.
On the fifth row we have the uncertainty due to non-perturbative parameters. %The central value is given by the the central replica of NNPDF31\_nnlo\_as\_0118~\cite{Ball:2017nwa}.
}
\end{center}
\end{figure}

%\begin{figure}
%\begin{center}
%\includegraphics[width=0.45\textwidth]{Figures/ratiosZW/RatioZWlowCOMBINEDtriple.pdf}
%\includegraphics[width=0.45\textwidth]{Figures/ratiosZW/RatioZWmedCOMBINEDtriple.pdf}%\\
%%\includegraphics[width=0.35\textwidth]{Figures/ratiosZW/ErrorLow.pdf}\hspace{1.8cm}
%%\includegraphics[width=0.35\textwidth]{Figures/ratiosZW/ErrorMed.pdf}
%%\vspace{-3cm}
%\caption{\label{fig:ratioZW} Ratio of $Z/W^-$ cross section for (left) $m_T\in [50,66]$ GeV, and (right) $m_T\in [66,99]$ GeV. In the central panel we quote the error coming from correlated scale variations.
 %In the lower panel we quote the error coming from uncorrelated scale variations.}
%\end{center}
%\end{figure}

It has been pointed out  since a long time that the observable 
\begin{align}\label{eq:ZW}
\left(
\frac{1}{\Delta \sigma^{Z}} \frac{d\sigma^{Z}}{\, dq_T}
\right)/\left(\frac{1}{\Delta \sigma^W}\frac{d\sigma^W}{\, dq_T} \right)\text{ vs. } q_T
\end{align}
can show a very small uncertainty because the PDF contributions tend to cancel and it can be useful for the measurements of the $W$ mass (some recent phenomenological work can be found in
 \cite{Bozzi:2011ww,Bozzi:2015hha}).
  The region of $q_T$ where this observable is actually measured 
corresponds to the region of validity of the TMD factorization theorem, so it is interesting to observe what is the impact of TMD.
In formula~\ref{eq:ZW} the numerator and the denominator are  weighted  by $\Delta\sigma^{W,Z}$ that is  the cross section integrated in the $q_T$ interval under study. 
This is preferable to normalizing  to the total cross  section $\sigma^{W,Z}$
 which should be extracted elsewhere.
 The  result of our prediction is shown in fig.~\ref{fig:ratioZW}-\ref{fig:ratioZWp} for $W^-$ and $W^+$ respectively.
We have considered two intervals of $m_T$, being the lower one useful to control  better TMD effects and several sources of error as in the previous section.
The scale uncertainty now is treated as in 
 \cite{Bizon:2019zgf}, considering separately the correlated and uncorrelated cases.

  The error from scale variations can be considered as correlated or uncorrelated as in \cite{Bizon:2019zgf}, see second and third row in fig.~\ref{fig:ratioZW}. The uncorrelated uncertainty is well below 1$\%$  for the central interval  and most of the  low interval of of $m_T$.
  On the other side the correlated uncertainty is below 2$\%$ only for $q_T\geq $ 4 GeV.
  A similar trend is shown when different sets of PDF are used. In this case the spread of the results can be at most 2$\%$  for $q_T\leq $ 4 GeV, and at most  1$\%$  for $q_T\geq $ 4 GeV.
  We find remarkable that  the PDF error results slightly inferior for the low $m_T$ interval, which confirms that this is an interesting case to study.
  Another interesting observation is that  the difference between the  HERA and NNPDF31 sets is always well below 1$\%$.
  The other sources of error that we have considered, give uncertainties less the 1$\%$ on all over the intervals that we have studied.

%***************************************************************************************************%
%%%%%%%%%%%%%%%%%
\subsection{Ratio  $p_T^{W^-}/p_T^{W^+}$}
\begin{figure}
\begin{center}
\includegraphics[width=0.35\textwidth]{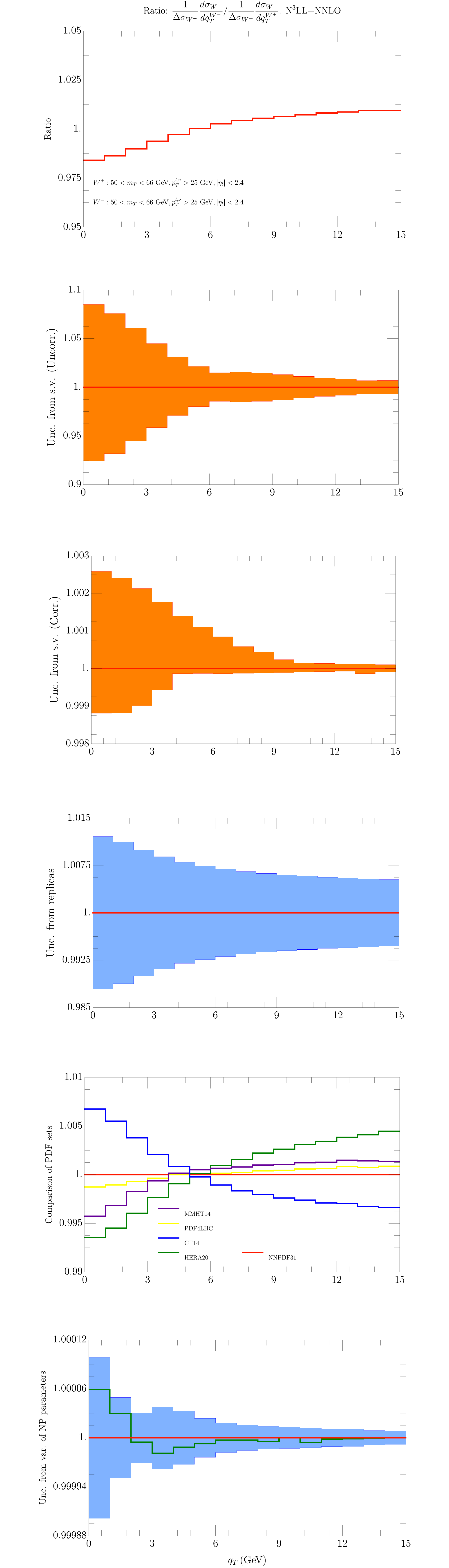}
\includegraphics[width=0.35\textwidth]{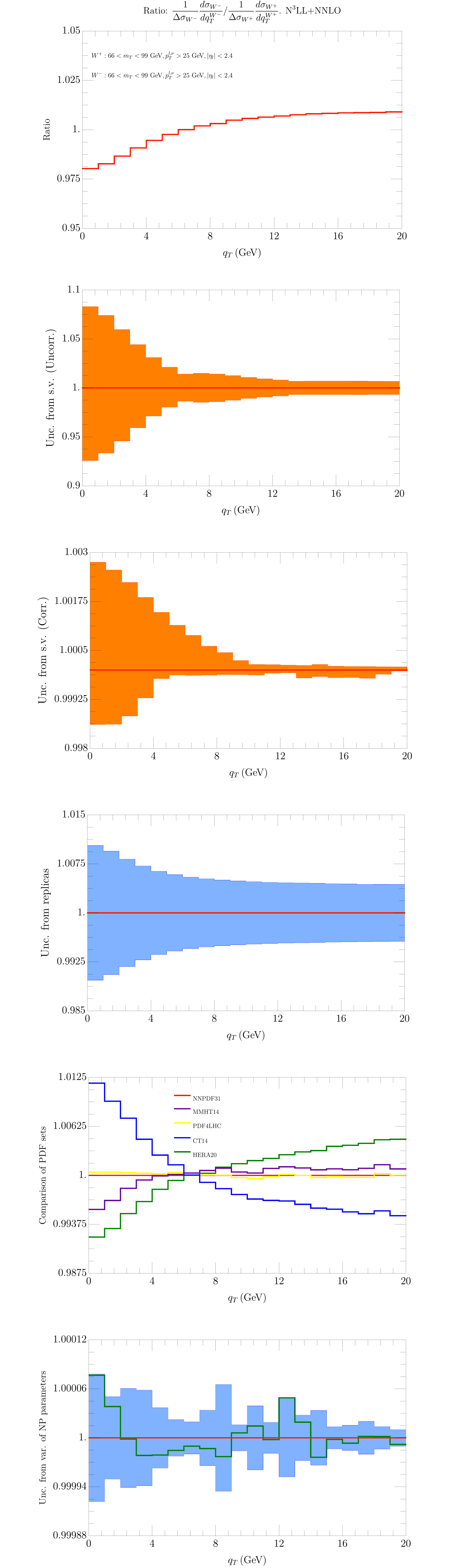}
%\vspace{-3cm}
\caption{\label{fig:ratioWW}
 Ratio of $W^-/W^+$ spectrum for $m_T\in [50,66]$ GeV (left column) , and $m_T\in [66,99]$ GeV (right column).
On the first row we have the  $W^-/W^+$ spectrum using the central replica of NNPDF31\_nnlo\_as\_0118~\cite{Ball:2017nwa}.
Uncorrelated and correlated  theoretical uncertainties are given in second and third row respectively. 
In the third line we have
the error  calculated as a variance in each bin of 100 replicas of the set NNPDF31\_nnlo\_as\_0118~\cite{Ball:2017nwa}.
The uncertainty is referred to the average value of each bin (red line). The value of the observable given by the central replica is undistinguishable from the red line.
On the fourth line we represent the value of each been with different sets of PDF.
On the fifth row we have the uncertainty due to non-perturbative parameters.
}
\end{center}
\end{figure}

%%%%%%%%%%%%%%%
The ratio of $W\pm$ cross sections  shown in fig.~\ref{fig:ratioWW}. We have considered again  two intervals of $m_T$ and it is nice to observe the similarity of these curves in the two intervals. 
The theoretical errors have been estimated and scale variations represent the biggest error. We have considered both correlated and uncorrelated scale variations and their value is very similar in this case.
The scale variations keeps below 2$\%$ for $q_T>4$ GeV and grows below this value. All other uncertainties keep below the 1$\%$.

%***************************************************************************************************%

%%%%%%%%%%%%%%%%%%%%%%%%%%%%%
\section{Comparisons with other groups and experiments}
\label{sec:comp}
%%%%%%%%%%%%%%%%%%%%%%%
\begin{figure}
\begin{center}
\includegraphics[width=0.45\textwidth]{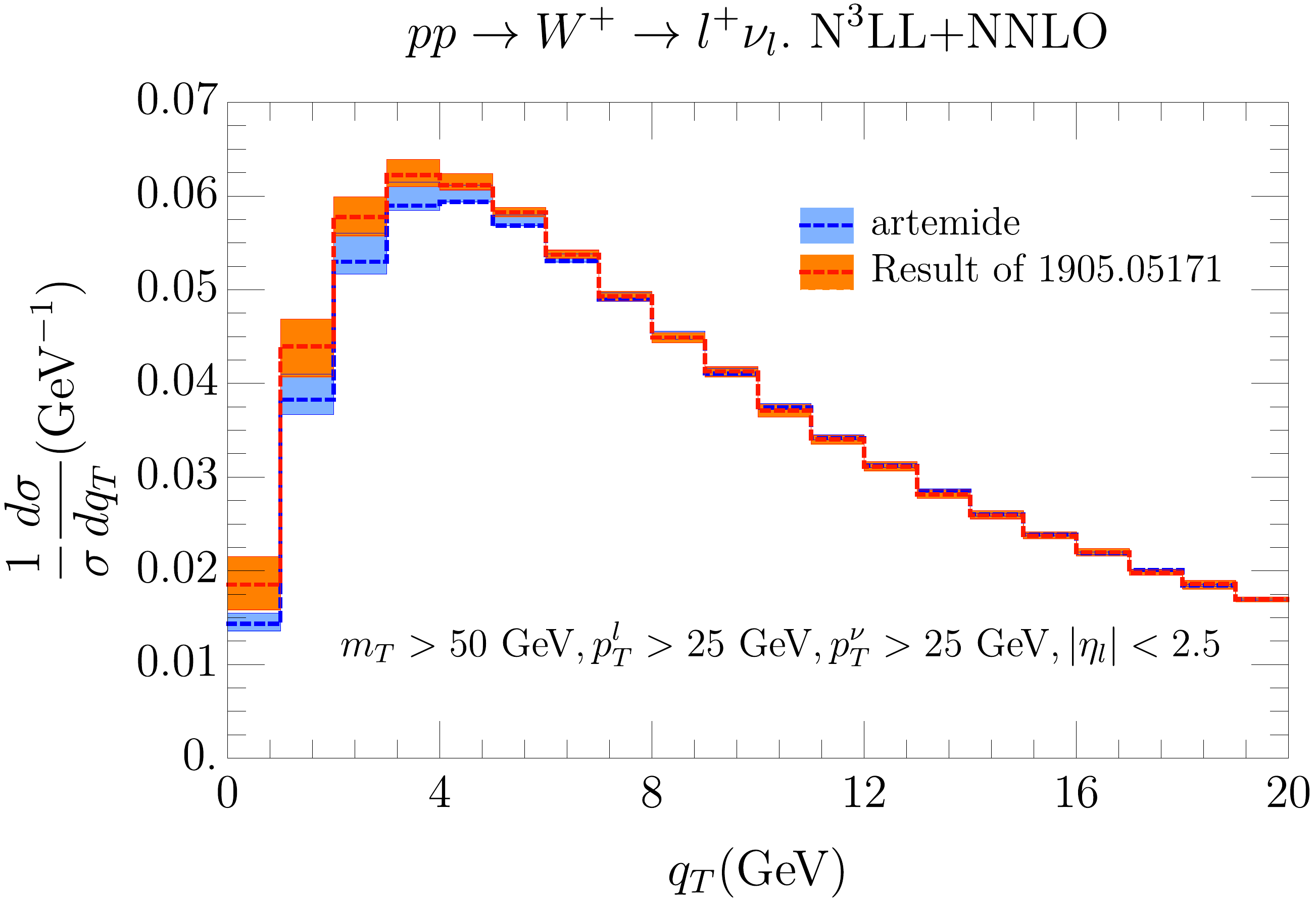}
\includegraphics[width=0.45\textwidth]{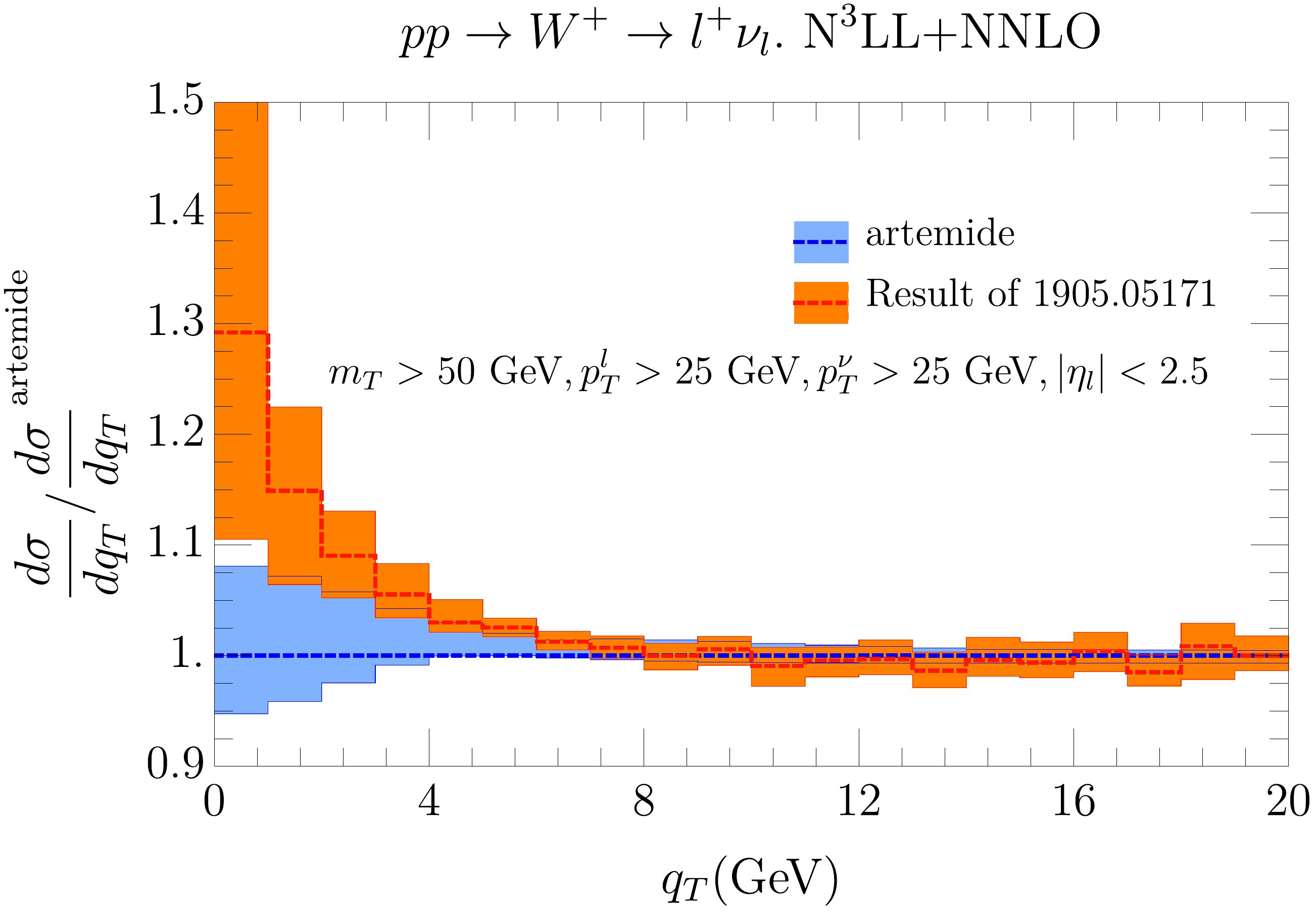}
%\vspace{-3cm}
\caption{\label{fig:Gehrmann} Comparison of our prediction with \cite{Bizon:2019zgf} including errors coming from scales variation. On the right the two cross sections are normalized to the central value of the prediction from Artemide.}
\end{center}
\end{figure}
The study of TMD in $W$-production has not been explored in its full potential in the literature, because the formulation of the TMD factorization theorem with full details is very recent.  
The main difficulty in establishing this kind of studies is that the TMD factorization only holds for $q_T\ll Q$, and in this  $W$-case  $q_T\ll m_T$. In this way we are interested in describing only a (relevant) part of the spectrum, the one which contains the peak of the distribution.

In this section we would like to compare the outcome of the results as coming from TMD factorization and recent TMD fits with theoretical and experimental results.

\begin{figure}
\begin{center}
\includegraphics[width=0.32\textwidth]{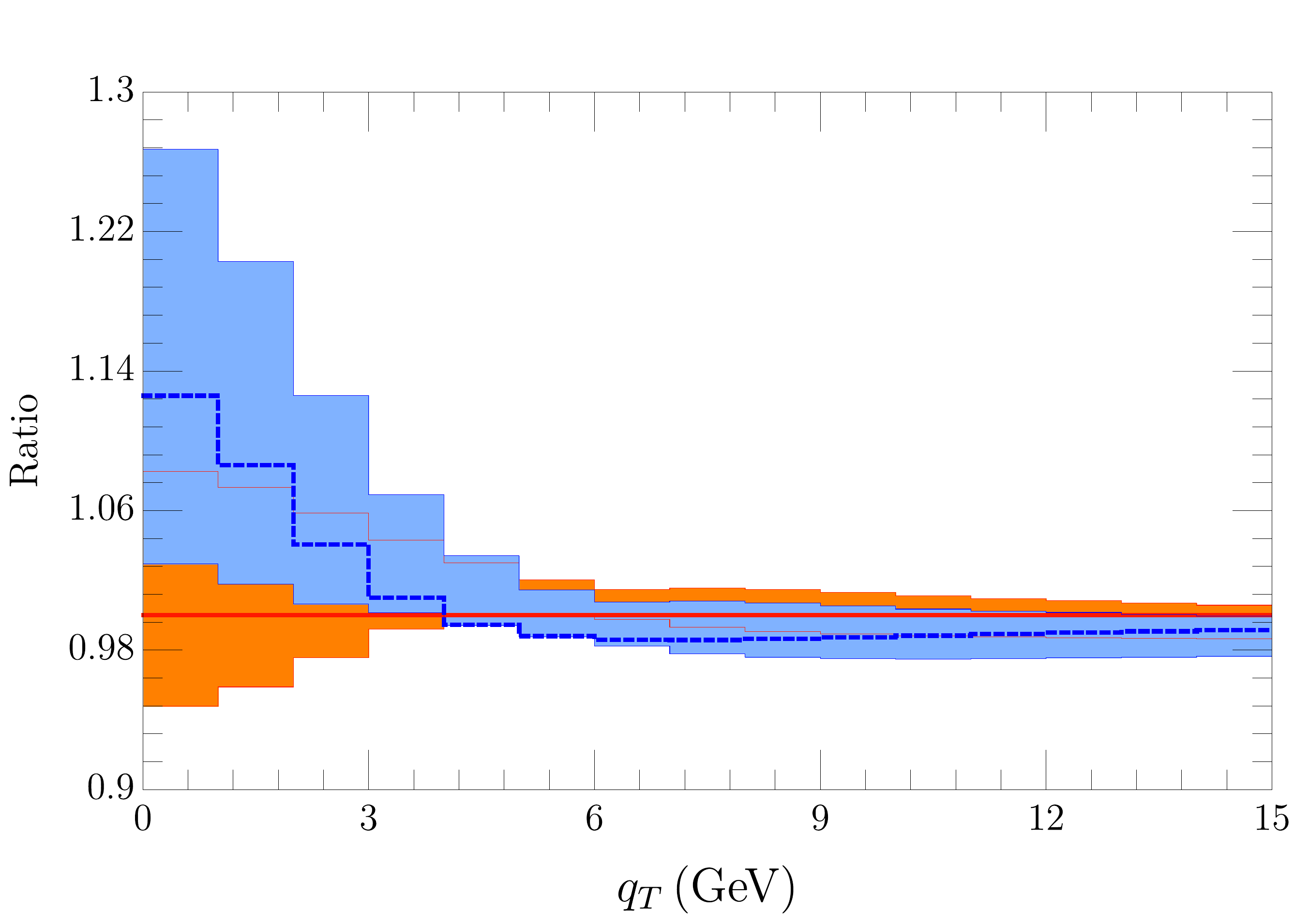}
\includegraphics[width=0.32\textwidth]{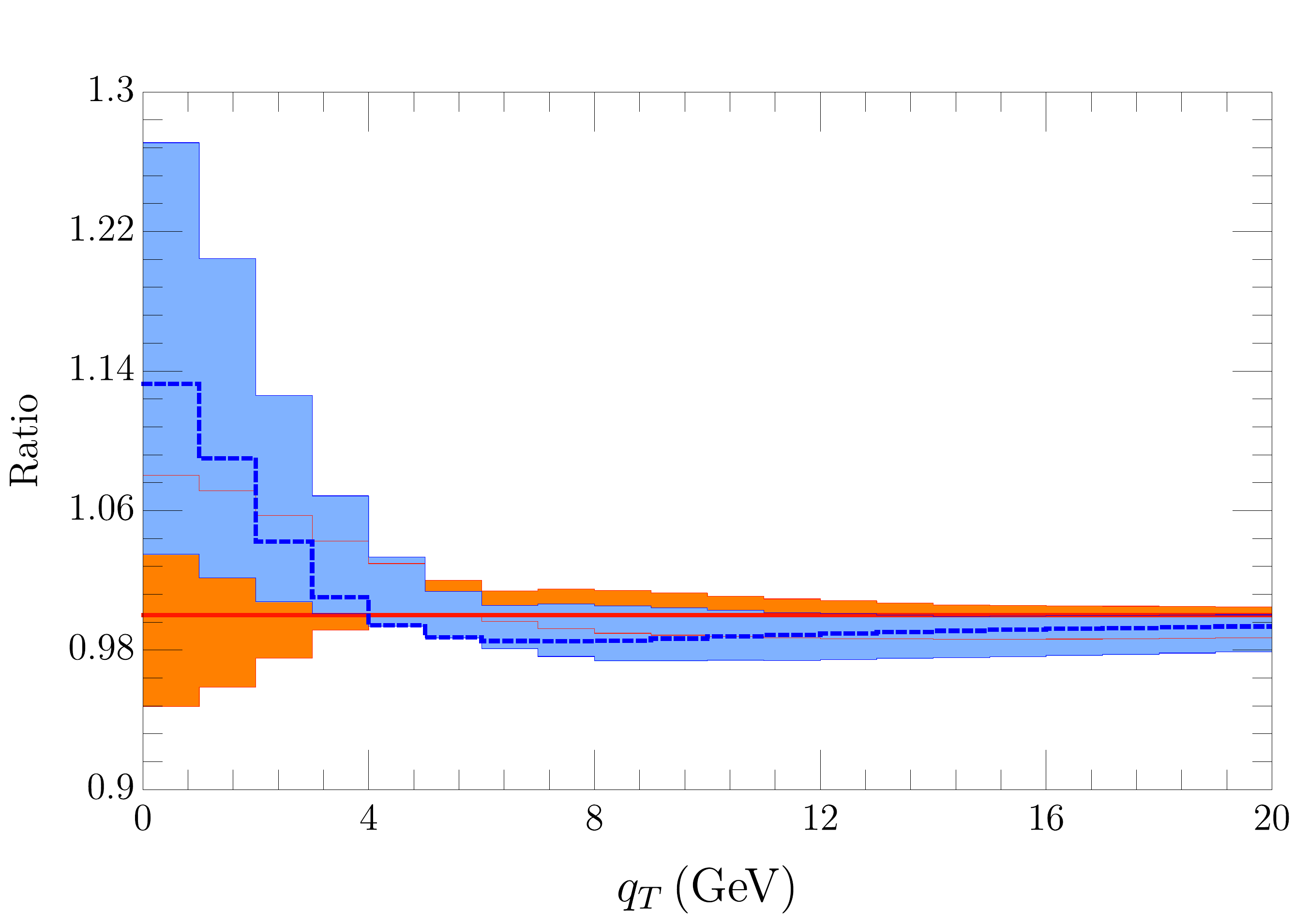}
\includegraphics[width=0.32\textwidth]{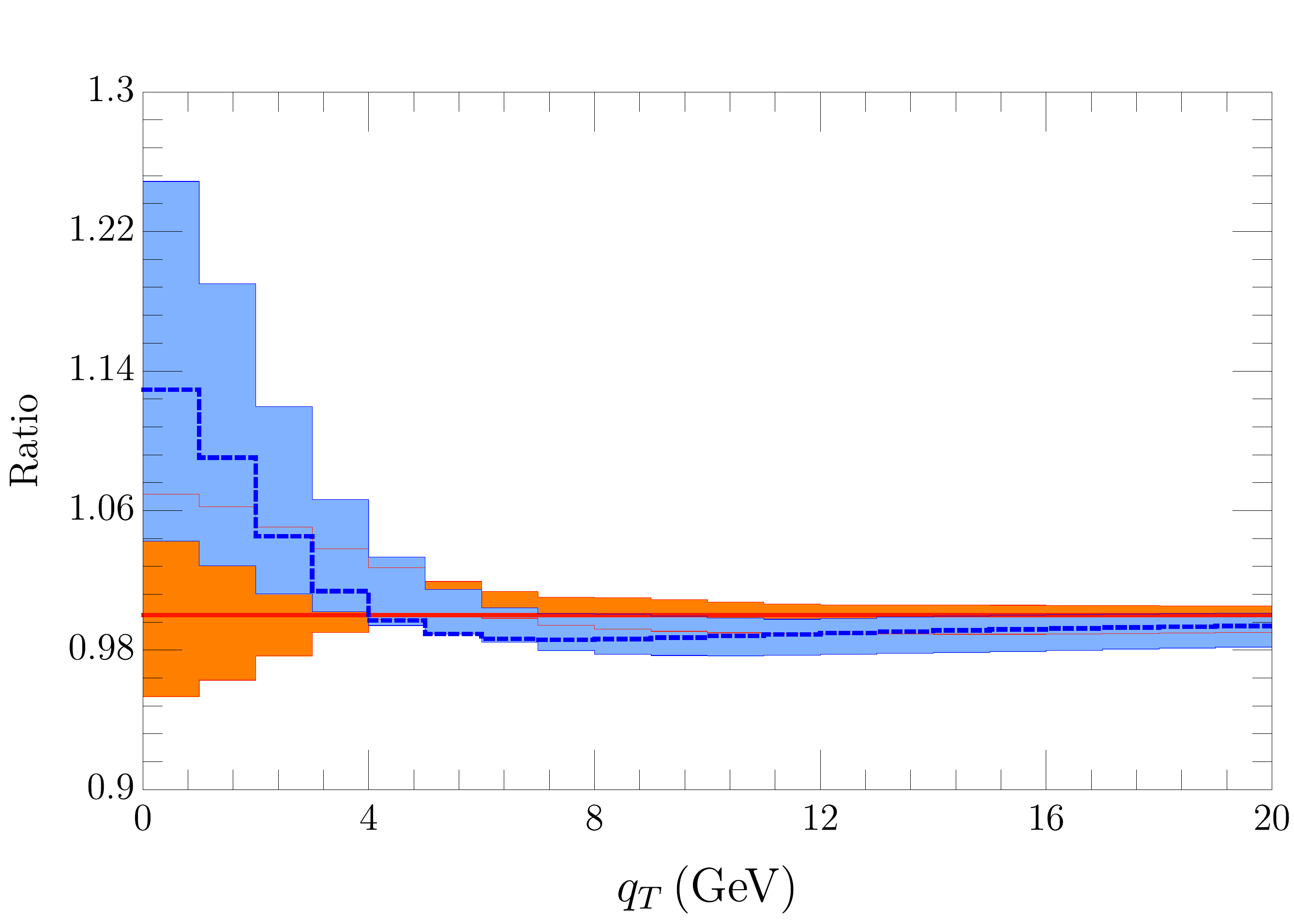}
\caption{\label{fig:la0} Ratio of $W^+$ cross section without $\lambda_i$ non-perturbative effects over the same cross section with the full model (blue band).  The scale errors are shown by the bands. The orange band  is ratio of the cross section in the full model over itself, and scalre error band is also shown. The left, central, right panels correspond respectively to the cases of $50\; {\rm GeV}< m_T<66; {\rm GeV}$,  $66\; {\rm GeV}< m_T<99; {\rm GeV}$, $99\; {\rm GeV}< m_T<120; {\rm GeV}$. }
\end{center}
\end{figure}

A theoretical prediction of the LHC case can be found in  \cite{Bizon:2019zgf} and the comparison with us is done in fig.~\ref{fig:Gehrmann}.
In this plot we show the cross section for the two groups  in the interval 0 GeV $<q_T <20$ GeV. The cross section is normalized to the total cross section as provided by  \cite{Bizon:2019zgf}, in order to have a reference value.
 The scale uncertainty  in the two curves is similar, although slightly reduced in our case. There can be multiple reasons for this, like a different choice of  scales or the usage of Monte Carlo calculation in  \cite{Bizon:2019zgf}. 
 One observes also a remarkable  difference  in the central values of the cross section for very low values of $q_T\leq5$ GeV. The origin of the difference may have multiple motives starting from the parametrization of the non-perturbative effects (which is based on a broader data analysis in our case) to the set of prescription and scale fixing used by the two groups. In general this difference is expected where non-perturbative effects are significative as it is shown by the figure, while it is much less significative for $q_T\geq 5 $ GeV.  In order to explore this we have considered a modified version of the TMD where the effect of the  constants $\lambda_i$ in eq.~(\ref{eq:fnp}) is nullified. This is achieved considering $f_{NP}(x,\vecb b)=\exp(-\lambda b^2)$ with $\lambda\simeq 10^{-3}$. The results are shown in fig.~\ref{fig:la0}  and they show to be consistent with fig.~\ref{fig:Gehrmann} where the TMD non-perturbative part was not considered, despite the fact that the kinematical cuts considered in the two figure are slightly different. We postpone anyhow a deeper study to a different work. 
%The plot shows an agreement for $q_T>8$ GeV, while for lower values of the transverse momentum TMD effects are evident.

%\begin{align}
%\label{eq:fnp}
%f_{NP}(x,\vecb b)=\exp \(-\frac{\la_1(1-x)+\la_2 x +x(1-x) \la_5}{\sqrt{1+\la_3 x^{\la_4}\vecb b^2}}\),
%\end{align}

\begin{figure}
\begin{center}
\includegraphics[width=0.45\textwidth]{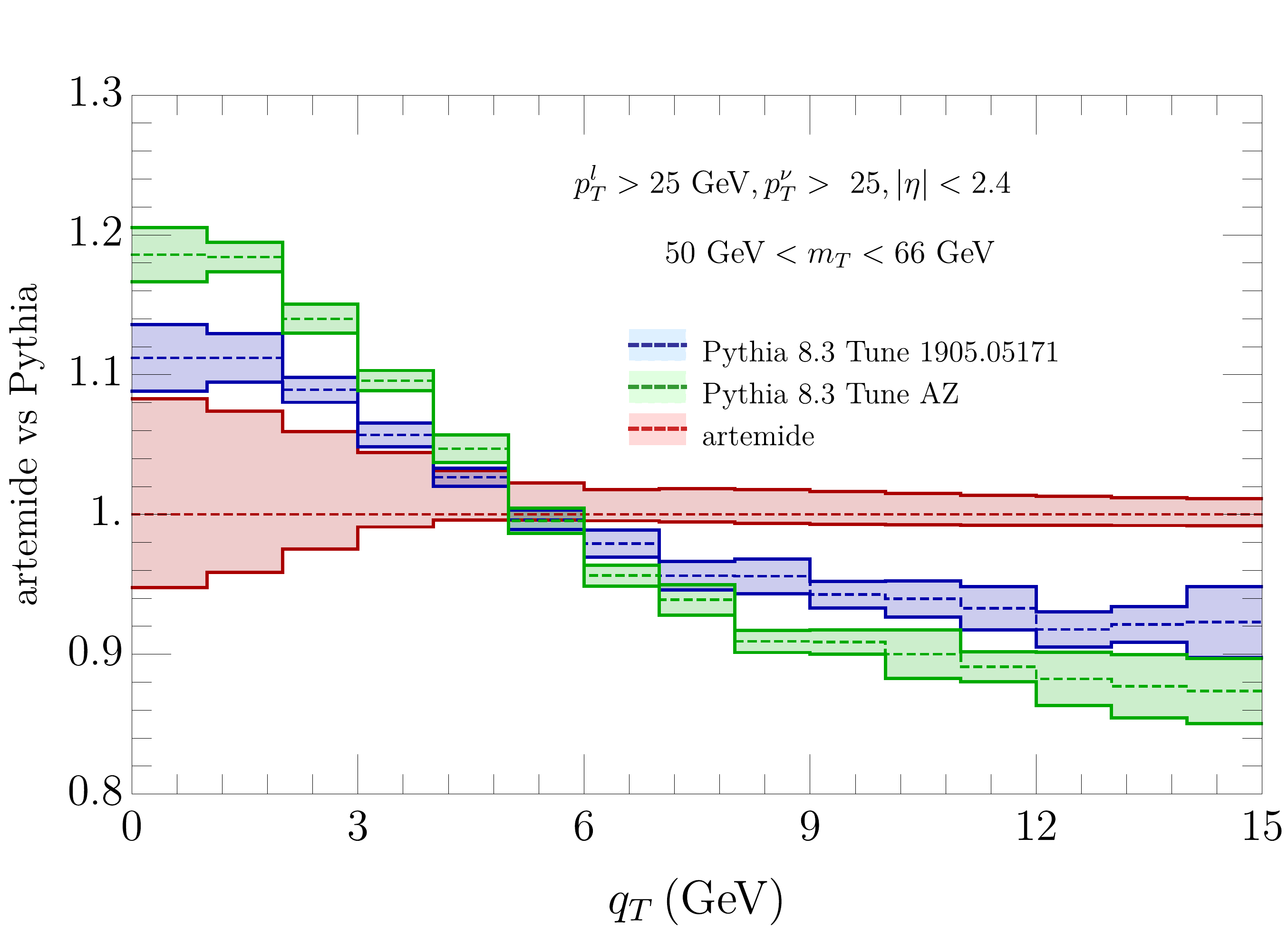}
\includegraphics[width=0.45\textwidth]{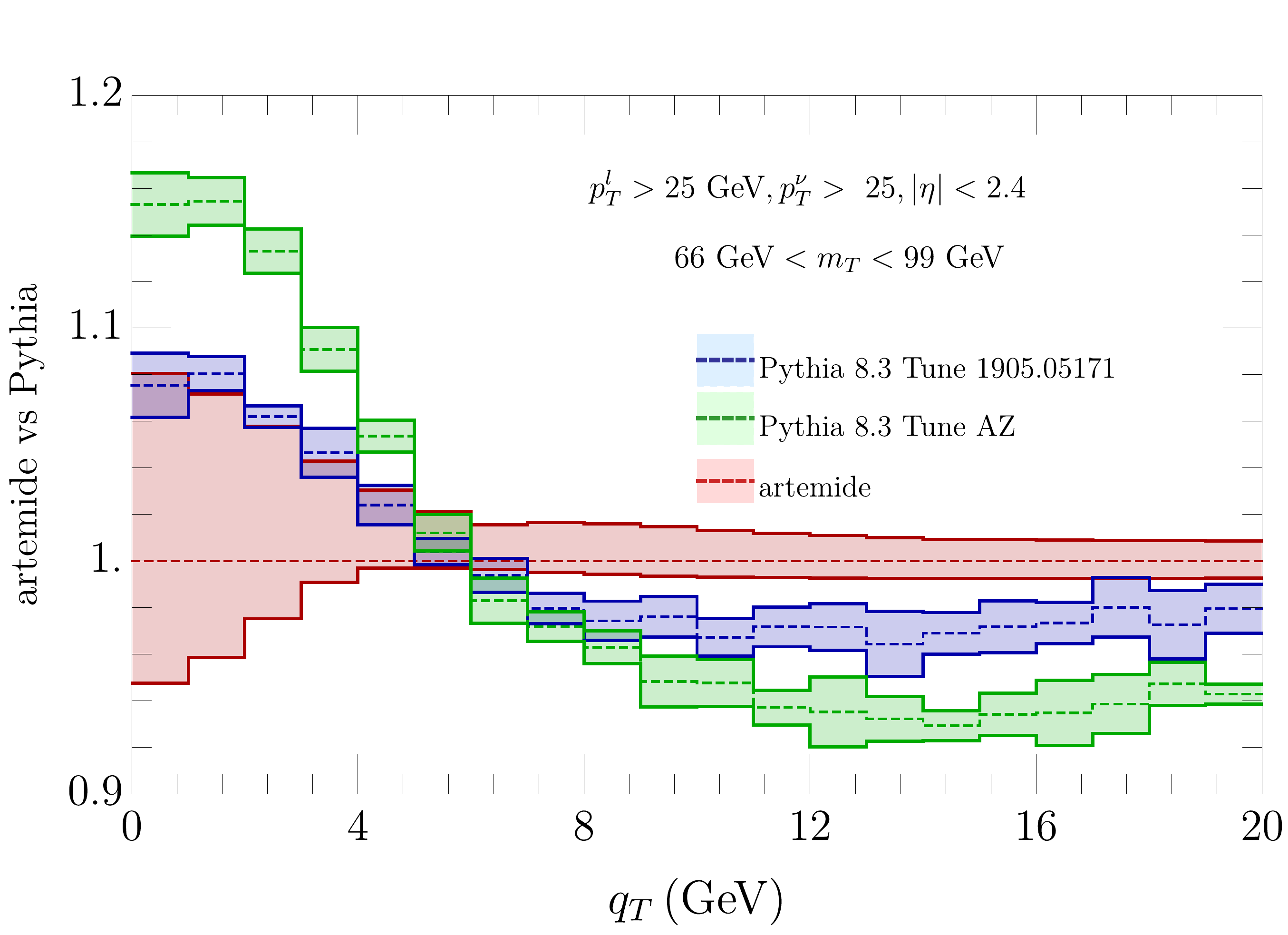}\\
\includegraphics[width=0.45\textwidth]{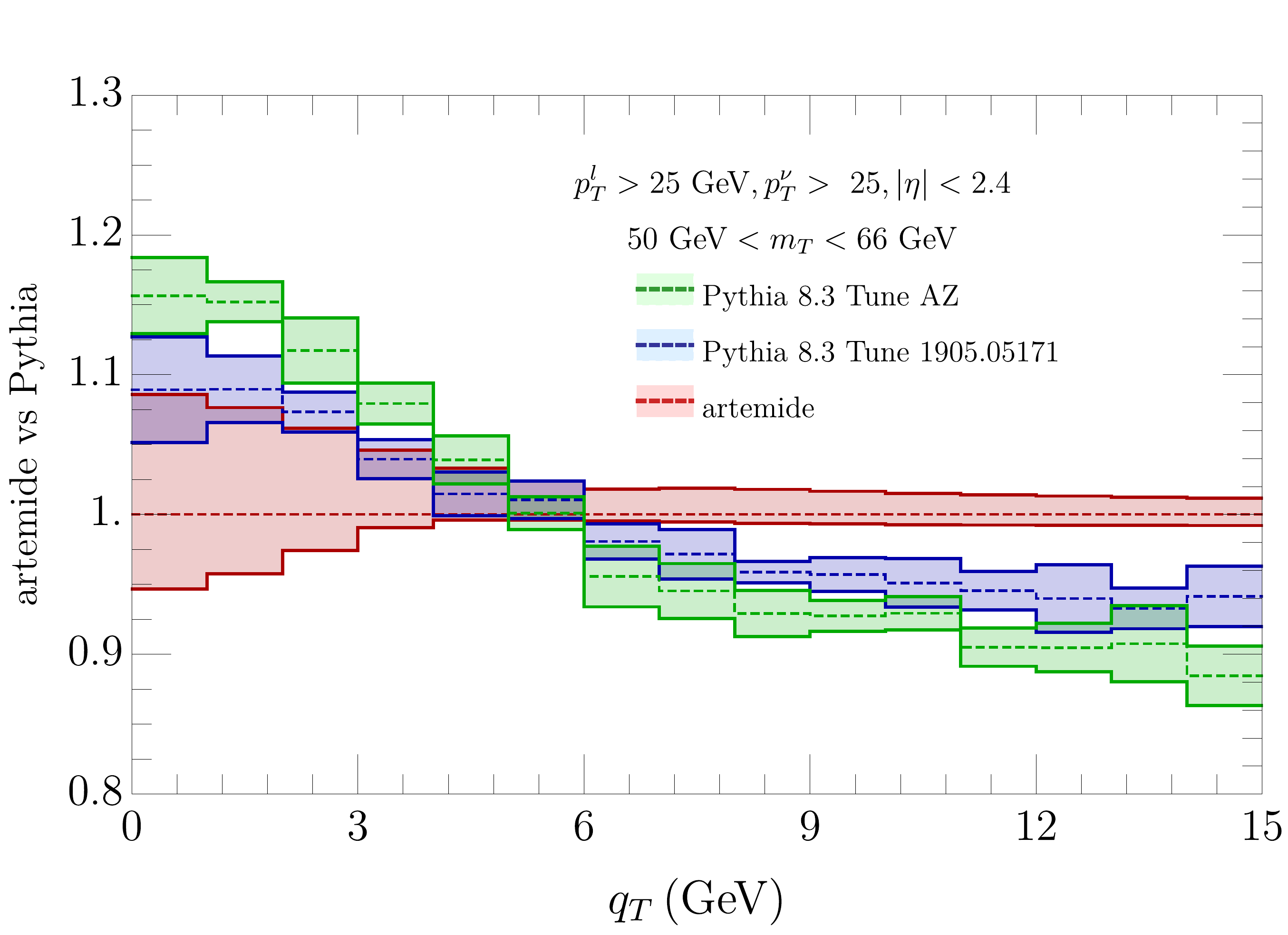}
\includegraphics[width=0.45\textwidth]{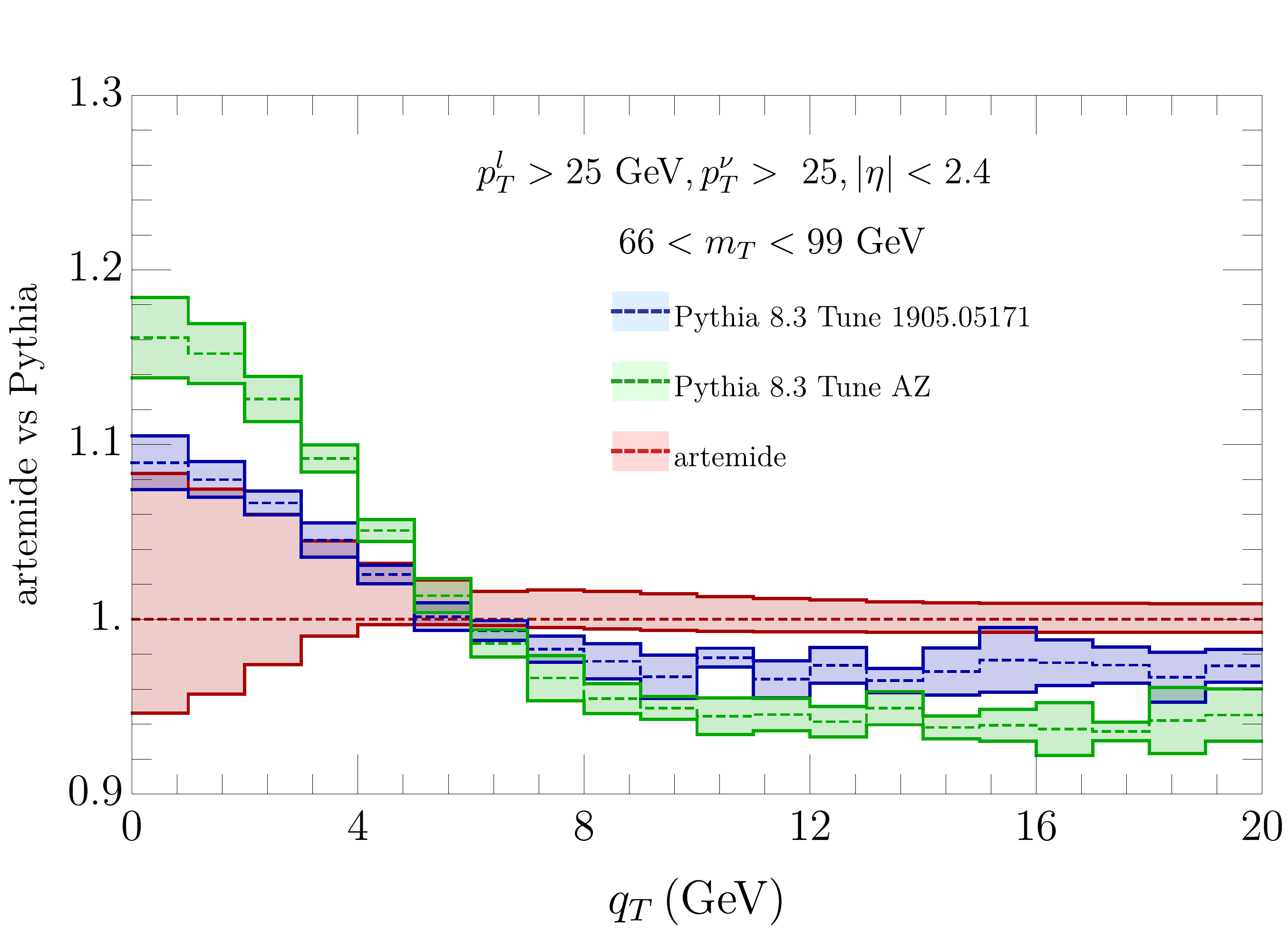}\\
%\includegraphics[width=0.45\textwidth]{Figures/CompPierwrt1.pdf}
%\vspace{-3cm}
\caption{\label{fig:Pythia1} Comparison of Artemide cross section with Pythia 8.3 AZ tune as in \cite{Bizon:2019zgf} (blue band) and as in the original ATLAS release (green band) for $W^+$ (top panels) and $W^-$ (low panels). The Artemide error comes from scale variations, the Pythia errors are commented in the text. }
\end{center}
\end{figure}

\begin{figure}
\begin{center}
\includegraphics[width=0.45\textwidth]{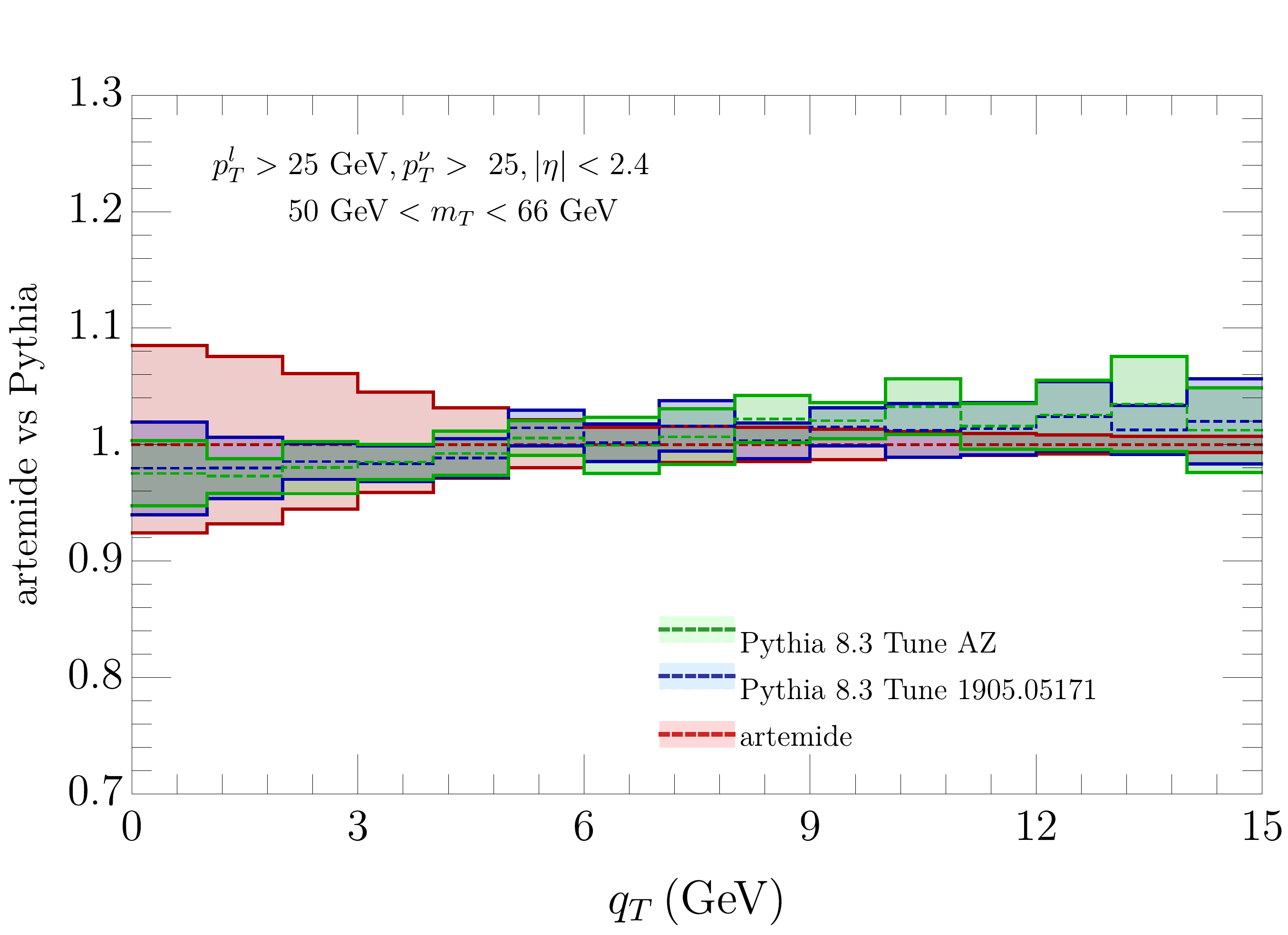}
\includegraphics[width=0.45\textwidth]{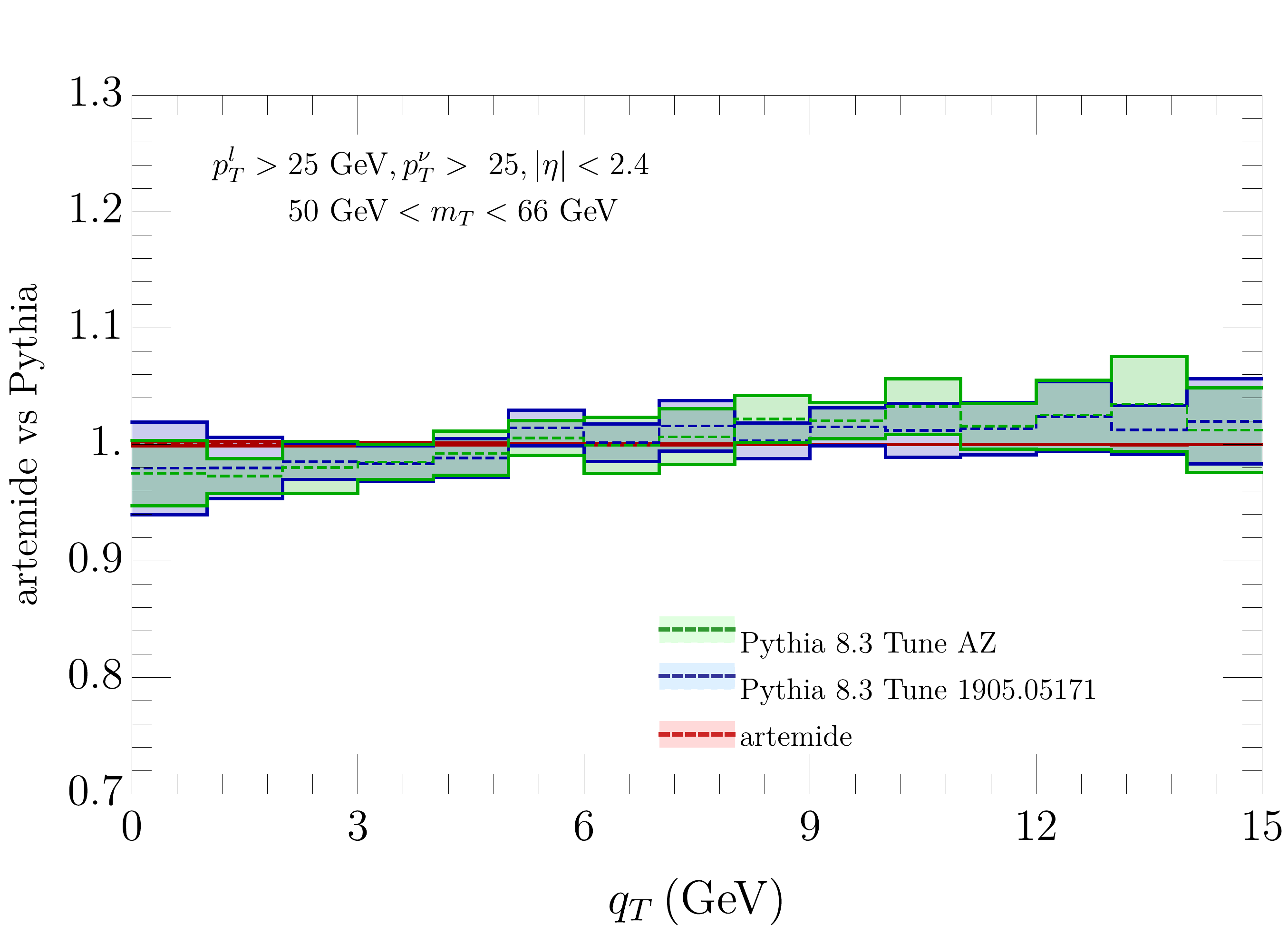}\\
\includegraphics[width=0.45\textwidth]{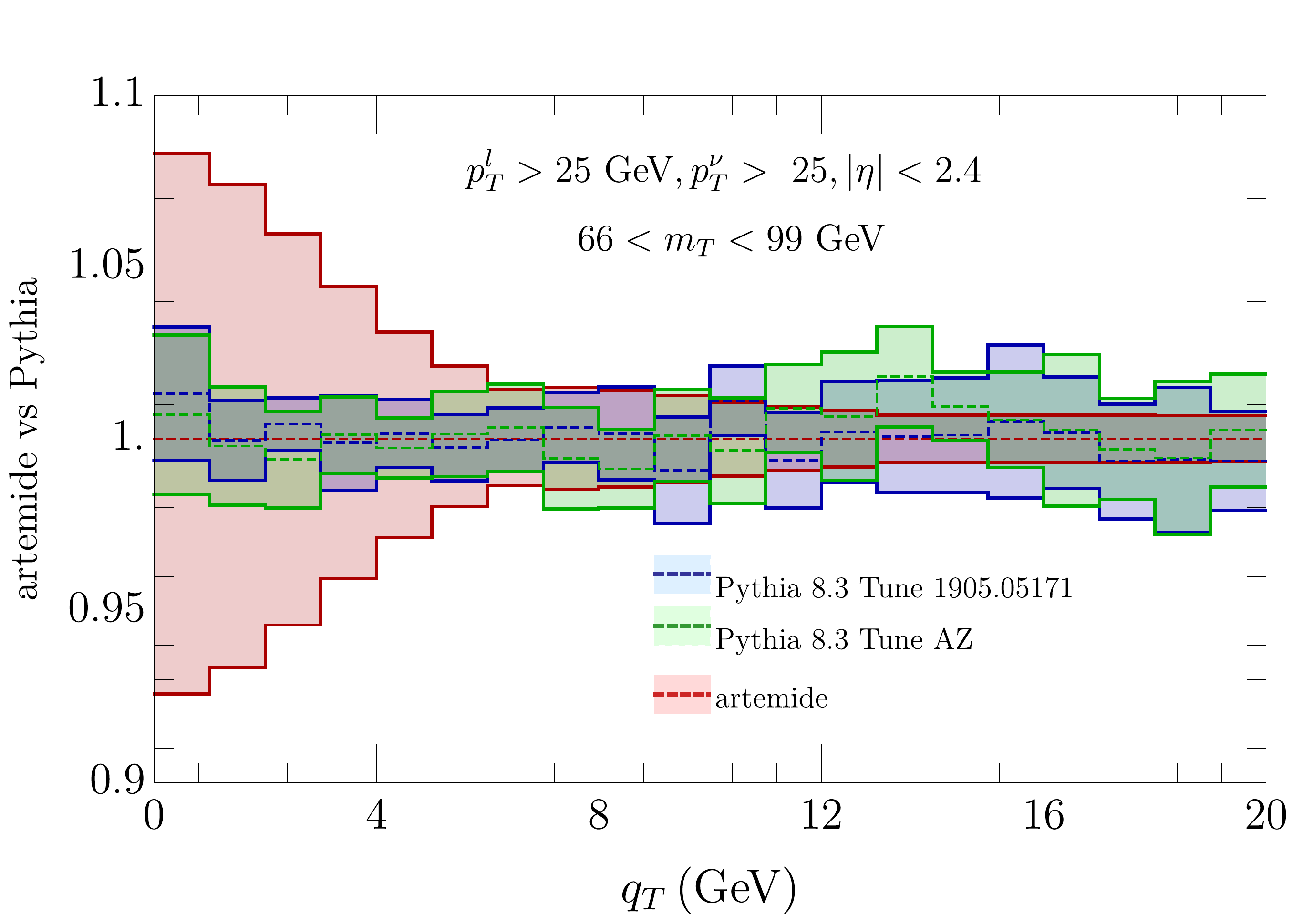}
\includegraphics[width=0.45\textwidth]{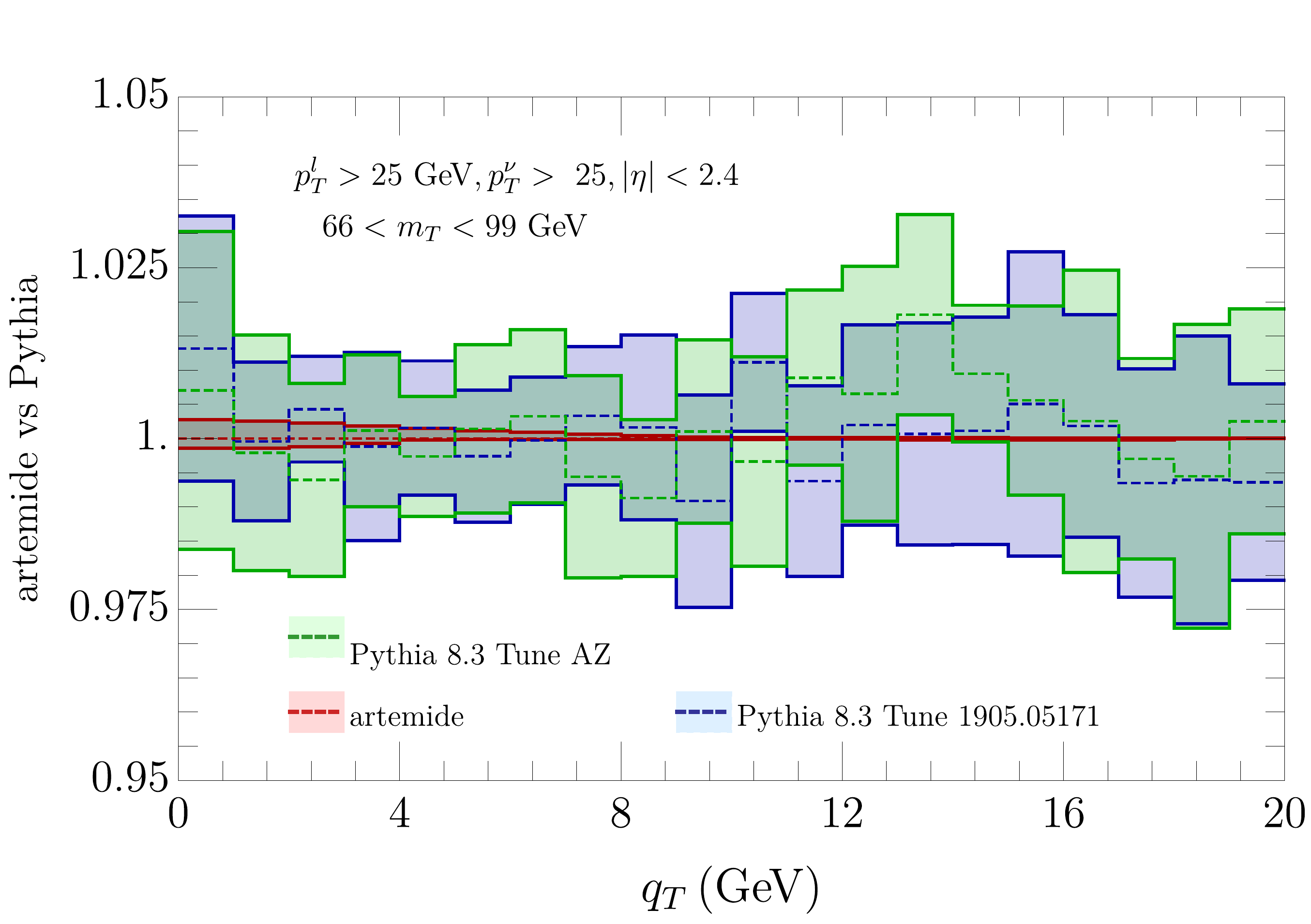}\\
%\includegraphics[width=0.45\textwidth]{Figures/CompPierwrt1.pdf}
%\vspace{-3cm}
\caption{\label{fig:Pythia2} Comparison of Artemide $p_T^{W^+}/p_T^{W^-}$ ratio with Pythia 8.3 AZ tune as in \cite{Bizon:2019zgf} (blue band) and as in the original ATLAS release (green band). Left panels show uncorrelated errors and right panels the correlated ones. The Artemide error comes from scale variations, the Pythia errors are commented in the text.}
\end{center}
\end{figure}

\begin{figure}
\begin{center}
\includegraphics[width=0.45\textwidth]{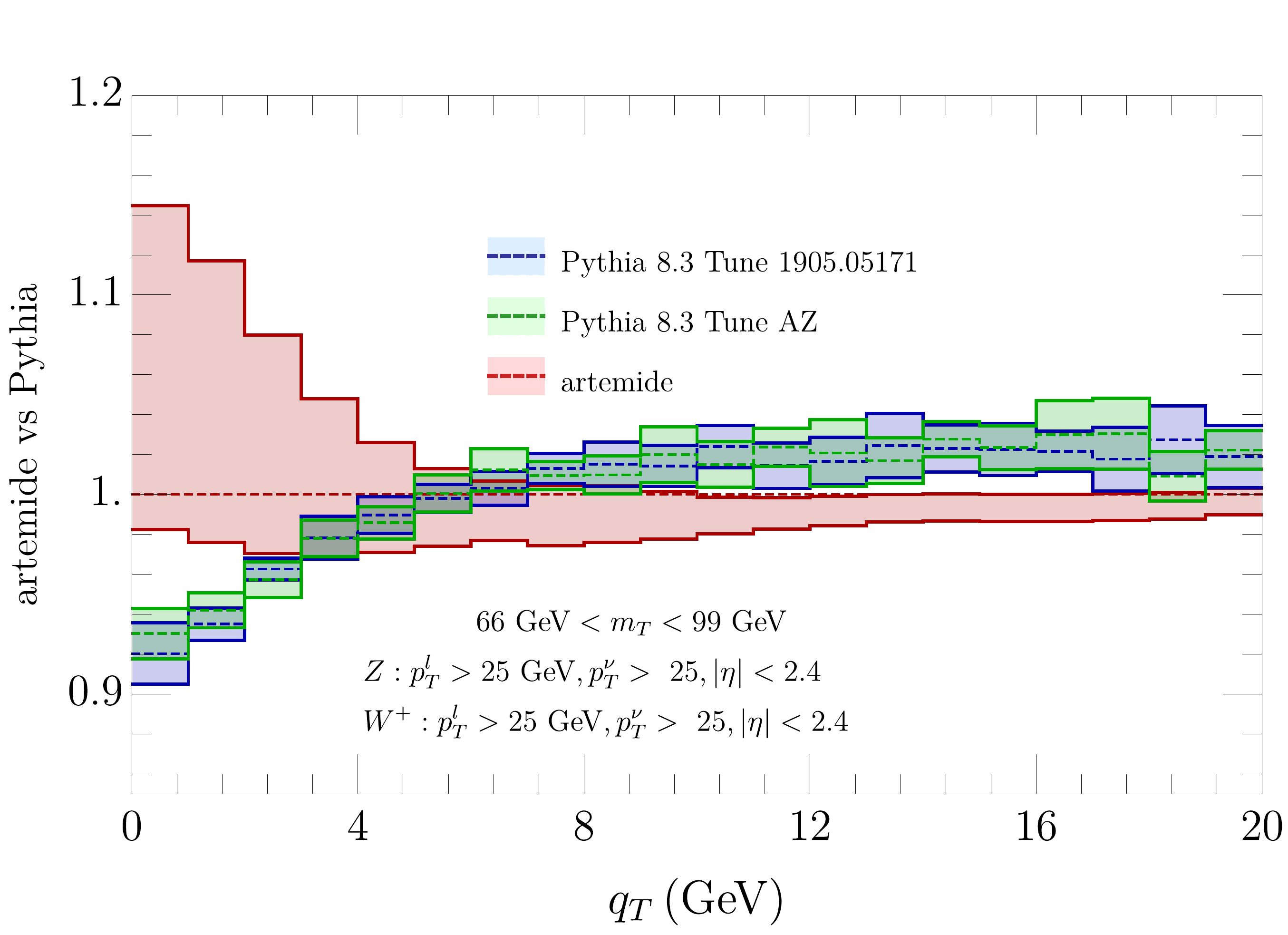}
\includegraphics[width=0.45\textwidth]{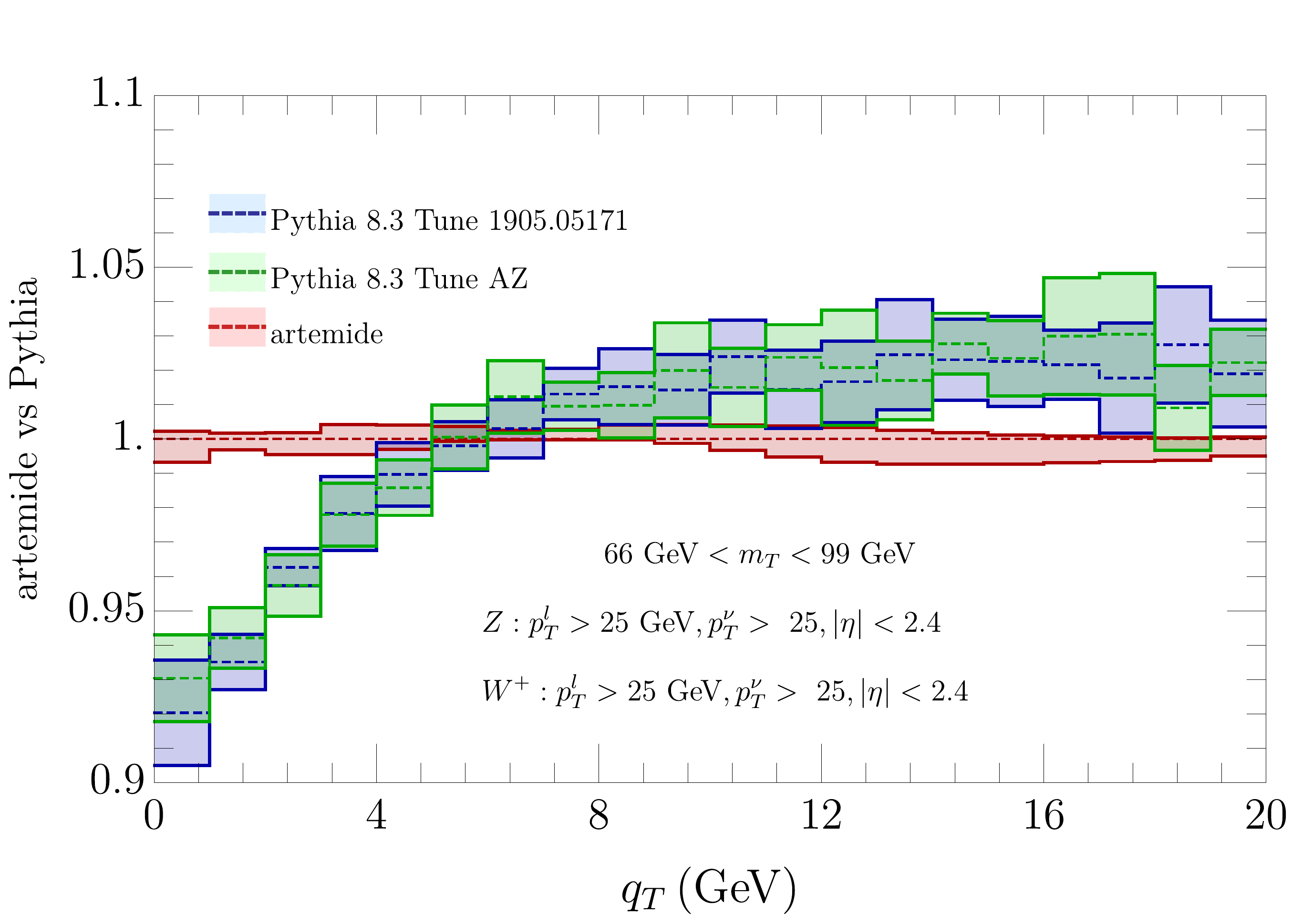}
%\includegraphics[width=0.45\textwidth]{Figures/CompPierwrt1.pdf}
%\vspace{-3cm}
\caption{\label{fig:Pythia3} Comparison of Artemide $p_T^Z/p_T^{W}$ ratio with Pythia 8.3 AZ tune as in \cite{Bizon:2019zgf}(blue band) and as in the original ATLAS release (green band).. Left panels show uncorrelated errors and right panels the correlated ones. The Artemide error comes from scale variations, the Pythia errors are commented in the text. }
\end{center}
\end{figure}

We have  considered also a comparison with Pythia 8.3 \cite{Sjostrand:2014zea}, with AZ tune as adapted by  ATLAS experiment. We have found two version of this tune one as used in  \cite{Bizon:2019zgf}\footnote{We thank Pier Monni for communicating  us that the AZ tune as provided by the Pythia collaboration does not coincide  exactly with the one used in ATLAS experiment and used actually in  \cite{Bizon:2019zgf} and one as provided by ATLAS experiment.
We have included both in our plots which are shown in  fig.~\ref{fig:Pythia1},~\ref{fig:Pythia2},~\ref{fig:Pythia3}. The pictures show the differential cross section normalized to their integration over the shown interval.
The errors from Pythia come from statistical uncertainty and we have checked that they are similar to the one obtained by the variation of the parameters of the tune.
} In the comparison of the cross sections fig.~\ref{fig:Pythia1} we observe  a general shift of the Pythia result with respect to ours,   whose sign depends on the  value of $q_T$, 
although the two estimates are mostly compatible within the errors (more for 66 GeV $<m_T<$ 99 GeV than in the less energetic interval). In the ratio of $W^\pm$ cross section \ref{fig:Pythia2} instead the agreement is complete, both considering correlate and uncorrelated errors, providing Artemide a smaller error band. Similar conclusion come from the $p_T^Z/p_T^W$ in fig.~\ref{fig:Pythia3}, although with more difference   for transverse momentum less than 5 GeV.

\begin{figure}
\begin{center}
\includegraphics[width=0.5\textwidth]{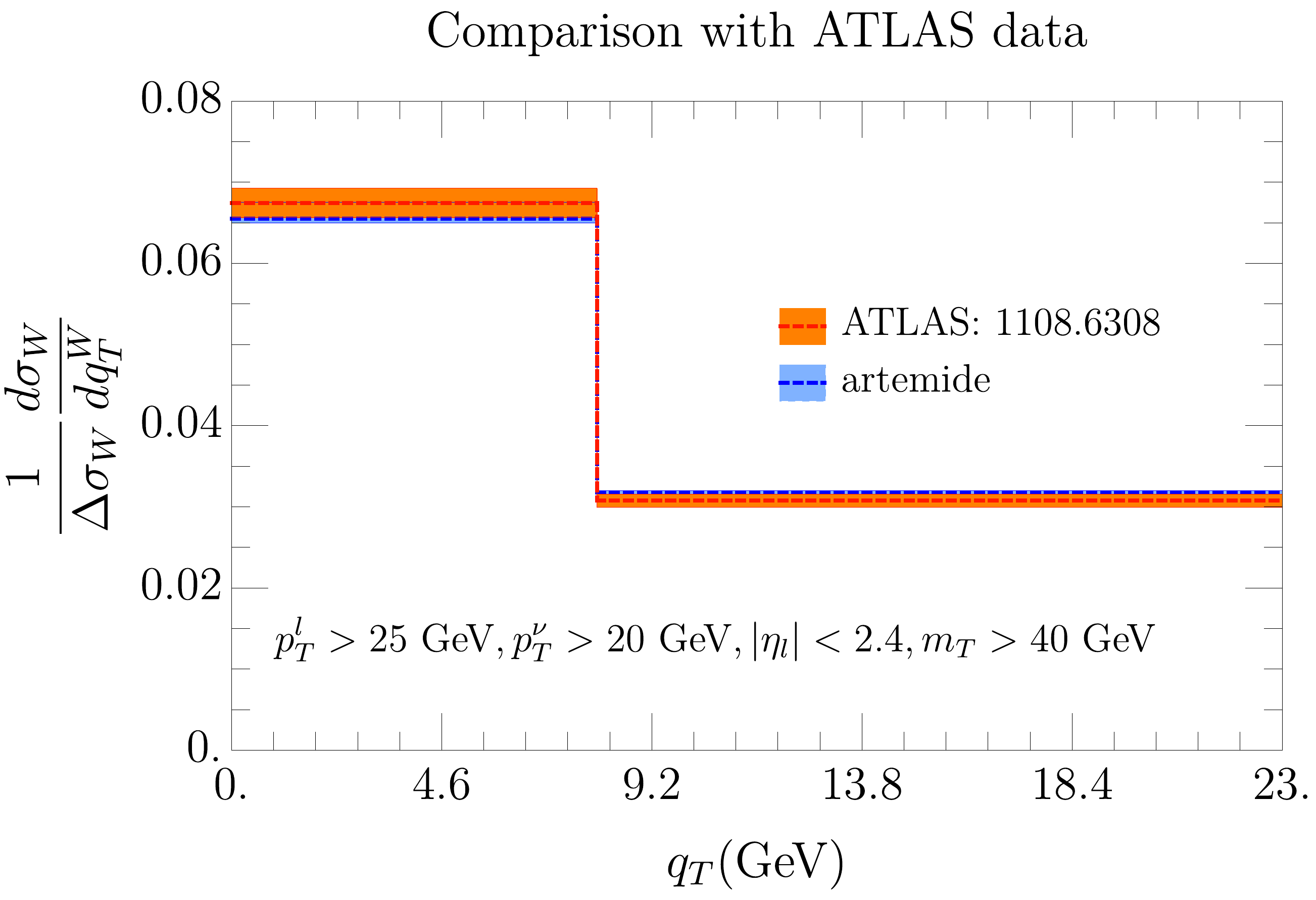}
%\vspace{-3cm}
\caption{\label{fig:ATLAS} Comparison of our prediction with data from ATLAS in% \cite{Aaboud:2017svj}
 \cite{Aad:2011fp}, table II,  including errors coming from scales variation. 
 }
\end{center}
\end{figure}

Current experiments at LHC at the moment have a limited number of data points. For ATLAS we have found  that for the low $q_T$ region,  one has only two points in   \cite{Aad:2011fp}, %\cite{Aaboud:2017svj}
 shown in fig.~\ref{fig:ATLAS}.
The cross section is normalized to its integrated value in the plotted $q_T$ interval. The theoretical and experimental  values are compatible within the errors. However it is evident that such a large binning does not allow a precise statement.

\begin{figure}
\begin{center}
\includegraphics[width=0.4\textwidth]{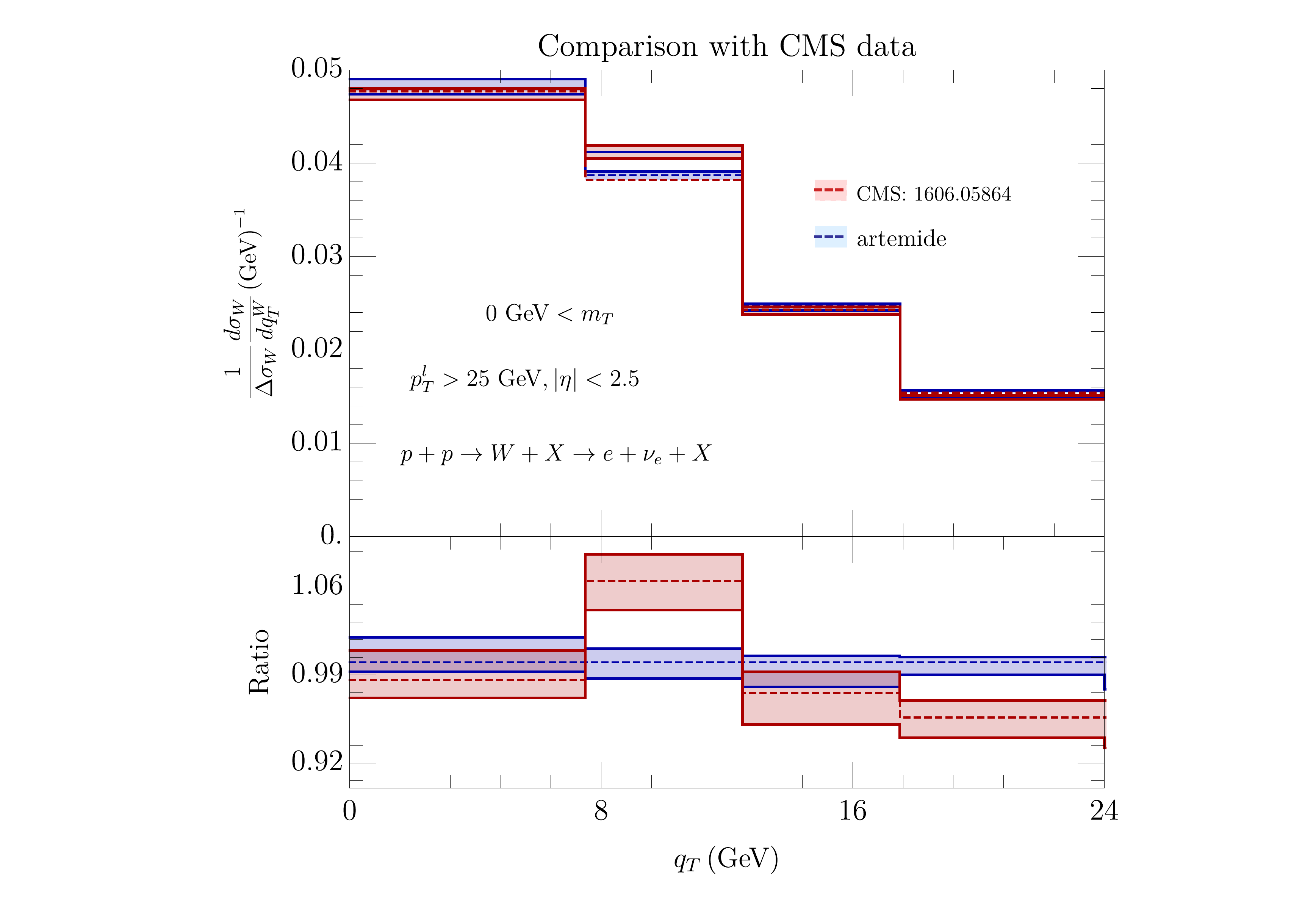}
\includegraphics[width=0.4\textwidth]{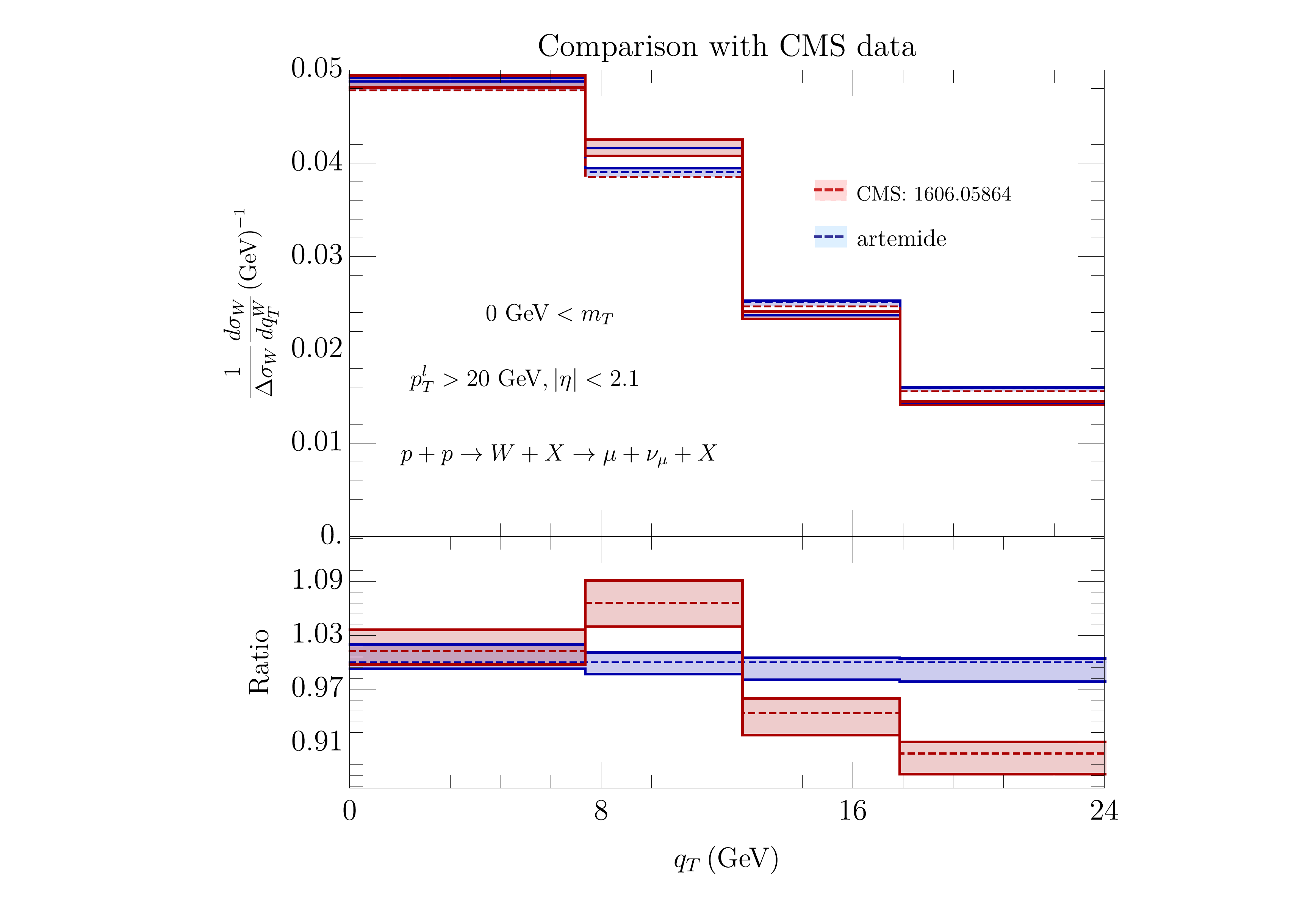}\\
\includegraphics[width=0.4\textwidth]{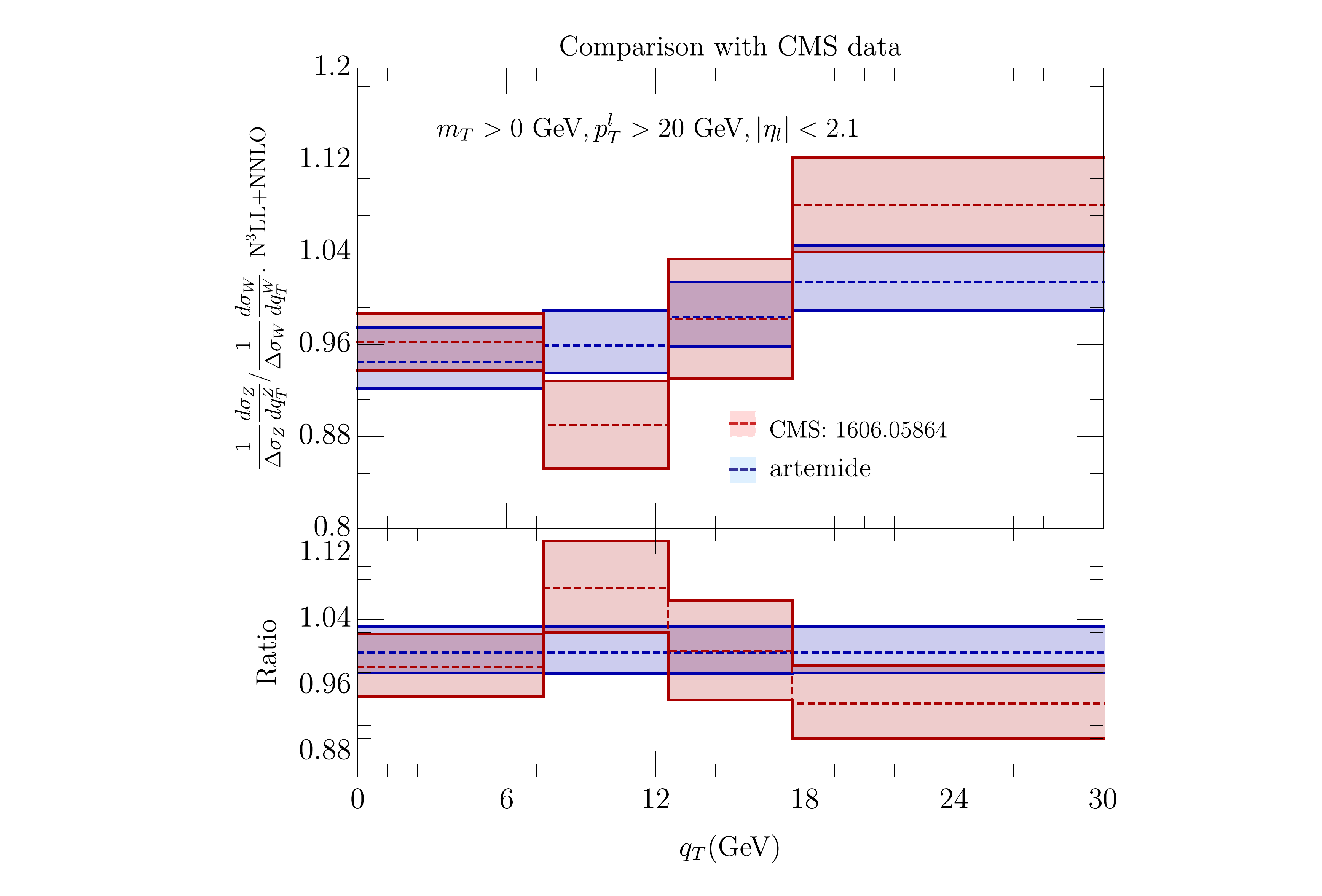}
\includegraphics[width=0.4\textwidth]{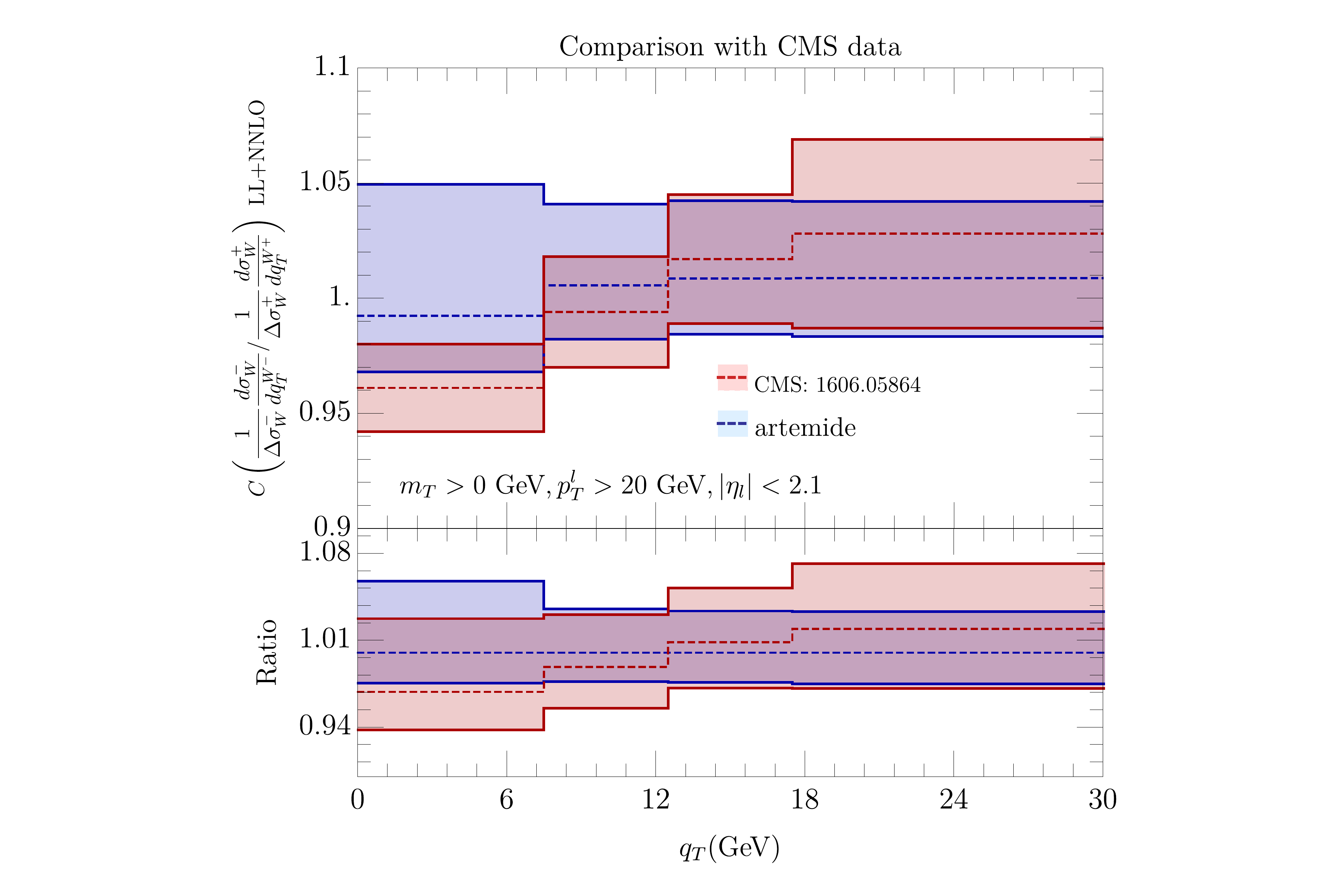}
%\vspace{-3cm}
\caption{\label{fig:cms} Comparison of our prediction with data from CMS in \cite{Khachatryan:2016nbe}.  Theoretical predictions include error coming from scales variation. The CMS data are always evaluated  for $p_T^l> 25$ GeV, $p_T^\nu> 20$ GeV, $|\eta_l|<2.4$.(Top left panel) $W$ boson normalized spectrum for electron final state, (top right panel) $W$ boson normalized spectrum for muon final state, (bottom left panel) $W/Z$ ratio of transverse momentum normalized spectrum,  (bottom right panel)  $W^+/W^-$ ratio of transverse momentum normalized spectrum.
 }
\end{center}
\end{figure}

The CMS experiment has published an analysis of data for the $s=\sqrt{8}$ TeV run in \cite{Khachatryan:2016nbe}. 
%{\bf Option 1}
%The experiment produces a spectrum totally inclusive in $m_T$, which cannot be  much used in the present TMD analysis because we cannot check the condition $q_T\ll m_T$.
%The conclusion of this section is so similar to the one of \cite{Hautmann:2020cyp}: a different binning and set of fiducial cuts is necessary to unveil the properties of TMD at LHC, which is certainly possible in  future analysis and runs.
%We explore this possibility in the rest of the paper.
%{\bf Option 2}
%
The data published by this collaboration are totally inclusive in $m_T$, however because of the cuts on leptons, the  values of $m_T\leq 20$ GeV give  small contributions. 
%Because the condition $q_T\ll m_T$ is not fully  accomplished in this case, in principle we expect some difference with the result from Artemide. However we find interesting to plot the TMD cross-section for the CMS to see how much we miss in our formalism. 
In fig.~\ref{fig:cms} we consider the cases of the $W$, $Z/W$  and $W^-/W^+$ transverse momentum spectrum comparing our results  with  \cite{Khachatryan:2016nbe}.
%We consider so two cases for our calculation. In the first one we consider the fiducial cuts exactly as  in \cite{Khachatryan:2016nbe} and we show it in fig.~\ref{fig:cms} (left).
%In the second case we consider the effect of putting $m_T>40$ GeV, fig.~\ref{fig:cms} (right), similarly to ATLAS. 
 The large bin  for $q_T>18$ GeV is the  most sensitive to power corrections which however do not seem to create particular problems.
 The first thing that one can observe is that there is a good agreement between data and our prediction when theoretical errors are included, which suggests the fact that most of the QCD corrections in  the CMS experiment are due to  the TMD region. 
 %The central value of our prediction however is not within the experimental error band (we comment one this later in this section.)  

 %A more complete prediction  from the theoretical side would include a matching to fixed order which hopefully can account for the residual difference that we observe.
%On the other side, putting a cut on $m_T>40$ GeV, increases very modestly this difference which suggests that actually the TMD region is responsible for most of the QCD corrections.

\begin{figure}
\begin{center}
\includegraphics[width=0.33\textwidth]{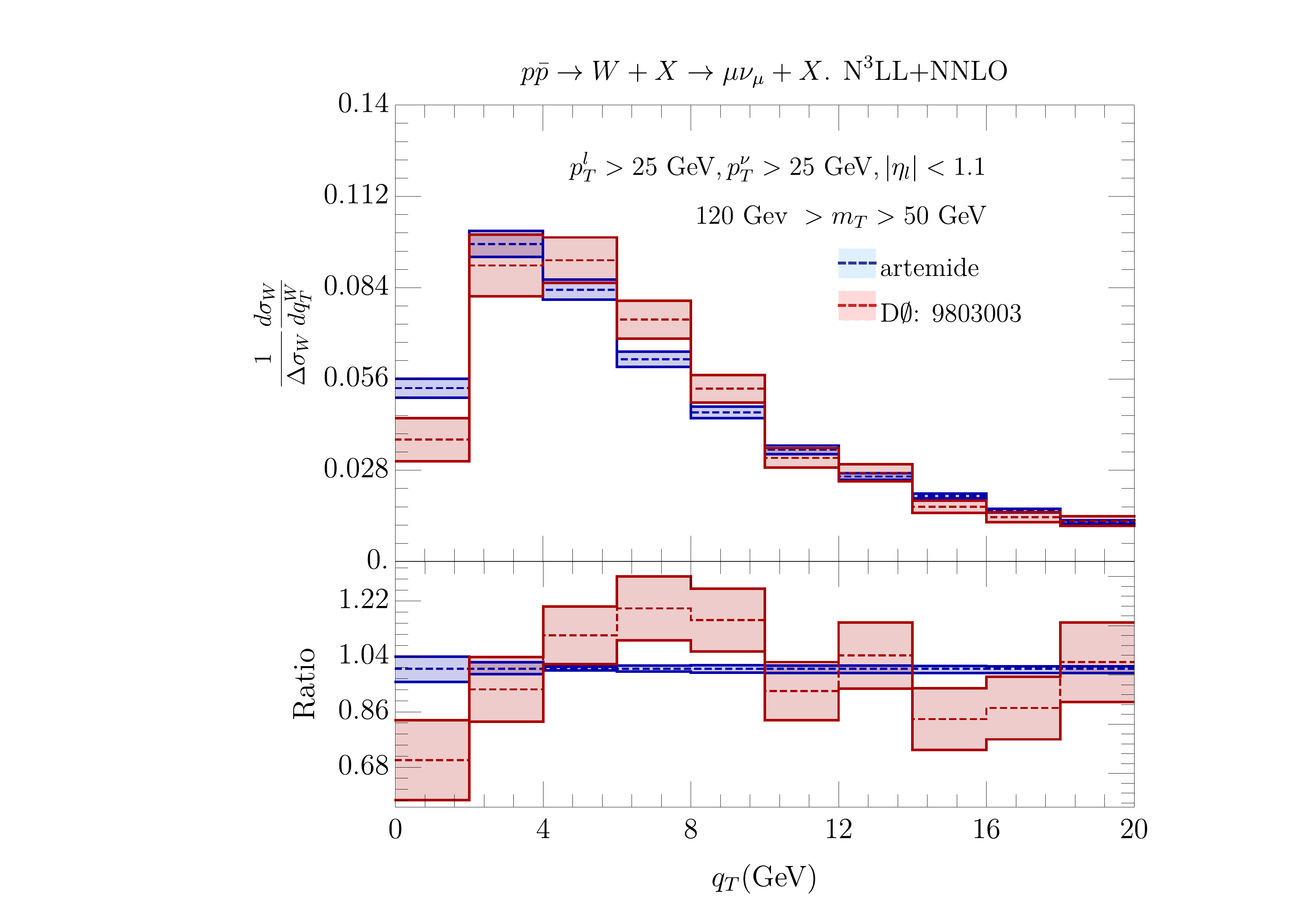}
\includegraphics[width=0.3\textwidth]{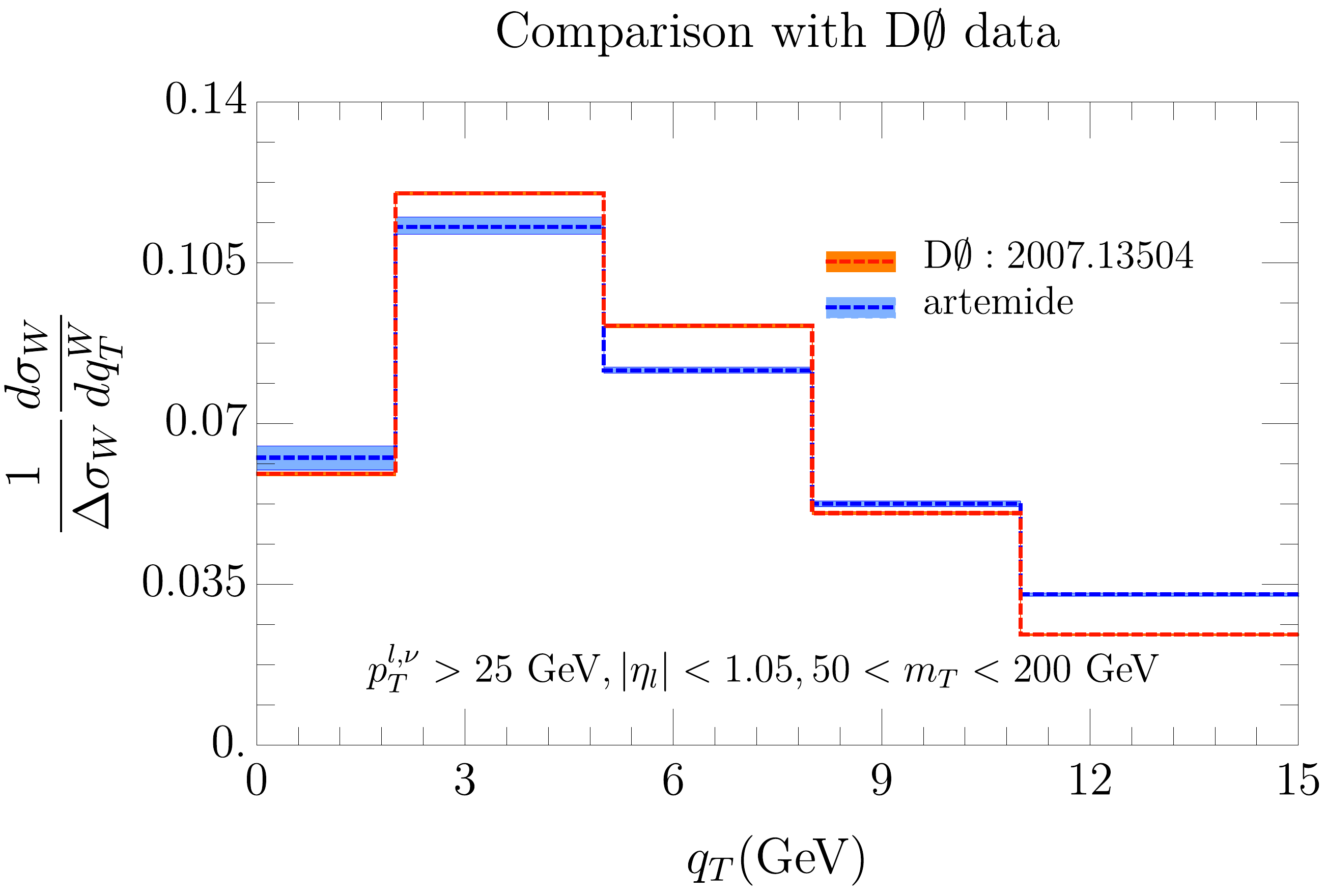}
\includegraphics[width=0.33\textwidth]{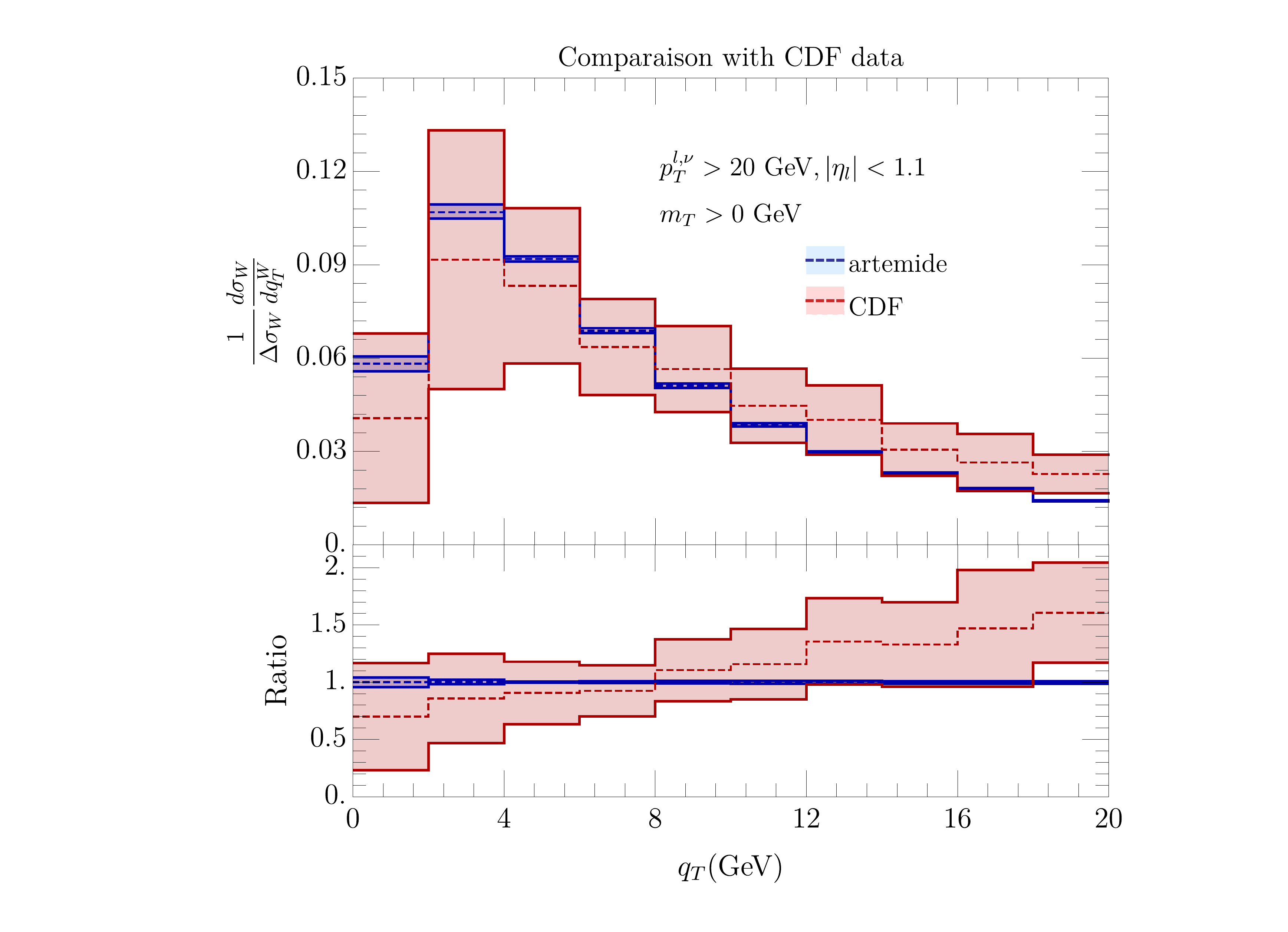}
%\vspace{-3cm}
\caption{\label{fig:D0} Comparison of our prediction with data from $D\slashed 0$ experiment at $\sqrt{s}=1.8$ TeV  \cite{Abbott:1998jy}  and$\sqrt{s}=1.96$ TeV \cite{Abazov:2020moo}
and with CDF experiment  at $\sqrt{s}=1.8$ TeV \cite{Abe:1991rk}. Theoretical predictions include errors coming from scales variation.}
\end{center}
\end{figure}

Going to TeVatron experiments we have found   data for $D\slashed 0$ and CDF and we show a comparison with our prediction in fig.~\ref{fig:D0}.
In the central plot of this  figure we do not show the experimental error, because the uncertainty of the bins are correlated through non-trivial correlation matrices.
The agreement with these experiments is in general greater than with LHC, mainly due to the larger experimental error as the following considerations remark.
\begin{table}[t]
\begin{center}
\begin{tabular}{|l|c|c|c|c|c|}\hline
 & CDF $\sqrt{s}=1.8$  TeV &  D0 $\sqrt{s}=1.8$  TeV & ATLAS & CMS $e\nu$ &CMS $\mu\nu$
\\\hline
Number of points& 10 &10&2& 4(3)&4\\\hline\hline
NNPDF31 &0.650& 1.845& 1.565 & 7.284 (1.694) &21.502 
\\\hline
HERA20 &0.617 & 2.009 &0.853& 6.024(0.310) &16.090
\\\hline
MMHT14 &0.667 & 2.166& 1.406& 7.465(1.505) &21.751
\\\hline
CT14 &0.677 &2.608 &1.324 &7.974 (1.482) &21.972
\\\hline
PDF4LHC &0.660 & 2.061 &1.405 & 7.733(1.605) &22.075
\\\hline
\end{tabular}
\end{center}
\caption{\label{tab:WPDFs} $\chi^2/N$ using the extraction of TMD of \cite{Scimemi:2019cmh}. For the CMS case $W\to e \nu$  in parenthesis we also report the $\chi^2$ with one less point as explained in text.}
\end{table}

\begin{table}[t]
\begin{center}
\begin{tabular}{|l|c|c|c|c|c|}\hline
 & CDF $\sqrt{s}=1.8$  TeV &  D0 $\sqrt{s}=1.8$  TeV & ATLAS & CMS $e\nu$ &CMS $\mu\nu$
\\\hline
Number of points& 10 &10&2& 4&4\\\hline\hline
NNPDF31 &0.540& 1.485& 0.463 & 1.674 &3.165 
\\\hline
HERA20 &0.469& 1.591& 0.271 & 1.563 &3.721 
\\\hline
\end{tabular}
\end{center}
\caption{\label{tab:WPDFs2} $\chi^2/N$ using the extraction of TMD of \cite{Scimemi:2019cmh} and theoretical errors (which include scale variation and PDF error coming from 1000 replicas).}
\end{table}

We have considered the cross sections $\chi^2$  for the different experiments with different sets of PDF  and the results are shown in table \ref{tab:WPDFs} (we also report the number of relevant points for each experiment).
The CDF result are  the ones with larger errors and  TMD predictions agrees with them while for the $D\slashed 0$  at $\sqrt{s}=$1.8 GeV, the agreement is worse. We have not considered the case of $D\slashed 0$  at $\sqrt{s}=$1.96  GeV, because the computation of the error in this case involves the knowledge of the W-spectrum  up to $q_T=600$ GeV, which we do not have.  In the case of LHC, in general we have a very limited number of points for each experiment.
ATLAS has a remarkable agreement with the HERA20 PDF sets, definitely better than with other sets. The CMS case is more elaborate. 
 For the case of  electronic decay of W at CMS, the high value of $\chi^2$ is basically driven by just one point out of four, as can be seen fig.~\ref{fig:cms}, left panel (it is the point in the bin 7.5 GeV$<q_T<$ 12.5 GeV). Removing this point, the $\chi^2$ is very similar to the ATLAS case. For the muon channel instead we cannot find a particular  point which is responsible for the high $\chi^2$.
  As a final remark we recall that  the CMS observable is totally  inclusive on $m_T$, which does not allow a perfect control of factorization hypothesis  at the theoretical level.
  %it can  be possible that CMS results are more sensitive to a more complex flavor structure of TMD that has not been considered up to now; one has to account that the number of points  at low $q_T$ extracted in each $W$-boson decay channel is 4, which is not a big number.
  We have checked the impact on the $\chi^2$  of the theoretical errors (scale uncertainty and PDF replicas) for the case of NNPDF and HERA20 and we have reported it in tab.~\ref{tab:WPDFs2}.
  In both cases the errors on each bin are considered uncorrelated to the rest. The scale uncertainty error is dominated by  the variation of $c_4$, the parameter which associated to the scale at which TMD are matched onto PDF. We observe that including this error we have a big reduction of the $\chi^2$ on all experiments.
In order to understand better this issue it would help to have experimental results with a definite interval of $m_T$ and also a TMD  extraction that includes the $W$ processes, that we postpone to a future work. 
Concerning this last point  we have compared the $\chi^2$ as coming from different extractions of TMD with the code {\it{artemide}}. The different extractions are obtained with the NNPDF31,  PDF set, and using different data sets as shown in table \ref{tab:PDFs2}. We find a substantial agreement among all extractions, which suggests a mild flavor dependence of the TMD.
In tab.~\ref{tab:WPDFs2}, the muon channel  in CMS is still not fully agreeing with the theoretical prediction. This fact, which is put in evidence here  for the first time, needs further study beyond the present work.

\begin{table}[t]
\begin{center}
\begin{tabular}{|l|c|c|c|c|c|}\hline
Ref. of Fit  and Data set & CDF $\sqrt{s}=1.8$  TeV &  D0 $\sqrt{s}=1.8$  TeV & ATLAS & CMS $e\nu$ &CMS $\mu\nu$
\\\hline\hline
 \cite{Scimemi:2019cmh} SIDIS+DY &0.650& 1.845& 1.565 & 7.284 &21.502 
 \\\hline
\cite{Bertone:2019nxa}  DY  &0.651 &2.003 &1.549 &7.783 &22.302
\\\hline
\cite{Bertone:2019nxa} DY (high energy) &0.627 &1.326 &1.999 &6.347 &20.923
\\\hline
 Case 4 of \cite{Hautmann:2020cyp} (LHC) &0.694 &2.312 &1.333 &7.681 &21.704
\\\hline
\end{tabular}
\end{center}
\caption{\label{tab:PDFs2} $\chi^2/N$ using extractions from different data sets. The PDF is NNPDF31. The first line is the same as in tab.~\ref{tab:PDFs}. For all extractions we have used the NNPDF31 PDF set.}
\end{table}

%Looking at the W differential cross section, the $\chi^2$ for  $D\slashed 0$ experiment are 18.5/10 at $\sqrt{s}=$1.8 GeV, and for CDF we have  6.5/10, which suggest a good agreement of our TMD with data.
%In the case of $D\slashed 0$ at $\sqrt{s}=$1.96 GeV, the experimental precision  is improved by about order of magnitude with respect to  $D\slashed 0$ at  $\sqrt{s}=$1.8 GeV, so that the central values of our code provide a much bigger $\chi^2$.
%Standing this latest result one could deduce that the flavor dependence of our TMD model is not sufficient to explain these data and a new fit with a more elaborate  flavor dependence is justified.

%  We find a $\chi^2$ of 3.15/2 for ATLAS.
%
 
%Comparing TeVatron with LHC we find that the first has a 2 GeV $q_T$ binning, while the experiments of the second facility consider much larger intervals.  The smaller binning is certainly preferable for a  TMD analysis.
%The conclusion of this section is so similar to the one of \cite{Hautmann:2020cyp}: a different binning and set of fiducial cuts is necessary to unveil the properties of TMD at LHC, which is certainly possible in  future analysis and runs.

%%%%%%%%%%%%%%%%%%%%
\section{Summary and outlook}
\label{sec:sum}
%%%%%%%%%%%%%%%%%%%
In this work we  have set the status of the current knowledge of the $W$-boson spectrum within the TMD factorization formulation of its cross section and using the latest extraction of these distributions in  \cite{Scimemi:2019cmh}. We have considered the $W$-production differential cross section,  the $p_T^Z/p_T^W$ and the 
 $p_T^{W^+}/p_T^{W^-}$ distributions as  functions of the boson transverse momentum $q_T$ in different intervals of $m_T$.  
 The kinematic of the processes has been deeply studied, providing details that were not treated in the literature (to our knowledge), and showing explicitly how fiducial cuts are implemented together with kinematical power corrections.
 The kinematical description so achieved has put us in the condition to use the TMD extraction made in  \cite{Scimemi:2019cmh} with the same level of precision. The perturbative inputs are  N$^3$LL for the evolution and NNLO for all the rest.
   Because of our experience with $Z$-boson production we have given particular attention to values of $m_T$ below ([50,66] GeV) and around the $W$-mass ([66,99] GeV).  In fact one expects to better control the non-perturbative QCD effects once the details of the observables below the $W$-mass are also known. It is remarkable that observables at low values for $m_T$ have been poorly studied despite the fact that there is no particular suppression of the cross section and  all errors look very similar.
   
An important chapter is the error analysis. We have considered as sources of error, scale variations, replicas uncertainty within the NNPDF31 PDF set, uncertainties due to different sets of PDF and the ones due to TMD non-perturbative parameterization.
This analysis is interesting because  it gives a first step in our understanding the errors. We consider it an important starting point that  can certainly be improved with  future work within TMD studies. For instance scale uncertainties can be checked including one more perturbative order (including very recent results at N$^3$LO, \cite{Li:2016ctv,Vladimirov:2016dll,Luo:2019hmp,Luo:2019szz,Ebert:2020yqt}), which however require a new TMD extraction. Then, the sets of PDF that we have checked are the ones analyzed also in  \cite{Scimemi:2019cmh}, which is still limited. More PDF sets should be used and included in the TMD extraction.  The actual impact of each source of error depends on the observable, and we have described it in sec.~\ref{sec:Wspeck}. 
 
 At present the available data provide already some information on the observables that we have studied. We have considered the cases of  TeVatron and LHC. The agreement  in the cross section is reasonable for CDF, D$\slashed 0$ at $\sqrt{s}=1.8 $ TeV, ATLAS, while it is not completely satisfactory CMS. For all these cross sections we have done a $\chi^2$ analysis. The message is that inferring and estimating QCD non-perturbative contributions to $W$ processes from neutral boson mediated ones has some effectiveness and that a more global treatment of  data is also worth studying. This raises also the question  of whether a more sophisticated TMD flavor dependence can improve the agreement with data. Other effects like QED contributions \cite{deFlorian:2018wcj,Cieri:2018sfk} will also be addressed in future studies.
 
In order to perform this study we have proposed to consider several intervals of $m_T$, for which we do not observe a particular suppression of  the cross section. Comparing experiments, we find a
 striking difference between LHC and TeVatron data in the $q_T$ binning of the final result. Definitely the 2 GeV binning of TeVatron allows a much better understanding of the QCD effects, despite the LHC precision. A similar binning  would be also desirable for LHC. This problem has already been discussed also in the case of Drell-Yan in \cite{Hautmann:2020cyp}.
% Finally comparing with the theoretical prediction of \cite{Bizon:2019zgf} in fig.~\ref{fig:Gehrmann} we have a general agreement  except in the region fo $q_T<$  5-8 GeV, where TMD effects  are non-negligeable.

\section*{Acknowledgements}
We thank Pier Monni,  Alexey Vladimirov, Chen Wang (for D$\slashed 0$ collaboration), Aram Apyan (CMS collaboration) for valuable communications and suggestions.
D.G.R. and I.S. are supported by the Spanish Ministry grant PID2019-106080GB-C21. D.G.R. acknowledges the support of the Universidad Complutense de Madrid through the predoctoral grant CT17/17-CT18/17.
This project has received funding from the European Union Horizon 2020 research and innovation program under grant agreement Num. 824093 (STRONG-2020). S.L.G. is supported by the Austrian Science Fund FWF under the Doctoral Program W1252-N27 Particles and Interactions. 

\appendix

\section{Explicit expression for the lepton tensor integration}
\label{app:iw}

The integral involved in the definition of the lepton tensor in \eq{LcutsW} is
\begin{align}
\label{P1}
I_W(Q^2,m_T^2,q_T)=\int \frac{d^3l}{2E}\frac{d^3l'}{2E'} \, \delta ^{(4)}(l+l'-q)\, \delta(Q^2-m_T^2-f(l,l'))\theta(\text{cuts})\[ll'-(ll')_T\].
\end{align}

Integrating the momentum of the neutrino with the help of the conservation momentum delta function 
we can rewrite eq.~\eqref{P1} as
\begin{align}
\label{eq:P2}
P_W(Q^2,m_T^2,q_T)=\int \frac{d^3l}{2E}  \delta((q-l)^2)\, \delta(Q^2-m_T^2-f(l,q))\theta(\text{cuts})\[l(q-l)-l_T(q_T-l_T)\].
\end{align}

We profit of the extra delta function $\delta(Q^2-m_T^2-f(l,q))$ to integrate over $\eta$. To find the result of this integration we should rewrite the function $f(l,q)$ in terms of the variables of the problem
\begin{align}
f(\eta,l_T)=2l_T \sqrt{Q^2+q_T^2} \cosh (\eta-y)-2l_T^2 -2l_T \sqrt{q_T^2+l_T ^2-2 q_T l_T \cos \phi},
\end{align}
where $\phi$ is the angle between transverse momenta of lepton and $W$ boson.

Thus, the delta function can be written as
\begin{align}
\label{eq:condeta}
\delta(Q^2-&m_T^2-f(l,q))= \frac{1}{2l_T \sqrt{Q^2+q_T^2}} \\
&\times \delta \[\cosh (\eta-y)-\(\frac{Q^2-m_T^2}{2l_T \sqrt{Q^2+q_T^2}} + \frac{l_T+\sqrt{q_T^2+l_T^2-2 q_T l_T \cos \phi}}{\sqrt{Q^2+q_T^2}}\)\] \nn,
\end{align}
giving a simple condition to perform the desired integral. Note that to do the integral we should rewrite the integral as
\begin{align}
\label{eq:jac}
\frac{d^3l}{2E}= \frac{dl^z}{2E} l_T dl_T d\phi= \frac{d\eta}{2} l_T dl_T d\phi=\frac{d \cosh (\eta-y)}{2\sqrt{\cosh^2(\eta-y)-1}} l_T d l_T d\phi.
\end{align}
Using the condition in \eq{condeta} we can rewrite the delta function $\delta ((q-l)^2)$ of \eq{P2} to obtain a condition that allows to make the integral over $l_T$ 
\begin{align}
\delta ((q-l)^2)=\delta \[m_T^2-2l_T^2+2l_T q_T \cos \phi -2 l_T \sqrt{l_T^2+q_T^2-2 q_T l_T \cos \phi}\].
\end{align}

To rewrite this delta we should find the zeros of its argument. We find
\begin{align}
l_T^{+}=\frac{m_T^2}{2} \frac{q_T \cos \phi+\sqrt{m_T^2+q_T^2}}{m_T^2+q_T^2(1-\cos^2 \phi)} >0, 
\end{align}
\begin{align}
l_T^{-}=\frac{m_T^2}{2} \frac{q_T \cos \phi-\sqrt{m_T^2+q_T^2}}{m_T^2+q_T^2(1-\cos^2 \phi)} <0, 
\end{align}
so the solution $l_T^-$ is never used because we imposed that $l_T>0$. Thus the delta function over $l_T$ can be rewritten as
\begin{align}
\delta ((q-l)^2)=A(Q^2,m_T^2,q_T,\cos \phi) \delta (l_T-l_T^+),
\end{align}
where 
\begin{align}
A(Q^2,m_T^2,q_T,\cos \phi)=\Bigg | \frac{\sqrt{q_T^2+l_T^2-2 q_T l_T \cos \phi}}{2(l_T+\sqrt{q_T^2+l_T^2-2 q_T l_T \cos \phi})(l_T-q_T\cos \phi + \sqrt{q_T^2+l_T^2-2 q_T l_T \cos \phi})}\Bigg |_{l_T=l_T^+}
\end{align}

For simplicity we can rewrite the term coming from the Jacobian in \eq{jac} as
\begin{align}
J(Q^2,q_T,\cos \phi)&=\frac{1}{\sqrt{\cosh^2(\eta-y)-1}}\\
&=\[-1+\(\frac{Q^2-m_T^2}{2l_T\sqrt{Q^2+q_T^2}}+\frac{l_T+\sqrt{q_T^2+l_T^2-2 q_T l_T \cos \phi}}{\sqrt{Q^2+q_T^2}}\)^2\]^{-1/2}. \nn
\end{align}

Finally the term that only depends on momenta in \eq{P2} is rewritten as
\begin{align}
M(Q^2,m_T^2,q_T,\cos \phi)&=l(q-l)-l_T(q_T-l_T)\\
&=\frac{Q^2-m_T^2}{2}+2l_T^2+l_T \sqrt{q_T^2+l_T^2-2 q_T l_T \cos \phi}-2q_T l_T \cos \phi. \nn
\end{align}

Thus, the final result is written in term of the auxiliary functions $J, A, M$ as a one-dimensional integral
\begin{align}
\label{P2f}
I_W(Q^2,m_T^2,q_T)&=\int_0^{2\pi} d\phi \frac{1}{4l_T\sqrt{Q^2+q_T^2}}\\
&\times J(Q^2,q_T,\cos \phi) A(Q^2,m_T^2,q_T,\cos \phi) M(Q^2,m_T^2,q_T,\cos \phi)\theta({\rm cuts}), \nn
\end{align}
that should be done numerically.

%%%%%%%%%%%%%%%%%%%%%%%
\section{Inputs from fits}
\label{sec:inputs}
%%%%%%%%%%%%%%%%%%%%%%%%%%%%%%%%%%

In this section we report some results from previous fit \cite{Scimemi:2019cmh}  that we have used in this paper.
The different PDF sets that have been reported in that fit and that we have used are listed in
tab.~\ref{tab:PDFs}. The results for the TMD constants are in tab.~\ref{tab:DY-PDF-chi}.

\begin{table}[t]
\begin{center}
\begin{tabular}{|l|l|c|c|}\hline
Short name & Full name & Ref. & LHAPDF id.
\\\hline\hline
NNPDF31 & NNPDF31\_nnlo\_as\_0118 & \cite{Ball:2017nwa} &303600 
\\\hline
HERA20 & HERAPDF20\_NNLO\_VAR & \cite{Abramowicz:2015mha} &61230 
\\\hline
MMHT14 & MMHT2014nnlo68cl & \cite{Harland-Lang:2014zoa} &25300
\\\hline
CT14 & CT14nnlo & \cite{Dulat:2015mca} & 13000
\\\hline
PDF4LHC & PDF4LHC15\_nnlo\_100 &  \cite{Butterworth:2015oua} & 91700
%\\\hline
%ABMP16 & ABMP16\_5\_nnlo & \cite{Alekhin:2017kpj} & 42560
\\\hline
\end{tabular}
\end{center}
\caption{\label{tab:PDFs} List of collinear PDF used as the boundary for unpolarized TMDPDF.}
\end{table}

\begin{table}[b]
\begin{center}
\small
\begin{tabular}{|l V{4} c V{4} l|ll|}
\hline
PDF set & $\chi^2/N_{pt}$ & Parameters for $\mathcal{D}$& Parameters for $f_1$ &
\\\hline\hline
HERA20 & 0.97 & \specialcellleft{$B_\text{NP}=2.29\pm 0.43$ \\ $c_0=(2.22\pm0.93)\cdot 10^{-2}$} 
& \specialcellleft{$\lambda_1=0.324\pm0.029$ \\ $\lambda_2=13.2\pm2.9$}
& \specialcellleft{$\lambda_3=(3.56\pm 1.59)\cdot 10^2$ \\ $\lambda_4=2.05\pm0.26$ \\ $\lambda_5=-10.4\pm 3.5$}
\\\hline
%HERA20(N$^3$LO) & 1.06 & \specialcellleft{$B_\text{NP}=1.94\pm 0.41$ \\ $c_0=(3.35\pm0.68)\cdot 10^{-2}$} 
%& \specialcellleft{$\lambda_1=0.326\pm0.024$ \\ $\lambda_2=10.1\pm1.6$}
%& \specialcellleft{$\lambda_3=(2.73\pm 0.91)\cdot 10^2$ \\ $\lambda_4=1.70\pm0.19$ \\ $\lambda_5=-6.5\pm 2.4$}
%\\\hline
%NNPDF31(N$^3$LO) & 1.13 & \specialcellleft{$B_\text{NP}=1.62\pm 0.24$ \\ $c_0=(3.42\pm1.04)\cdot 10^{-2}$} 
%& \specialcellleft{$\lambda_1=0.282\pm0.017$ \\ $\lambda_2=9.7\pm1.3$}
%& \specialcellleft{$\lambda_3=(3.17\pm 0.83)\cdot 10^2$ \\ $\lambda_4=2.42\pm0.13$ \\ $\lambda_5=-6.1\pm 1.6$}
%\\\hline
NNPDF31 & 1.14 & \specialcellleft{$B_\text{NP}=1.86\pm 0.30$ \\ $c_0=(2.96\pm1.04)\cdot 10^{-2}$} 
& \specialcellleft{$\lambda_1=0.253\pm0.032$ \\ $\lambda_2=9.0\pm3.0$}
& \specialcellleft{$\lambda_3=(3.47\pm 1.16)\cdot 10^2$ \\ $\lambda_4=2.48\pm0.15$ \\ $\lambda_5=-5.7\pm 3.4$}
\\\hline
MMHT14 & 1.34 & \specialcellleft{$B_\text{NP}=1.55\pm 0.29$ \\ $c_0=(4.70\pm1.77)\cdot 10^{-2}$} 
& \specialcellleft{$\lambda_1=0.198\pm0.040$ \\ $\lambda_2=26.4\pm4.9$}
& \specialcellleft{$\lambda_3=(26.8\pm 13.2)\cdot 10^3$ \\ $\lambda_4=3.01\pm0.17$ \\ $\lambda_5=-23.4\pm 5.4$}
\\\hline
PDF4LHC & 1.53 & \specialcellleft{$B_\text{NP}=1.93\pm 0.47$ \\ $c_0=(3.66\pm2.09)\cdot 10^{-2}$} 
& \specialcellleft{$\lambda_1=0.218\pm0.041$ \\ $\lambda_2=17.9\pm4.5$}
& \specialcellleft{$\lambda_3=(9.26\pm 8.38)\cdot 10^2$ \\ $\lambda_4=2.54\pm0.17$ \\ $\lambda_5=-15.5\pm 4.7$}
\\\hline
CT14 & 1.59 & \specialcellleft{$B_\text{NP}=2.35\pm 0.61$ \\ $c_0=(2.27\pm1.33)\cdot 10^{-2}$} 
& \specialcellleft{$\lambda_1=0.277\pm0.029$ \\ $\lambda_2=24.9\pm2.9$}
& \specialcellleft{$\lambda_3=(12.4\pm 3.2)\cdot 10^3$ \\ $\lambda_4=2.67\pm0.13$ \\ $\lambda_5=-23.8\pm 2.9$}
\\\hline
\end{tabular}
\end{center}
\caption{\label{tab:DY-PDF-chi} Values of $\chi^2$ and NP parameters obtained in the fit of DY set of the data with different PDF inputs. Each set of PDF provide the corresponding  value of $\alpha_s(M_Z)$.}
\end{table}

\section{Errors  propagation from TMD parameters}
\label{app:errors}

An alternative way to present the errors coming from non-perturbative parametrization of the TMD is the following.
We take from \cite{Scimemi:2019cmh} the different values for the non-pertubative parameters coming from fits using several PDF sets. We have a data array of seven non-pertubative parameters and five values for each one with its respective uncertainty as it is shown in table \ref{tab:DY-PDF-chi}. We calculate the mean value of each parameter and its uncertainty associated doing the square sum of the uncertainties and taking the square-root of the sum. We show the values in table~\ref{tab:lerr}. In order to quantify the uncertainty in our cross section due to the uncertainties in our non-pertubative parameters we use \eq{Uncdsigma} 

\begin{eqnarray}
\label{eq:Uncdsigma}
\begin{aligned}
\Delta\left(\dfrac{d\sigma}{dq_T}\right)&=\sqrt{\sum_{ij}\dfrac{\partial}{\partial \theta_i}\left(\dfrac{d\sigma}{dq_T}\right)\bigg|_{\theta_i=\bar{\theta}_i}\dfrac{\partial}{\partial\theta_j}\left(\dfrac{d\sigma}{dq_T}\right)\bigg|_{\theta_j=\bar{\theta}_j}V_{ij}}\\
&\simeq\dfrac{1}{2}\sqrt{\sum_{ij}\delta\left(\dfrac{d\sigma}{dq_T}\right)_i\delta\left(\dfrac{d\sigma}{dq_T}\right)_jC_{ij}},
\end{aligned}
\end{eqnarray}
where $\theta_i=\{\text{B}_{\text{NP}},\text{c}_0,\lambda_1,\lambda_2,\lambda_3,\lambda_4,\la_5\}$ are the non-perturbative parameters and $\bar{\theta}_i$ their mean values collected in table \ref{tab:DY-PDF-chi}; $V_{ij}$ and $C_{ij}$ are the covariance and correlation matrices respectively of our non-perturbatibe parameters and $\delta\left(\frac{d\sigma}{dq_T}\right)_i$ is the difference of the differential cross section evaluated in the extremes of the interval of the non-perturbative parameter $i$ and keeping the rest of non-perturbative parameters fixed at their mean value. In order to go from the first line to the second line, we have approximated the derivative of the cross section with respect to $\theta_i$ by  
\begin{eqnarray}
\dfrac{\partial}{\partial \theta_i}\left(\dfrac{d\sigma}{dq_T}\right)\bigg|_{\theta_i=\bar{\theta}_i}\simeq\dfrac{1}{2\delta\bar{\theta}_i}\left(\dfrac{d\sigma}{dq_T}\left(\bar{\theta}_i+\delta\bar{\theta}_i,\bar{\theta}_{j\neq i}\right)-\dfrac{d\sigma}{dq_T}\left(\bar{\theta}_i-\delta\bar{\theta}_i,\bar{\theta}_{j\neq i}\right)\right),
\end{eqnarray}
where $\delta\bar{\theta}_i$ is the uncertainty associated to the mean value of the non-perturbative parameter $\theta_i$. Using the relation between the correlation and covariance matrices $C_{ij}=\frac{V_{ij}}{\delta\bar{\theta}_i\delta\bar{\theta}_j}$ we arrive to \eq{Uncdsigma}. 

The correlation matrix at NNLO for the $\lambda_i$ isprovided in \cite{Scimemi:2019cmh} and, as a reference, we have considered  the central PDF of NNPDF31\_nnlo\_as\_0118~\cite{Ball:2017nwa}.

\begin{table}[h]
\begin{center}
\begin{tabular}{||c|c|c|c|c|c|c||}\hline
$B_{NP}$ & $c_0$ & $\lambda_1$& $\lambda_2$& $\lambda_3$& $\lambda_4$& $\lambda_5$\\\hline
2.01$\pm$0.19 & $\left(\text{3.42}\pm\text{0.70}\right)\cdot\text{10}^{\text{-2}}$& 0.248$\pm$0.015& 18.3$\pm$1.5& $\left(\text{8.2}\pm\text{2.7}\right)\cdot\text{10}^{\text{2}}$& 2.484$\pm$0.085& -15.6$\pm$1.7\\
\hline
\end{tabular} 
\caption{\label{tab:lerr}Intervals for the  TMD non-perturbative parameters calculated as explained in the text.}
\end{center}
\end{table}

The result of this error estimate is very similar to the one described in the main part of the paper, which is a confirmation of the Gaussian  distribution of TMD parameter errors for this case.

%***************************************************************************************************%
\cleardoublepage
\bibliographystyle{JHEP}
\normalbaselines %Fixes spacing of bibliography
%\addcontentsline{toc}{chapter}{Bibliography} %adds Bibliography to your table of contents
%\bibliography{notes} %your bibliography file - change the path if needed
%***************************************************************************************************%
\bibliography{TMD_ref}
\end{document}